\documentclass[aps,floats]{revtex4-2}
\usepackage{amsmath,amssymb}
\usepackage{graphicx,epsfig}
\usepackage[greek,english]{babel}
\usepackage{bbold}

\usepackage{amsfonts}

\usepackage{mathtools}
\usepackage{makecell,tabularx}
\setcellgapes{1pt}
\makegapedcells
 
\begin{document}
\bibliographystyle {plain}

\pdfoutput=1
\def\oppropto{\mathop{\propto}} 
\def\opsimeq{\mathop{\simeq}}
\def\opoverderline{\mathop{\overline}}
\def\operarrow{\mathop{\longrightarrow}}
\def\opsim{\mathop{\sim}}

\def\opmin{\mathop{\min}} 
\def\opmax{\mathop{\max}} 
\def\oplim{\mathop{\lim}}

%%%%%%%%%%%%%%%%%%%%%%%%%%%%%%%%%%%%%%%%%%%%%%%%%%%%%%%%%%%%%%%%%%%%%%%%%%%%
\title{ Large deviations for the Pearson family of ergodic diffusion processes 
 \\ involving a quadratic diffusion coefficient and a linear force } 

%%%%%%%%%%%%%%%%%%%%%%%%%%%%%%%%%%%%%%%%%%%%%%%%%%%%%%%%%%%%%%%%%%%%%%%%%%%%

\author{C\'ecile Monthus}
\affiliation{Universit\'e Paris-Saclay, CNRS, CEA, Institut de Physique Th\'eorique, 91191 Gif-sur-Yvette, France}

%%%%%%%%%%%%%%%%%%%%%%%%%%%%%%%%%%%%%%%%%%%%%%%%%%%%%%%%%%%%%%%%%%%%%%%%%%%%

\begin{abstract}

The Pearson family of ergodic diffusions with a quadratic diffusion coefficient and a linear force are characterized by explicit dynamics of their integer moments and by explicit relaxation spectral properties towards their steady state. Besides the Ornstein-Uhlenbeck process with a Gaussian steady state, the other representative examples of the Pearson family are the Square-Root or the Cox-Ingersoll-Ross process converging towards the Gamma-distribution, the Jacobi process converging towards the Beta-distribution, the reciprocal-Gamma process (corresponding to an exponential functional of the Brownian motion) that converges towards the Inverse-Gamma-distribution, the Fisher-Snedecor process, and the Student process, so that the last three steady states display heavy-tails. The goal of the present paper is to analyze the large deviations properties of these various diffusion processes in a unified framework. We first consider the level 1 concerning time-averaged observables over a large time-window $T$ :  we write the first rescaled cumulants for generic observables and we identify the specific observables whose large deviations can be explicitly computed from the dominant eigenvalue of the appropriate deformed-generator. The explicit large deviations at level 2 concerning the time-averaged density are then used to analyze the statistical inference of model parameters from data on a very long stochastic trajectory in order to obtain the explicit rate function for the two inferred parameters of the Pearson linear force.

\end{abstract}

\maketitle

%%%%%%%%%%%%%%%%%%%%%%%%%%%%%%%%%%%%%%%%%%%%%%%%

\section{Introduction   }

The Pearson family of ergodic diffusions (see  \cite{pearson_wong,diaconis,autocorrelation,pearson_class,pearson2012,PearsonHeavyTailed,pearson2018}
and references therein) contains the one-dimensional diffusion processes
 with a quadratic diffusion coefficient $D(x)$ and a linear force $F(x)$. 
Besides their simple steady states introduced by Pearson in 1895
in order to have a simple family of histograms to analyze non-gaussian biological data \cite{pearson1895},
Pearson diffusions enjoy very specific dynamical properties,
in particular the explicit dynamics of their integer moments and
the explicit spectral decomposition of their propagators.
Via changes of variables, they actually encompass 
most of the ergodic diffusions with explicit relaxation spectra. 

The goal of the present paper is to study whether Pearson diffusions also display specific properties
from the point of view of their large deviation properties. Indeed, the theory of large deviations 
 (see the reviews \cite{oono,ellis,review_touchette} and references therein)
 has become the unifying language of statistical physics, in particular in the field of nonequilibrium 
(see the reviews with different scopes \cite{derrida-lecture,harris_Schu,searles,harris,mft,sollich_review,lazarescu_companion,lazarescu_generic,jack_review}, 
the PhD Theses \cite{fortelle_thesis,vivien_thesis,chetrite_thesis,wynants_thesis,chabane_thesis,duBuisson_thesis} 
 and the Habilitation Thesis \cite{chetrite_HDR}).
In particular, the statistics of time-averaged observables over a large time-window $T$
have been studied for many Markov processes via the deformed-Markov-generator approach
 \cite{peliti,derrida-lecture,sollich_review,lazarescu_companion,lazarescu_generic,jack_review,vivien_thesis,lecomte_chaotic,lecomte_thermo,lecomte_formalism,lecomte_glass,kristina1,kristina2,jack_ensemble,simon1,simon2,tailleur,simon3,Gunter1,Gunter2,Gunter3,Gunter4,chetrite_canonical,chetrite_conditioned,chetrite_optimal,chetrite_HDR,touchette_circle,touchette_langevin,touchette_occ,touchette_occupation,garrahan_lecture,Vivo,c_ring,c_detailed,chemical,derrida-conditioned,derrida-ring,bertin-conditioned,touchette-reflected,touchette-reflectedbis,c_lyapunov,previousquantum2.5doob,quantum2.5doob,quantum2.5dooblong,c_ruelle,lapolla,c_east,chabane,us_gyrator,duBuisson_gyrator} with the construction of the corresponding Doob's conditioned process.
Since these large deviations at the level 1 concerning time-averaged observables 
 are unfortunately not always explicit,
it is interesting to consider the large deviations at higher levels in order to obtain explicit rate functions.
The level 2 for the empirical density over a large time-window $T$
is explicit for Markov processes with vanishing steady currents satisfying detailed-balance.
For non-equilibrium steady-states with non-vanishing steady currents breaking detailed-balance,
the level 2.5 concerning the joint distribution of the empirical density and the empirical flows
is the appropriate level 
where rate functions can be written explicitly for discrete-time Markov chains 
 \cite{fortelle_thesis,fortelle_chain,review_touchette,c_largedevdisorder,c_reset},
for continuous-time Markov jump processes 
\cite{fortelle_thesis,fortelle_jump,maes_canonical,maes_onandbeyond,wynants_thesis,chetrite_formal,BFG1,BFG2,chetrite_HDR,c_ring,c_interactions,c_open,barato_periodic,chetrite_periodic,c_reset}
and for diffusion processes 
\cite{wynants_thesis,maes_diffusion,chetrite_formal,engel,chetrite_HDR,c_reset,c_lyapunov}.
An interesting direct application of these explicit large deviations at higher levels is
 the statistical inference of model parameters from data on a very long
trajectory \cite{c_inference}.

The paper is organized as follows.
After the introduction of the general notations for Pearson diffusions in section \ref{sec_nota},
we emphasize their very specific properties from three complementary perspectives,
namely for the dynamics of the integer moments in section \ref{sec_mk},
for the spectral decomposition of the propagators in section \ref{sec_spectral}, 
and for the associated quantum supersymmetric Hamilonians in section \ref{sec_susy}.
In section \ref{sec_mapping}, we mention how the Pearson diffusions 
can be mapped onto other interesting diffusion processes
with additive or multiplicative noise that inherit their nice properties after appropriate translation.
We then study in detail the large deviations properties at various levels for Pearson diffusions.
In section \ref{sec_level1}, the statistics of various time-averaged observables over the time-window $[0,T]$
are analyzed via the large deviations at level 1 : we first write the first rescaled cumulants for generic observables
and we then determine the specific observables characterized by explicit large deviations.
In section \ref{sec_level2}, the explicit large deviations at level 2 for the empirical density 
seen during a large time-window $[0,T]$ are used to analyze
the statistical inference of the Pearson parameters
from the data of a long stochastic trajectory.
Finally, this general framework is applied to the five representative examples of Pearson diffusions
with linear or quadratic diffusion coefficient $D(x)$,
namely the Square-Root or the Cox-Ingersoll-Ross process converging to the Gamma-distribution in section  \ref{sec_gamma}, the reciprocal-Gamma process converging towards the Inverse-Gamma-distribution in section \ref{sec_kesten},
 the Fisher-Snedecor process in section \ref{sec_fisher},
the Jacobi process converging to the Beta-distribution in section \ref{sec_jacobi},
and the Student process in section \ref{sec_student}.
Our conclusions are summarized in section \ref{sec_conclusion}.

%%%%%%%%%%%%%%%%%%%%%%%%%%%%%%%%%%%%%%%%%%%%%%%%

\section{ General notations for the Pearson family of diffusion processes   }

\label{sec_nota}

In the whole paper, it will be essential to stress the very specific properties of Pearson diffusions
with respect to other ergodic one-dimensional diffusion processes
on intervals $ ]x_L,x_R[$ with vanishing-current boundary conditions.
So let us first introduce the notations for this general diffusion before the description of 
the additional specific properties of the Pearson family.

\subsection{ Ergodic diffusion process on the interval $ ]x_L,x_R[$ with vanishing-current boundary conditions}

\label{subsec_geneDiff}

The Fokker-Planck dynamics 
for the probability distribution $P_t(x)$ to be at position $x$ at time $t$
corresponds to
the continuity equation
\begin{eqnarray}
 \partial_t P_t(x)    && =   -  \partial_{x}   J_t(x)
  \nonumber \\
 J_t(x) && \equiv F(x)   P_t(x) - D (x)  \partial_{x} P_t(x)
\label{fokkerplanck}
\end{eqnarray}
 where the current $J_t(x) $ associated to $P_t(x) $
involves the Fokker-Planck force $F(x) $ and the diffusion coefficient $D(x)$.

\subsubsection{ Discussion of the vanishing-current boundary conditions 
at the boundaries $x_L$ and $x_R$ of the interval $ ]x_L,x_R[ $  }

The conservation of the total probability $\int_{x_L}^{x_R} dx P_t(x)=1 $ on the interval $ ]x_L,x_R[$
\begin{eqnarray}
0= \partial_t \int_{x_L}^{x_R} dx P_t(x)    && =   - \int_{x_L}^{x_R} dx  \partial_{x}   J_t(x)
= - \left[J_t(x) \right]_{x=x_L}^{x=x_R} = J_t(x_L)-J_t(x_R)
\label{conservnorma}
\end{eqnarray}
can be satisfied via two types of boundary conditions:

(1) the case of periodic boundary conditions, where $x_L$ and $x_R$ are identified $x_L \equiv x_R$,
and where Eq. \ref{conservnorma} is thus automatically satisfied,
corresponds to the geometry of a one-dimensional ring with no real physical boundaries.
Diffusion processes on periodic rings
have been much studied recently as the simplest geometry where a non-equilibrium steady current is possible
(see \cite{derrida-ring,bertin-conditioned,c_lyapunov,us_kemeny}  and references therein),
but will not be discussed further here, since Pearson diffusions are not defined on a periodic ring.

(2) the case of vanishing-current boundary conditions at the two boundaries $x_L$ and $x_R$
\begin{eqnarray}
J_t(x_L) && =0
\nonumber \\
J_t(x_R) && =0
\label{jboundaries}
\end{eqnarray}
is usually called 'reflecting boundary conditions' although this vocabulary can be somewhat misleading,
since Eq. \ref{jboundaries} can correspond to very different physical situations :

(2-a) The terminology  'reflecting boundary conditions' seems appropriate
when the diffusive particle is really able to reach the boundaries at the finite positions $x_L$ and $x_R$ 
and is then prevented from leaving the interval $]x_L,x_R[$ only by a true 'action' of the physical walls at the boundaries :
 the simplest example is the free Brownian particle on a finite interval with reflecting walls.
 
(2-b)  When the boundaries are at infinities $x_L=-\infty$ and $x_R=+\infty$,
 as for instance for the Ornstein-Uhlenbeck process defined on the full line $]-\infty,+\infty[$,
 the terminology 'reflecting boundary conditions' does not seem appropriate, since there are no reflecting physical walls.
 
(2-c) When the boundary $x_L$ or $x_R$ is finite but cannot be really reached by the diffusive particle
as a consequence of the specific forms of the force $F(x)$ and of the diffusion coefficient $D(x)$ on the interval,
as in some Pearson examples that will be described later,
the terminology  'reflecting boundary conditions' does not seem very appropriate either.

In order to cover these various possibilities (2-a) (2-b) (2-c) that will occur in Pearson diffusions,
we will thus use the terminology "vanishing-current boundary conditions"
for Eq. \ref{jboundaries} that are the only boundary conditions that will be considered in the present paper.

%%%%%%%%%%%%%%%%%%%%%%%%%%%%%%%%%%%%%%%%%%%%%%%%%

\subsubsection{ Steady state $P_*( x) $ associated to the vanishing steady current $ J_*(x) =0$ (Detailed-Balance)  }

In the steady version of the Fokker-Planck Eq. \ref{fokkerplanck},
the steady-current $J_*(x)$ satisfying $\partial_{x}   J_*(x) =0$ cannot depend on $x$
 and should vanish at the two boundaries (Eq. \ref{jboundaries}),
so that it vanishes identically on the whole interval $ x \in ]x_L,x_R[$
\begin{eqnarray}
0= J_*(x) =  F(x)   P_*( x ) - D (x)  \frac{d P_*( x) }{dx}
\label{zFPJsteady}
\end{eqnarray}
i.e. one cannot avoid the detailed-balance.
The normalized steady state $P_*(x)$ can be written as the Boltzmann distribution 
\begin{eqnarray}
  P_*(x)  = \frac{ e^{ -U(x)} }{Z}
 \label{steadyeq}
\end{eqnarray}
in the potential $U(x)$ determined by the ratio of the force $F(x)$ and of the diffusion coefficient $D(x)$
\begin{eqnarray}
U'(x) && = - \frac{F(x)}{D(x)}
\nonumber \\
U(x) && \equiv - \int_{x_{ref}}^x dy \frac{F(y)}{D(y)} 
 \label{Ux}
\end{eqnarray}
where $x_{ref} \in ]x_L,x_R[$ is some reference position,
while the normalization $Z$ corresponds to the partition function of the interval $ ]x_L,x_R[$ 
\begin{eqnarray}
Z=  \int_{x_L}^{x_R} dx e^{ -  U(x) } 
 \label{partitioneq}
\end{eqnarray}

%%%%%%%%%%%%%%%%%%%%%%%%%%%%%%%%%%%%%%%%%%%%%%%%%

\subsubsection{ Fokker-Planck generator ${\cal L}_x $ and its adjoint operator ${\cal L}^{\dagger}_x $  }

Using the potential $U(x)$ of Eq. \ref{Ux}, the differential operator $ {\cal L}_x$ that 
governs the 
Fokker-Planck Eq. \ref{fokkerplanck}
\begin{eqnarray}
 \partial_t P_t(x)  ={\cal L}_x P_t(x)
\label{FPgenerator}
\end{eqnarray}
 reads 
\begin{eqnarray}
{\cal L}_x
  =     \partial_{x}  \bigg( - F(x)    + D (x)  \partial_{x} \bigg)  
  = \partial_{x} \bigg[ D(x) \bigg(  U'(x)     +   \partial_{x} \bigg)  \bigg]
\label{generator}
\end{eqnarray}
while the adjoint operator is given by
\begin{eqnarray}
  {\cal L}^{\dagger}_x 
 \equiv   F(x)  \partial_{x}    +   \partial_{x} \bigg( D (x)  \partial_{x}   \bigg)
 = \bigg( - U'(x)  +   \partial_{x} \bigg) D(x) \partial_{x}  
\label{adjoint}
\end{eqnarray}

%%%%%%%%%%%%%%%%%%%%%%%%%%%%%%%%%%%%%%%%%%%%%%%%%

\subsubsection{ Langevin Stochastic Differential Equations associated to the Fokker-Planck dynamics  }

The Fokker-Planck dynamics of Eq. \ref{FPgenerator}
is associated to the following
Langevin Stochastic Differential Equations involving a Brownian motion $dB(t)$
\begin{eqnarray}
dx(t) &&=  F_I( x (t) ) \ dt + \sqrt{ 2 D ( x (t) ) } \ dB(t)
\ \ \ \ \ \ \ \ \ \ \ \ \ \ [{\rm Ito \ Interpretation}]
\nonumber \\
dx(t) && =  F_S( x (t) ) \ dt + \sqrt{ 2 D ( x (t) ) }  \ dB(t)
\ \ \ \ \ \ \ \ \ \ \ \ \ \ [{\rm Stratonovich \ Interpretation}]
 \label{langevin}
\end{eqnarray}
where the force depends on the interpretation whenever the diffusion coefficient $D(x)$ depends on the position $x$ :
the Ito force $F_I(x) $ and the Stratonovich force $ F_S(x) $ 
can be computed in terms of the Fokker-Planck force $F(x)$ and in terms of the derivative of the diffusion coefficient $D(x)$ via
\begin{eqnarray}
    F_I(x) && = F(x)+ D' (x)
\nonumber \\
   F_S(x) && = F(x)+  \frac{D' (x)}{2} 
\label{fokkerplancklangevin}
\end{eqnarray}

The Ito force $F_I(x)  $ is especially useful when one wishes 
to rewrite the adjoint operator of Eq. \ref{adjoint} with all the derivatives on the right as
\begin{eqnarray}
  {\cal L}^{\dagger}_x 
=   F(x)  \partial_{x}    +   \partial_{x} \bigg( D (x)  \partial_{x}   \bigg)
= \bigg(F(x)+ D' (x) \bigg) \partial_x  + D(x)  \partial_{x}^2 
\equiv    F_I(x) \partial_x  + D(x)  \partial_{x}^2
\label{adjointIto}
\end{eqnarray}
while the Stratonovich force $F_S(x)  $ is more convenient when one makes changes
of variables in the Langevin dynamics, since one can use the standard rules of calculus
(instead of the specific Ito rules of calculus if one uses the Ito force).

%%%%%%%%%%%%%%%%%%%%%%%%%%%%%%%%%%%%

\subsection{ Pearson diffusions : linear force $F(x)$ and quadratic $D(x) $
on the maximal interval  $]x_L,x_R[$ where $D(x) \geq 0$  }

\subsubsection{ General form of Pearson diffusions   }

The Pearson family of diffusion processes 
is characterized by a positive quadratic diffusion coefficient $D(x) $
and by a linear force $F(x)$ in the Fokker-Planck dynamics of Eq. \ref{fokkerplanck}
\begin{eqnarray}
 D(x) && =a x^2+b x +c \geq 0
\nonumber \\
F(x)  && = \lambda - \gamma x 
\label{fpearson}
\end{eqnarray}
Their ratio directly determines the derivative $U'(x)$ of the potential $U(x)$ of Eq. \ref{Ux}
\begin{eqnarray}
U'(x) && = - \frac{F(x)}{D(x)} = \frac{  \gamma x- \lambda }{a x^2+b x +c}
\label{uprimepearson}
\end{eqnarray}
 that governs the steady state $P_*(x)$ of Eq. \ref{steadyeq}.
The Ito force $F_I(x)$ and the Stratonovich forces $F_S(x)$ of Eq. \ref{fokkerplancklangevin}
are also linear but they involve different 
coefficients than the Fokker-Planck force $F(x)$ 
\begin{eqnarray}
    F_I(x) && = F(x)+ D' (x) = (\lambda+b) -  (\gamma-2a) x \equiv  \lambda_I - \gamma_I x 
\nonumber \\
   F_S(x) && = F(x)+  \frac{D' (x)}{2} = (\lambda +\frac{b}{2})  -  (\gamma-a) x \equiv  \lambda_S - \gamma_S x 
\label{fokkerplancklangevinpearson}
\end{eqnarray}

In addition, the Pearson diffusion processes are defined on the maximal interval $]x_L,x_R[$
where the quadratic diffusion coefficient $D(x)$ remains positive $D(x) \geq 0$,
so that when the boundaries $x_L$ and/or $x_R$ are not infinite, 
the diffusion coefficient vanishes at the boundaries
\begin{eqnarray}
    x_R=+\infty \ \ \ {\rm or } \ \ x_R \ {\rm finite \ with} \ \ D(x_R)=0 
\nonumber \\
   x_L=-\infty \ \ \ {\rm or } \ \ x_L \ {\rm finite \ with} \ \ D(x_L)=0 
\label{Dvanishboundaries}
\end{eqnarray}
It should be stressed that this choice is not a detail 
and is actually essential for the most important dynamical properties of Pearson diffusions 
as will be recalled in the three following sections.

\subsubsection{ The six representative examples of the Pearson family   }

%%%%%%%%%%%%%%%%%%%%%%%%%%%%%%%

\begin{table}[!h]
\setcellgapes{3pt}
\begin{tabular}{|p{7cm}|c|c|}
\hline
Normalized steady state $P_*(x)$ for $x \in ]x_L,x_R[$
& Diffusion Coefficient $D(x)=a x^2+b x +c$   
& Linear Force $F(x) =\lambda - \gamma x$ 
\\
\hline  
(ii)  Gamma-distribution for $x \in ]0,+\infty[ \ \ \ \ \ \ \ \ \ \ \ \ $
  $P_*(x) = \frac{\gamma^{\alpha}}{ \Gamma(\alpha)} x^{\alpha-1}e^{- \gamma x} 
  \ \ \ \ \ \ \ \ \ \ \ \ \ \ \ \ \ \ \ \ \ \ \ \ \ \ \ \ \ \ \ \ \ \ \ \ \ \ \ \ $
 with parameters $\alpha>0$ and $\gamma>0$  
&  $D(x)=x$
& $F(x)  = (\alpha-1) - \gamma x $
 \\
\hline 
(iii)  Inverse-Gamma-distribution for $x \in ]0,+\infty[$ 
  $P_*(x) = \frac{\lambda^{\mu}}{\Gamma(\mu) x^{1+\mu} } e^{- \frac{\lambda}{x}} 
   \ \ \ \ \ \ \ \ \ \ \ \ \ \ \ \ \ \ \ \ \ \ \ \ \ \ \ \ \ \ \ \ \ \ \ \ \ \ \ \ $
with parameters $\lambda>0$ and $\mu>0$  
&  $D(x)=x^2$
& $F(x)  =  \lambda - (\mu+1)  x $
  \\
\hline
(iv)  Fisher-Snedecor-distribution for $x \in ]0,+\infty[$  
$P_*(x) = \frac{\Gamma(\alpha+\mu)}{ \Gamma(\alpha) \Gamma(\mu) } \frac{ x^{\alpha-1} }{(1+x)^{\alpha+\mu} }
 \ \ \ \ \ \ \ \ \ \ \ \ \ \ \ \ \ \ \ \ \ \ \ \ \ \ \ \ \ \ \ \ \ \ \ \ \ \ \ \ $
with parameters $\alpha>0$ and $\mu>0$  
&  $D(x)=x (x+1)$
& $F(x)  = (\alpha-1) - (\mu+1) x $
  \\
\hline
(v)   $ $  Beta-distribution 
for $x \in ]0,1[ \ \ \ \ \   \ \ \ \ \ \ \ \ \ \ \ $   
$P_*(x) = \frac{\Gamma(\alpha+\beta)}{ \Gamma(\alpha) \Gamma(\beta) } x^{\alpha-1} (1-x)^{\beta-1}
 \ \ \ \ \ \ \ \ \ \ \ \ \ \ \ \ \ \ \ \ \ \ \ \ \ $
with parameters $\alpha>0$ and $\beta>0$  
&  $D(x)=x(1-x)$
& $F(x)  = (\alpha-1) - (\alpha+\beta-2) x $
  \\
\hline
(vi)  Student-distribution for $x \in ]-\infty,+\infty[$ 
$P_*(x) =\frac{\Gamma(\frac{\mu+1}{2}) }
{\Gamma(\frac{1}{2})\Gamma(\frac{\mu}{2}) \left( 1+x^2 \right)^{\frac{1+\mu}{2}}}
 \ \ \ \ \ \ \ \ \ \ \ \ \ \ \ \ \ \ \ \ \ \ \ \ \ \ \ \ \ \ \ \ \ \ \ \ \ \ \ \ $
with parameter $\mu>0$   
&  $D(x)= 1+x^2$
& $F(x)  =  - (1+\mu) x $
  \\
\hline
\end{tabular}
\caption{ The five representative examples (ii-vi) of the Pearson family that will be considered in the present paper.
} 
\label{tablePearson}
\end{table}

Since the Pearson family is closed under affine transformations $x \to \eta x+\zeta$ 
describing rescaling and/or translations of the space coordinate $x$,
one can choose some coefficients of the diffusion coefficient $D(x)$ 
to focus on representative examples. 
The most important discussion
is on the number of roots of the diffusion coefficient $D(x)$ and on their positions
in the complex plane, that leads to the following six standard representative examples,
where the diffusion coefficient $D(x)$ is given,
while the corresponding steady state $P_*(x)$ contains two parameters related
to the two parameters of the linear force (see the table \ref{tablePearson}) :

(i) When $a=0$ and $b=0$ in Eq. \ref{fpearson}, 
the diffusion coefficient is simply constant $D(x)=c$ and 
the linear force leads to the Ornstein-Uhlenbeck process with a Gaussian steady state.
Since this case is already much studied on its own independently of the Pearson family,
with its supplementary specific properties as a Gaussian process,
it will not be discussed in the present paper.
Its large deviations properties have already been much studied 
even in higher dimensions where non-equilibrium steady states are possible
(see the two recent papers \cite{us_gyrator,duBuisson_gyrator}, the PhD thesis \cite{duBuisson_thesis}, 
and references therein).

(ii) When $a=0$ and $b \ne 0$ in Eq. \ref{fpearson}, 
the diffusion coefficient is simply linear. 
The representative example is the Square-Root or the Cox-Ingersoll-Ross process 
with $D(x)=x$ 
converging towards a Gamma-distribution 
on $ ]0,+\infty[$ that will be discussed in section \ref{sec_gamma}.

(iii) When $a \ne 0$ and the two roots of $D(x)$ are real and coincide,
the representative example is the process with $D(x)=x^2$ 
converging towards a heavy-tailed Inverse-Gamma-distribution
on $ ]0,+\infty[ $ that will be discussed in section \ref{sec_kesten}.

(iv) When $a > 0$ and the two roots of $D(x)$ are real and different,
the representative example is the process with $D(x)=x(1+x)$ 
converging towards a heavy-tailed Fisher-Snedecor-distribution 
 on $ ]0,+\infty[ $ that will be discussed in section \ref{sec_fisher}.

(v) When $a < 0$ and the two roots of $D(x)$ are real and different,
the representative example is the Jacobi process with $D(x)=x(1-x)$ 
converging towards a Beta-distribution on $ ]0,1[ $ that will be discussed in section \ref{sec_jacobi}.

(vi) When $a \ne 0$ and the two roots of $D(x)$ are complex-conjugate in Eq. \ref{fpearson},
the representative example is the process with $D(x)=1+x^2$ 
converging towards a heavy-tailed Student-distribution
 on $ ]-\infty,+\infty[ $ that will be discussed in section \ref{sec_student}.
 
Let us stress the possible behaviors of the normalized steady state $P_*(x)$ near finite or infinite boundaries
with the examples $x_L=0$ and $x_R=+\infty$ for the representative examples given in the table \ref{tablePearson} :

$\bullet$ Near the finite boundary $x_L=0$, the normalized steady state $P_*(x)$
 is dominated by the essential singularity $e^{- \frac{\lambda}{x} } $ in the representative example (iii), while the three other cases 
(ii) (iv) (v) display the following normalizable power-law behavior parametrized by the exponent $\alpha>0 $
\begin{eqnarray}
\text{Cases (ii) (iv) (v) } : \ \ P_*(x) \oppropto_{x \to 0^+} x^{\alpha-1}  
\begin{cases}
\text{ vanishing for } \ \ \alpha >1 
 \\
\text{ finite for } \ \ \alpha =1
\\
\text{ diverging for } \ \ 0<\alpha <1
\end{cases}
\label{zeropower}
\end{eqnarray}

$\bullet$ Near the infinite boundary $x_R=+\infty$, the normalized steady state $P_*(x)$ 
is dominated by the exponential decay $e^{- \gamma x } $ in the representative example (ii), 
while the three other cases the cases (iii) (iv) and (vi) display 
the following normalizable power-law decay parametrized by the exponent $\mu>0 $
\begin{eqnarray}
\text{Cases (iii) (iv) (vi) } : \ \ P_*(x) \opsimeq_{x \to + \infty} \frac{1}{x^{1+\mu} } \ \ \ {\rm with } \ \ \mu>0
\label{infinitypower}
\end{eqnarray}

In summary, the first nice property of the Pearson family
is that the corresponding steady states $P_*(x)$ are well-known probability distributions
that appear in many other contexts in probability and statistics, 
as well as in many applications in physics, biology and finance.
 However, the most important properties of Pearson diffusions are
 the explicit spectra governing their dynamical properties,
 as recalled in the next three sections.

%%%%%%%%%%%%%%%%%%%%%%%%%%%%%%%%%%%%%%%%%

\section{ Dynamics of the moments $ m_k(t) =  \int_{x_L}^{x_R} dx x^k P_t(x)$ for Pearson diffusions}

\label{sec_mk}

In this section, we recall why the integer moments 
$ m_k(t) =  \int_{x_L}^{x_R} dx x^k P_t(x)$ 
enjoys very specific dynamical properties for Pearson diffusions.

\subsection{ Reminder on the dynamics of averaged values $w^{av}[t] =\int_{x_L}^{x_R} dx w(x) P_t(x)$ 
for a diffusion with $F(x)$ and $D(x)$} 

For an observable $w(x)$ of a general diffusion process described in subsection \ref{subsec_geneDiff}, it is natural to consider
the average $w^{av}[t] $ of $w(x(t))$ over the possible positions $x(t)$ at time $t$
distributed with the probability density $P_t(x) $
\begin{eqnarray}
w^{av}[t] \equiv \overline{ w(x(t)) } = \int_{x_L}^{x_R} dx w(x) P_t(x)
 \label{Oav}
\end{eqnarray}
Its dynamics can be analyzed using 
the Fokker-Planck dynamics of Eq. \ref{fokkerplanck} for $P_t(x)$ 
and performing integrations by parts with the boundary conditions of Eq. \ref{jboundaries}
\begin{eqnarray}
\partial_t w^{av}[t] && = \int_{x_L}^{x_R} dx w(x) \partial_t P_t(x)
= - \int_{x_L}^{x_R} dx w(x) \partial_x J_t(x)  
=\int_{x_L}^{x_R} dx J_t(x) w'(x)
\nonumber \\
&& =\int_{x_L}^{x_R} dx \bigg(  F(x)   P_t(x) - D (x)  \partial_{x} P_t(x) \bigg) w'(x)
\nonumber \\
&& =\int_{x_L}^{x_R} dx    P_t(x) \bigg(  F(x)  w'(x) + \partial_{x} [ D(x) w'(x) ] \bigg)
 - \bigg[ P_t(x)  D(x) w'(x) \bigg]_{x=x_L}^{x=x_R}
 \nonumber \\
&& =\int_{x_L}^{x_R} dx    P_t(x)  \bigg( {\cal L}^{\dagger}_x  w(x) \bigg)
 - \bigg[ P_t(x)  D(x) w'(x) \bigg]_{x=x_L}^{x=x_R}
 \label{Oavdyn}
\end{eqnarray}
The first term is a bulk contribution that corresponds to the average over $P_t(x)$
of the observable $\bigg( {\cal L}^{\dagger}_x  w(x) \bigg) $ 
that involves the action of the adjoint operator 
${\cal L}^{\dagger}_x $ of Eqs \ref{adjoint} and \ref{adjointIto} on the observable $w(x)$.
The second term is a contribution from the two boundaries at $x=x_L$ and $x=x_R$
that does not vanish in general when the only hypothesis is 
the vanishing-current boundary conditions of Eq. \ref{jboundaries}.

%%%%%%%%%%%%%%%%%%%%%%%%%%%%%%%%

\subsection{ Simplifications for the Pearson family : closed dynamical equations for the successive moments $m_k(t)$ }

For the Pearson family of diffusion processes,
the property of Eq. \ref{Dvanishboundaries} yields 
that the dynamics of Eq. \ref{Oavdyn} does not contain boundary terms
and thus reduces to the bulk contribution  
that involves the adjoint operator 
${\cal L}^{\dagger}_x $ of Eqs \ref{adjoint} and \ref{adjointIto}
\begin{eqnarray}
\partial_t w^{av}[t] && =\int_{x_L}^{x_R} dx    P_t(x) \bigg( {\cal L}^{\dagger}_x  w(x) \bigg)
 = \int_{x_L}^{x_R} dx    P_t(x) \bigg(F_I(x) w'(x)  + D(x)  w''(x) \bigg) 
 \label{OavdynPearson}
\end{eqnarray}
In addition, since the linear Ito force $ F_I(x)$ multiplies the first derivative $w'(x)$
and the quadratic diffusion coefficient $D(x)$ multiplies the second derivative $w''(x)$,
the moment of order $k$ corresponding to the observable $w(x)=x^k$ 
\begin{eqnarray}
m_k(t) \equiv  \int_{x_L}^{x_R} dx x^k P_t(x)
 \label{mktk}
\end{eqnarray}
satisfies the dynamics of Eq. \ref{OavdynPearson}
\begin{eqnarray}
\partial_t m_k(t) &&  =  \int_{x_L}^{x_R} dx    P_t(x)  \bigg[ \bigg(  \lambda_I - \gamma_I x \bigg) k x^{k-1} 
+ \bigg(a x^2+b x +c  \bigg) k (k-1) x^{k-2}   \bigg] 
\nonumber \\
&& = - k \bigg(  \gamma_I - a (k-1) \bigg) m_k(t)
+ k \bigg( b  (k-1)  + \lambda_I  \bigg) m_{k-1}(t)
+ c k (k-1) m_{k-2}(t)
 \label{dynPearsonmk}
\end{eqnarray}
that only involves the moment $m_k(t) $ itself and the two lower moments of order $(k-1)$ and $(k-2)$,
or only the lower moment of order $(k-1)$ when $c=0$.

The homogeneous dynamics of Eq. \ref{dynPearsonmk} 
\begin{eqnarray}
\text {Homogeneous dynamics :} \ \ \ \ \ \  \partial_t m_k(t) &&   = - k \bigg(  \gamma_I - a (k-1) \bigg) m_k(t)
\equiv - \epsilon_k m_k(t)
 \label{dynPearsonmkHomo}
\end{eqnarray}
involves the following explicit rate $\epsilon_k$
with very different physical meanings for positive and negative $\epsilon_k$
\begin{eqnarray}
 \epsilon_k \equiv k \bigg( \gamma_I - a (k-1)  \bigg)
\ \ \ i.e. \ \  \begin{cases}
\text{ exponential relaxation as $e^{-  t \epsilon_k}  $ if $\epsilon_k >0$ } \\
 \text{ exponential growth as $e^{  t (- \epsilon_k)}  $  if $\epsilon_k <0$}
\end{cases}
 \label{ekmk}
\end{eqnarray}

The steady version of Eq. \ref{dynPearsonmk} yield that the 
moments $m_k^*$ of the steady state $P_*(x)$
\begin{eqnarray}
m_k^* \equiv  \int_{x_L}^{x_R} dx x^k P_*(x)
 \label{mksteady}
\end{eqnarray}
satisfy the following simple recurrence when they exist 
\begin{eqnarray}
0 = k \bigg( a (k-1) - \gamma_I \bigg) m_k^*
+ k \bigg( b  (k-1)  + \lambda_I  \bigg) m_{k-1}^*
+ c k (k-1) m_{k-2}^*
 \label{recmksteady}
\end{eqnarray}
that explains why the steady moments $m_k^*$ of Pearson diffusions
have very simple expressions in terms of the Gamma-function,
as will be recalled later in the sections devoted to the various representative examples.

Let us now analyze more precisely the dynamics of Eq. \ref{dynPearsonmk}.
starting with the two first moments $k=1$ and $k=2$.

%%%%%%%%%%%%%%%%%%%%%%%%%%%%%%%%%%%%%%%%%%%%%

\subsubsection{ Explicit dynamics of the first moment $m_1(t)=\int_{x_L}^{x_R} dx x P_t(x)$}

For $k=1$, the rate of Eq. \ref{ekmk} reduces to
\begin{eqnarray}
 \epsilon_1 = \gamma_I 
 \label{e1m1}
\end{eqnarray}
and Eq. \ref{dynPearsonmk} gives 
the following closed dynamics 
for the first moment $m_1(t) $ 
 using the normalization $m_0(t)=\int_{x_L}^{x_R} dx  P_t(x)=1$
\begin{eqnarray}
\partial_t  m_1(t)    =  - \epsilon_1 m_1(t)  + \lambda_I  \equiv - \epsilon_1 m_1(t)  + \lambda_I 
 \label{dynPearsonm1}
\end{eqnarray}
The solution
\begin{eqnarray}
  m_1(t)    =  \frac{\lambda_I}{\gamma_I} + \left(  m_1(0)  - \frac{\lambda_I}{\gamma_I} \right) e^{-  t \epsilon_1}  
\ i.e.   \begin{cases}
\text{ exponential relaxation as $e^{-  t \epsilon_1}  $ towards $m_1^*=\frac{\lambda_I}{\gamma_I}  $
for $\epsilon_1 >0$ } \\
 \text{ exponential growth as $e^{  t (- \epsilon_1)}  $ towards $m_1^*=+\infty$ if $\epsilon_1 <0$}
\end{cases}
 \label{dynPearsonm1integ}
\end{eqnarray}
shows that the essential role of the sign of the rate $\epsilon_1=\gamma_I$ :
 if $ \epsilon_1 >0$, then the first moment $m_1(t)$ converges with rate $\epsilon_1$ towards 
 the finite steady value $m_1^* =\frac{\lambda_I}{\gamma_I} $,
 while if $\epsilon_1<0$, then the first moment $m_1(t)$ grows exponentially as $e^{  t (-\epsilon_1)}  $
  and the first moment of the steady state is infinite $m_1^*=+\infty$.
 
 %%%%%%%%%%%%%%%%%%%%%%%%%%%%%%%%%%

\subsubsection{ Explicit dynamics of the second moment $m_2(t)=\int_{x_L}^{x_R} dx x^2 P_t(x)$}

For $k=2$, the rate of Eq. \ref{ekmk} reads
\begin{eqnarray}
 \epsilon_2 = 2 (  \gamma_I -a )  
 \label{e2m2}
\end{eqnarray}
and Eq. \ref{dynPearsonmk} gives the following dynamics
for the second moment $m_2(t)$
\begin{eqnarray}
\partial_t  m_2(t) &&   = - \epsilon_2 m_2(t)+ 2 \bigg( b   + \lambda_I  \bigg) m_1(t)+ 2 c  
 \label{OavdynPearsonkm2}
\end{eqnarray}
One can plug the solution of Eq. \ref{dynPearsonm1integ} for the first moment $m_1(t)$
and integrate to obtain the solution 
\begin{eqnarray}
 m_2(t)  &&  = e^{- t \epsilon_2} \left( m_2(0)+ \int_0^t d \tau e^{  \tau \epsilon_2 } \left[
 2c + 2 ( b   + \lambda_I  )   \frac{\lambda_I}{\gamma_I} + 2 ( b   + \lambda_I  )\left(  m_1(0)  - \frac{\lambda_I}{\gamma_I} \right) e^{-  \tau \epsilon_1}   \right] \right)
 \nonumber \\
 && =  e^{-  t \epsilon_2 }  m_2(0)
 + \frac{1- e^{-  t \epsilon_2 } }{ \epsilon_2 } 
 \left[2c + 2 ( b   + \lambda_I  )   \frac{\lambda_I}{\gamma_I} \right]
 +  \frac{ e^{- t \epsilon_1} - e^{-  t \epsilon_2 }}{ \epsilon_2-\epsilon_1 } 2 ( b   + \lambda_I  )\left(  m_1(0)  - \frac{\lambda_I}{\gamma_I} \right)   
 \nonumber \\
&&   \opsimeq_{t \to +\infty}
  \begin{cases}
   \frac{2c + 2 ( b   + \lambda_I  )   \frac{\lambda_I}{\gamma_I} }{ \epsilon_2 } 
\frac{c +  ( b   + \lambda_I  )   \frac{\lambda_I}{\gamma_I}}{ (  \gamma_I -a ) } 
 = m_2^* 
& \text{for $\epsilon_1 >0$ and $\epsilon_2 >0 $} \\
 +\infty  = m_2^*
 & \text{ otherwise}
\end{cases}
 \label{dynPearsonm2integ}
\end{eqnarray}
where the signs of the two rates $\epsilon_1$ and $\epsilon_2$ determine whether the asymptotic value $m_2^*$
remains finite or diverges.

%%%%%%%%%%%%%%%%%%%%%%%%%%%%%%%%%%%%%%%%%%%%%%

\subsubsection{ Dynamics of the successive integer moment $m_k(t)$ of order $k =3,4,..$}

It is now clear how the dynamics of the successive integer moment $m_k(t)$
 can be solved recursively via this pedestrian method :
one can plug the solutions found previously for $m_{k-1}(t) $ and $m_{k-2}(t) $
 into the dynamical Eq. \ref{dynPearsonmk} for $m_k(t)$ and
 the new rate 
 that will appear in the solution for $m_k(t)$ with respect to the rates already present in 
 $m_{k-1}(t) $ and $m_{k-2}(t) $ is $\epsilon_k$ of Eq. \ref{ekmk}. 
 Taking into account the inhomogeneous term containing
  the two previous moments $m_{k-1}(t) $ and $m_{k-2}(t) $,
one finally obtains that the solution for $m_k(t)$ 
can be written as a linear combination of $e^{- t \epsilon_j}$ with $j=0,1,..,k$ with coefficients $M_{kj}$
that should be computed in terms of the initial condition at $t=0$
\begin{eqnarray}
 m_k(t)  = \sum_{j=0}^k M_{kj} e^{- t \epsilon_j}    \opsimeq_{t \to +\infty}
  \begin{cases}
  \text{relaxation towards the finite steady value $m_k^*=M_{k0} $ if $\epsilon_j >0$ for $1 \leq j \leq k$} \\  
  \text{exponential growth towards $m_k^*=\infty $  otherwise}
\end{cases}
 \label{OavdynPearsonkm2integ}
\end{eqnarray}
so that it will converge towards a finite steady value $m_k^*$ only if the $k$ rates $(\epsilon_1,\epsilon_2,..,\epsilon_k)$ are strictly positive and will diverge otherwise.

%%%%%%%%%%%%%%%%%%%%%%%%%%%%%%%%%%%%%%%%%%%%%%%%%

\subsubsection{ Closed dynamics for the 
Laplace transform ${\hat P}_t(s) $ or the Fourier transform ${\tilde P}_t(q) $ in Appendix \ref{app_laplace}}

As recalled in Appendix \ref{app_laplace}, 
in order to summarize the full hierarchy of the dynamical equations
for the moments $m_k(t)$ of Eq. \ref{dynPearsonmk} with $k \in \mathbb{N}$,
one can write closed dynamical equations for the Laplace transform ${\hat P}_t(s) $ of Eq. \ref{laplace}
(or for the Fourier transform ${\tilde P}_t(q) $ of Eq. \ref{fourier}) : in specific examples of Pearson diffusions,
this method is very useful to obtain explicit expressions in terms of special functions (see \cite{c_flux} for instance)
but will not be discussed further in the present paper.

%%%%%%%%%%%%%%%%%%%%%%%%%%%%%%%%%

\section{ Spectral properties for the propagator $P_t(x \vert x_0)$ of Pearson diffusions} 

\label{sec_spectral}

In this section, we discuss the specific 
spectral properties for the propagator $P_t(x \vert x_0)$ for Pearson diffusions
in order to make the link with the explicit rates $\epsilon_k$ of Eq. \ref{ekmk} that appear in the dynamics
of the moments $m_k(t)$ discussed in the previous section.

\subsection{ Reminder on the spectral properties for a general diffusion with $F(x)$ and $D(x)$} 

For a general diffusion process discussed in subsection \ref{subsec_geneDiff},
the propagator $P_t(x \vert x_0)$ satisfies the forward dynamics with respect to its final position $x$ 
that involves the differential operator $ {\cal L}_x$ of Eq. \ref{generator}
\begin{eqnarray}
 \partial_t P_t(x \vert x_0)      =    {\cal L}_x P_t(x \vert x_0)
\label{forward}
\end{eqnarray}
while it satisfies the backward dynamics 
with respect to its initial position $x_0$ 
that involves the adjoint operator ${\cal L}^{\dagger}_{x_0}$ of Eq. \ref{adjoint}
\begin{eqnarray}
 \partial_t P_t(x \vert x_0)    =  {\cal L}_{x_0}^{\dagger} P_t(x \vert x_0)
\label{backward}
\end{eqnarray}
However it is well known that whenever there is detailed-balance, it is possible to make a similarity transformation
towards a symmetric operator \cite{gardiner,vankampen,risken} as described in the next subsection.

%%%%%%%%%%%%%%%%%%%%%%%%%%%%%%%%%%%%%

\subsubsection{ Similarity transformation towards a quantum hermitian Hamiltonian $H= H^{\dagger}$ } 

The standard change of variables involving the steady state of Eq. \ref{steadyeq} with its potential $U(x)$ of Eq. \ref{Ux}
\begin{eqnarray}
P_t(x\vert x_0) = \sqrt{  \frac{ P_*(x) }{P_*(x_0) } } \psi_t(x \vert x_0) = 
 e^{  \frac{ U(x_0)-U(x)}{2}}  \psi_t(x \vert x_0)
\label{ppsi}
\end{eqnarray}
transforms the forward dynamics of Eq \ref{forward} for $ P_t(x\vert x_0) $
 into the euclidean Schr\"odinger equation for the quantum propagator $\psi_t(x \vert x_0)$
\begin{eqnarray}
-  \partial_t \psi_t(x\vert x_0)  = H \psi_t(x \vert x_0)
\label{schropsi}
\end{eqnarray}
where the quantum hermitian Hamiltonian 
\begin{eqnarray}
 H = H^{\dagger} =  - \frac{ \partial  }{\partial x} D(x) \frac{ \partial  }{\partial x} +V(x)
\label{hamiltonien}
\end{eqnarray}
corresponds to an effective position-dependent 'mass' 
whenever the diffusion coefficient  $ D(x)=\frac{1}{2 m(x)}$ depends explicitly on the position $x$,
while the scalar potential $V(x)$ reads
\begin{eqnarray}
V(x) && \equiv D(x)  \frac{ [U'(x)]^2 }{4 }  -D(x) \frac{U''(x)}{2} -D'(x) \frac{U'(x)}{2}
\nonumber \\
&& = \frac{ F^2(x) }{4 D(x) } + \frac{F'(x)}{2}
\label{vfromu}
\end{eqnarray}
The discussion of the supersymmetric structure of the Hamiltonian of Eq. \ref{hamiltonien}
is postponed to the next section \ref{sec_susy}.

%%%%%%%%%%%%%%%%%%%%%%%%%%%%%%%%%%%%%%%%%%

\subsubsection{ Spectral decompositions of the quantum propagator $\langle x \vert e^{- H t} \vert x_0 \rangle $  } 

As explained in textbooks on quantum mechanics, there are three kinds of
spectra for a quantum Hamiltonian like $H$ of Eq. \ref{hamiltonien} :

$\bullet$ (a) an infinite series of discrete energies $E_n$ labelled by $n=0,1,..,+\infty$
associated to eigenstates $\phi_n(x)$
\begin{eqnarray}
E_n \phi_n(x) = H   \phi_n(x) 
\label{hsusyeigen}
\end{eqnarray}
satisfying the orthonormalization
\begin{eqnarray}
\delta_{n n' } = \langle \phi_n \vert \phi_{n'} \rangle 
= \int_{x_L}^{x_R} dx \phi_n^*(x) \phi_{n'}(x)
\label{orthophin}
\end{eqnarray}
The simplest example is the well-known quantum harmonic oscillator.

$\bullet$ (b) an energy continuum $E(q)$ labelled by the continuous wave-number $q$
associated to eigenstates $\varphi_q(x)$
\begin{eqnarray}
E(q) \varphi_q(x) = H   \varphi_q(x)
\label{hsusyeigencontinum}
\end{eqnarray}
satisfying the orthonormalization analogous to Eq. \ref{orthophin}
but where the Kronecker $\delta_{nn'} $ is replaced by the Dirac delta function $\delta(.)$
\begin{eqnarray}
\delta(q-q') = \langle \varphi_q \vert \varphi_{q'} \rangle 
= \int_{x_L}^{x_R} dx \varphi_q^*(x) \varphi_{q'}(x)
\label{orthophincontinuum}
\end{eqnarray}
Note that an energy continuum is possible only if the interval $]x_L,x_R[$ is infinite,
the simplest example being the free Hamiltonian $H= - D \partial_x^2$ for $] x \in -\infty,+\infty[$, 
where the eigenstates reduce to the orthonormalized plane-waves $\varphi_q(x) =\frac{e^{i q x}}{\sqrt{2 \pi}}$ 
with $] q \in -\infty,+\infty[$ and energies $E(q)=D q^2$.

$\bullet$ (c) a certain number of discrete energies $E_n$ labelled by $n=0,1,.,n_{max}$
associated to eigenstates $\phi_n(x)$ satisfying the orthonormalization of Eq. \ref{orthophin},
followed by an energy continuum $E(q)$ labelled by a continuous wave-number $q$
associated to eigenstates $\varphi_q(x)$ satisfying the orthonormalization of Eq. \ref{orthophincontinuum},
and that are orthogonal to the discrete eigenstates $ \phi_n(x)$
\begin{eqnarray}
0 = \langle \varphi_q \vert \phi_n \rangle 
= \int_{x_L}^{x_R} dx \varphi_q^*(x) \phi_n(x)
\label{orthophincontinuumorthodiscrete}
\end{eqnarray}
The simple example is the free Hamiltonian with a delta-attractive-potential at the origin.

In this case (c), the spectral decomposition of the quantum propagator associated to the Hamiltonian 
reads
\begin{eqnarray}
\psi_t(x \vert x_0)  \equiv  \langle x \vert e^{- H t} \vert x_0 \rangle
= \sum_{n=0}^{n_{max}} e^{- t E_n} \langle x \vert \phi_n \rangle \langle\phi_n \vert x_0 \rangle
+ \int dq e^{-t E(q)}  \langle x \vert \varphi_q \rangle \langle \varphi_q \vert x_0 \rangle
\label{psispectralfull}
\end{eqnarray}
while the case (a) only involves the first discrete contribution with $n_{max}=+\infty$,
and while the case (b) only involves the second integral contribution.

%%%%%%%%%%%%%%%%%%%%%%%%%%%%%%%%%%%

\subsubsection{ Spectral decompositions of the Fokker-Planck propagator $P_t(x \vert x_0)$ } 

To simplify the notations in the following, 
we will focus for the time being on the case 
where the spectrum of the Hamiltonian $H$ 
contains only an infinite series of discrete energies $E_n$ labelled by $n=0,1,..,+\infty$,
and where the corresponding eigenstates $\phi_n(x)$ are real-valued,
so that the spectral decomposition of the quantum propagator of Eq. \ref{psispectralfull}
reads 
\begin{eqnarray}
\psi_t(x \vert x_0)  \equiv  \langle x \vert e^{- H t} \vert x_0 \rangle
= \sum_{n=0}^{+\infty} e^{- t E_n}  \phi_n (x)  \phi_n (x_0) 
\label{psispectral}
\end{eqnarray}
The corresponding spectral decomposition 
of the Fokker-Planck propagator $P_t(x \vert x_0) $
reads
via the change of variables of Eq. \ref{ppsi}
\begin{eqnarray}
P_t(x \vert x_0)  \equiv \sqrt{ \frac{P_*(x)}{P_*(x_0) }  } \psi_t(x \vert x_0) 
&& = \sum_{n=0}^{+\infty} e^{- t E_n} \bigg(\sqrt{ P_*(x) } \langle x \vert \phi_n \rangle \bigg) 
\bigg( \langle\phi_n \vert x_0 \rangle \frac{1}{ \sqrt{ P_*(x_0) }}\bigg) 
\nonumber \\
&& = \sum_{n=0}^{+\infty} e^{- t E_n}  r_n(x) l_n(x_0)
= P_*(x) + \sum_{n=1}^{+\infty} e^{- t E_n}  r_n(x) l_n(x_0) 
\label{FPspectralq}
\end{eqnarray}
where
\begin{eqnarray}
r_n(x) && \equiv \sqrt{ P_*(x) } \phi_n(x) = \frac{ e^{ - \frac{ U(x)}{2} } }{\sqrt Z} \phi_n(x)
\nonumber \\
 l_n ( x_0 )  && \equiv \phi_n(x_0) \frac{1}{ \sqrt{ P_*(x_0) }}
=  \phi_n(x_0) e^{  \frac{ U(x_0)}{2} } \sqrt Z
\label{rnlnphin}
\end{eqnarray}
are the right eigenvectors $r_n(x)$
and the left eigenvectors $l_n(x_0)$ associated to the energy $E_n$
\begin{eqnarray}
 -E_n r_n(x) &&= {\cal L}_x r_n(x)  
 \nonumber \\
  -E_n l_n(x_0) && = {\cal L}^{\dagger}_{x_0} l_n(x_0)   
 \label{fokkerplanckeigen}
\end{eqnarray}
satisfying the orthonormalization inherited from Eq. \ref{orthophin}
\begin{eqnarray}
\delta_{n n' } = \langle l_n \vert r_{n'} \rangle = \int_{x_L}^{x_R} dx l_n(x) r_{n'}(x)
 \label{ortholr}
\end{eqnarray}
The vanishing eigenvalue $E_0=0$ is associated to the convergence towards the steady state 
$P_*(x) $ for any initial condition $x_0$
\begin{eqnarray}
  r_0(x) &&= P_*(x)
 \nonumber \\
  l_0(x_0) && =    1
 \label{r0l0}
\end{eqnarray}
and to the positive quantum ground-state
\begin{eqnarray}
  \phi_0(x) =   \sqrt{ P_*(x) } = \frac{ e^{ - \frac{ U(x)}{2} } }{\sqrt Z}
 \label{phi0}
\end{eqnarray}

The vanishing-current boundary conditions inherited from Eq. \ref{jboundaries}
\begin{eqnarray}
  j_n(x_L)=0=j_n(x_R)
 \label{BCjn}
\end{eqnarray}
involve the current $j_n(x)$ associated to the right eigenvector $r_n(x)$
\begin{eqnarray}
  j_n(x) && \equiv \bigg( F(x)  - D (x)  \partial_{x} \bigg) r_n(x) 
  = - D(x) \bigg(  U'(x)  +  \partial_{x} \bigg)  r_n(x) 
  = - D(x) \bigg(  U'(x) r_n(x) +  r_n'(x) \bigg) 
 \label{jnright}
\end{eqnarray}
that can be translated for the quantum eigenvector $\phi_n(x)$ via Eq. \ref{rnlnphin}
\begin{eqnarray}
  j_n(x)  && = - D(x) \bigg(  U'(x)  +  \partial_{x} \bigg)  \phi_n(x)   \frac{e^{-\frac{U(x)}{2}}}{\sqrt{Z} } 
 = - D(x)  \frac{e^{-\frac{U(x)}{2}}}{\sqrt{Z} }  \bigg(  \frac{U'(x)}{2}  \phi_n(x)+ \phi_n'(x) \bigg)    
 \label{jnquantum}
\end{eqnarray}
and for the left eigenvector $l_n(x)$ into
\begin{eqnarray}
  j_n(x)  && = - D(x) \bigg(  U'(x)  +  \partial_{x} \bigg)  l_n(x) \frac{e^{-U(x)}}{Z}
  = -   D(x) \frac{e^{-U(x)}}{Z}   l_n'(x) 
  = - D(x) P_*(x)  l_n'(x) 
 \label{jnleft}
\end{eqnarray}

In particular, let us stress that
 at a finite boundary $x_L=0$ where the diffusion coefficient $D(x_L=0)$ and the steady state $P_*(x_L=0)$
remain finite, the vanishing-current boundary condition $j_n(x_L=0)=0$ leads to the simplified conditions
\begin{eqnarray}
\text{ If $x_L=0$ with finite $D(0)$ and finite $P_*(0)$ } :    j_n(0)  =0
\to  \begin{cases}
U'(0) r_n(0) +  r_n'(0) =0 \text{ for the right eigenvectors } 
\\
\frac{U'(0)}{2}  \phi_n(0)+ \phi_n'(0) =0  \text{ for the quantum eigenvectors } 
\\
  l_n'(0) =0 \text{ for the left eigenvectors } 
\end{cases}
\ \  \label{nonsingularboundary}
\end{eqnarray}
while for Pearson diffusions where the diffusion coefficient vanishes $D(0)=0$ 
and where the steady state $P_*(0)$ can be vanishing, diverging, or finite as stressed in Eq. \ref{zeropower},
the analysis of boundary conditions will be completely different, as will be explained in detail in the further subsection 
\ref{subsec_BCnormaLeft}.

In summary, one can equivalently study :

(a) the spectral properties of the operator ${\cal L}_x $ with the right eigenvectors $r_n(x)$

(b) the spectral properties of the adjoint operator ${\cal L}_x^{\dagger} $ with the left eigenvectors $l_n(x)$

(c) the spectral properties of the quantum Hamiltonian $H$ with its eigenvectors $\phi_n(x)$.

In practice, it is simpler to focus 
either on the quantum eigenvectors $\phi_n(x)$ as will be discussed in the next section \ref{sec_susy},
or on the left eigenvectors $l_n(x)$, as we discuss in the remainder of the present section.

%%%%%%%%%%%%%%%%%%%%%%%%%%%%%%%%%%%%%%%%%%

\subsubsection{Properties of the left eigenvectors $l_n(x)$ with their physical meaning as 
the observables with the simplest dynamics} 

The translation 
of the orthonormalization of Eqs \ref{ortholr} \ref{orthophin} using \ref{rnlnphin}
\begin{eqnarray}
\delta_{nm} =  \int_{x_L}^{x_R} dx l_m(x) l_n(x) P_*(x)
\label{ortholeftalone}
\end{eqnarray}
means that the left eigenvectors $l_n(.)$ 
are an orthogonal family with respect to the steady state $P_*(x)$.

To understand the physical meaning of the left eigenvectors $l_n(x)$ for $n >0$,
it is useful to consider them as observables  
and to analyze the corresponding averaged values at time $t$ of Eq. \ref{Oav}
 \begin{eqnarray}
l_n^{av}[t] \equiv \int_{x_L}^{x_R} dx l_n(x) P_t(x)
 \label{lnav}
\end{eqnarray}
Using the boundary conditions of Eq. \ref{BCjn} with Eq. \ref{jnleft},
one obtains that their dynamics of Eq. \ref{Oavdyn} do not contain boundary contributions,
while the bulk contribution involving the adjoint operator ${\cal L}^{\dagger}_x $ 
simplifies using the eigenvalue equation
of Eq. \ref{fokkerplanckeigen}
\begin{eqnarray}
\partial_t l_n^{av}[t] &&  =\int_{x_L}^{x_R} dx    P_t(x)  {\cal L}^{\dagger}_x  l_n(x)
=- E_n \int_{x_L}^{x_R} dx    P_t(x)   l_n(x)
= -E_n l_n^{av}[t]
 \label{lnavdyn}
\end{eqnarray}
The solution corresponds to the exponential relaxation towards zero 
with the single energy $E_n$
\begin{eqnarray}
l_n^{av}[t] = l_n^{av}[0] e^{- t E_n} 
 \label{lnavdynintegrate}
\end{eqnarray}
In conclusion, the left eigenvectors $l_n(x)$ for $n>0$ are 
 very simple observables, whose relaxation dynamics towards zero involves 
 a single energy $E_n$ instead of the whole spectrum that would a priori appear 
 for a general observable.

%%%%%%%%%%%%%%%%%%%%%%%%%%%%%%%%

\subsection{ Pearson diffusions : simplifications for the left eigenvectors $l_n(x)$  }

 %%%%%%%%%%%%%%%%%%%%%%%%%%%%%%%%

\subsubsection{ Eigenvalue equations for the left eigenvectors $l_n(x)$ of Pearson diffusions : possibility of polynomial solutions } 
 
The eigenvalue equation of Eq. \ref{fokkerplanckeigen} for the left eigenvector $l_n(x)$
reads using the adjoint operator ${\cal L}^{\dagger}_{x} $ for Pearson diffusions of Eq. \ref{adjointIto}
\begin{eqnarray}
  -E_n l_n(x)  = {\cal L}^{\dagger}_{x} l_n(x)   
&&  =  F_I(x) l_n'(x)  + D(x)  l_n''(x)
  \nonumber \\
  && = \bigg(  \lambda_I - \gamma_I x \bigg)  l_n'(x)
+ \bigg(a x^2+b x +c  \bigg)  l_n''(x)
 \label{eigenleft}
\end{eqnarray}
Since $l_0(x)=1$, it is natural to try the following polynomial form of degree $n$ for $l_n(x)$ for $n>0$
\begin{eqnarray}
 l_n(x) = \sum_{k=0}^n K_{nk} x^k = K_{nn} x^n + K_{n,n-1} x^{n-1} + .. + K_{n,1} x + K_{n,0}
 \label{leftpolynomial}
\end{eqnarray}
 to see whether one can satisfy the boundary conditions and the normalizability.
 
 %%%%%%%%%%%%%%%%%%%%%%%%%%%%%%%%%%%%%%%%

\subsubsection{ Boundary conditions and normalizability of the polynomial solutions $l_n(x)$ for Pearson diffusions} 

\label{subsec_BCnormaLeft}

The vanishing-current boundary conditions of Eq. \ref{BCjn} 
when written in terms of the left eigenvectors $l_n(x)$
via \ref{jnleft}
\begin{eqnarray}
0 && = - j_n(x_L) =  D(x_L) P_*(x_L)  l_n'(x_L) 
\nonumber \\
0 && = - j_n(x_R)=  D(x_R) P_*(x_R)  l_n'(x_R) 
 \label{BCjnLEFT}
\end{eqnarray}
can be discussed as follows for the polynomial of Eq. \ref{leftpolynomial} for $l_n(x)$.

When the boundary $x_L$ is finite, i.e. for $x_L=0$ in the 
four representative examples (ii-iii-iv-v) of the table \ref{tablePearson},
the derivative of the polynomial of Eq. \ref{leftpolynomial} reduces to the constant $l_n'(x=0)=K_{n,1}$.
One can check that the product of the diffusion coefficient $D(x)$ (that always vanishes as $x$ or as $x^2$ in the limit $x \to 0$)
and of the normalizable steady state $P_*(x)$ (that may vanish, be finite or diverge in the limit $x \to 0$ as stressed in Eq. \ref{zeropower}) always vanishes
 in the limit $x \to x_L=0$, so that the boundary condition of Eq. \ref{BCjnLEFT}
 is always satisfied for any $n$
 \begin{eqnarray}
x_L=0 \ \ :  \lim_{x \to x_L=0} \left[ D(x) P_*(x) K_{n,1} \right] =0 \ \ \ \text { : the boundary condition is always satisfied } 
 \label{BCjnLEFTxl0pearson}
\end{eqnarray}
  Similarly when the boundary $x_R$ is finite, i.e. for $x_R=1$ in the 
representative example (v) of the table \ref{tablePearson},
the product of the diffusion coefficient $D(x)$ and of the steady state $P_*(x)$ vanishes
 in the limit $x \to x_R=1$, so that the boundary condition of Eq. \ref{BCjnLEFT}
 is again always satisfied.

When the right boundary is infinite $x_R=+\infty$ as in the 
four representative examples (ii-iii-iv-vi) of the table \ref{tablePearson},
then the derivative of the polynomial of Eq. \ref{leftpolynomial} is dominated by the power $l_n'(x) \simeq n K_{nn} x^{n-1}$
for $x \to + \infty$, while the diffusion coefficient $D(x)$ diverges as $x$ or $x^2$ for $x \to +\infty$.
So one needs to discuss whether the decay of the normalizable steady state $P_*(x)$ for $x \to +\infty$
is sufficient to compensate these power-laws or not.
In the Pearson representative example (ii) where $P_*(x) \propto x^{\alpha-1}e^{- \gamma x}$ 
is dominated by the exponential decay, 
the boundary condition of Eq. \ref{BCjnLEFT} is always satisfied for any $n$
 \begin{eqnarray}
\text{Case (ii)}  : \lim_{x \to x_R=+\infty} \left[ D(x) P_*(x) n K_{nn} x^{n-1}\right] =0 \ \text { : the boundary condition 
at $x_R=+\infty$ is always satisfied } 
 \label{BCjnLEFTxRexpo}
\end{eqnarray}
In the Pearson representative examples (iii-iv-vi) of the table \ref{tablePearson}
where the diffusion coefficient diverges as $D(x) \propto x^2$ in the limit $x \to +\infty$,
while the normalizable steady state decays only as the power-law $P_*(x) \propto x^{-1-\mu}$ of Eq. \ref{infinitypower}
with $\mu>0$,
the discussion is as follows
 \begin{eqnarray}
\text{ Cases (iii-iv-vi) } :   D(x) P_*(x) l_n'(x)
 \oppropto_{x \to +\infty} x^{n-\mu} 
  \begin{cases} 
 0 \ \text{for } n<\mu :  \text{ the boundary condition at $x_R=+\infty$ is satisfied  }
\\
\ne 0 \ \text{for }  n \geq \mu   :  \text{ the boundary condition at $x_R=+\infty$ is not satisfied  }
\end{cases}
 \label{BCjnLEFTxRpower}
\end{eqnarray}

Besides the boundary conditions,
the polynomial solutions $l_n(x)$ of Eq. \ref{leftpolynomial}
should satisfy the normalization of Eq. \ref{ortholeftalone} with respect to the steady state $P_*(x)$.
Since $l_n(x)$ is a polynomial of order $n$, 
it is possible to satisfy the normalization of Eq. \ref{ortholeftalone} for $l_n(x)$
only if the steady moment $m_{2n}^* $ of order $(2n)$ is finite
\begin{eqnarray}
1 =  \int_{x_L}^{x_R} dx  l_n^2(x) P_*(x) \ \ \ \text{possible only if $m_{2n}^* \equiv \int_{x_L}^{x_R} dx  x^{2n} P_*(x)<\infty$}
\label{ortholeftn2}
\end{eqnarray}
For the Pearson representative examples (iii-iv-vi) of the table \ref{tablePearson}
where the steady state decays only as the power-law $P_*(x) \propto x^{-1-\mu}$ for $x \to + \infty$,
the steady moment $m_{2n}^* $ is finite only for $2n < \mu$
\begin{eqnarray}
\text{ Cases (iii-iv-vi) : the polynomial $l_n(x)$ is normalizable with respect to $P_*(x)$ only if $n < \frac{\mu}{2}$}
\label{ortholeftn2bis}
\end{eqnarray}
So this normalizability condition $n < \frac{\mu}{2}$ is more restrictive than the requirement $n<\mu $ 
from the boundary conditions of Eq. \ref{BCjnLEFTxRpower}.
Let us stress this very specific conclusion for Pearson diffusions that will be again useful in further sections:
\begin{eqnarray}
\text{ Pearson : if $l_n(x)$ is normalizable with respect to $P_*(x)$, then the boundary conditions are automatically satisfied   }
\ \ \ \ \ 
\label{normalizability}
\end{eqnarray}
i.e. in practice for the discrete spectrum of Pearson diffusions, one only needs to focus on the normalizability of eigenvectors
without worrying about boundary conditions anymore. 

Putting everything together, the conclusions of this discussion concerning boundary conditions
and normalizability of polynomial solutions for Pearson diffusions can be summarized as follows:
 \begin{eqnarray}
  \begin{cases} 
\text{Cases (ii) and (v)   :  an infinite number of polynomial eigenvectors $l_n(x)$ with $n=0,1,2,..,+\infty$ }
\\
\text{Cases (iii-iv-vi) with } \displaystyle P_*(x) \oppropto_{x \to + \infty}  \frac{1}{x^{1+\mu} } : \text{ finite number of polynomial eigenvectors $l_n(x)$ with } 0 \leq n < \frac{\mu}{2}
\end{cases}
 \label{boundstatesln}
\end{eqnarray}

%%%%%%%%%%%%%%%%%%%%%%%%%%%%%%%%%%%%%%%%%%%%%%%%%%%%%%%

\subsubsection{ Explicit discrete energies $E_n >0$ associated to the polynomial left eigenvectors $l_n(x)$ }

When the polynomial $l_n(x)$ of Eq. \ref{leftpolynomial} is a valid normalizable eigenvector
satisfying the boundary conditions at $x_L$ and $x_R$ (see Eq. \ref{boundstatesln}),
the corresponding energy $E_n >0$ is 
determined by plugging the highest monomial $x^n$ into Eq. \ref{eigenleft} so that
it only involves the quadratic coefficient $a$ of $D(x)$ and the coefficient $\gamma_I = \gamma-2a$ of the Ito force
\begin{eqnarray}
   E_n   =   \gamma_I  n + a   n (1-n) = (\gamma-a) n - a n^2 \ \ \ \ \ \ \text{ if $l_n(x)$ is a valid normalizable eigenvector }
 \label{eigendiscrete}
\end{eqnarray}
where one recognizes the expression of rate $\epsilon_n$ Eq. \ref{ekmk} discussed the previous section.
However, it is very important to stress here that 
the discrete energy $E_n$ of Eq. \ref{eigendiscrete} is a valid eigenvalue 
only if it is associated to a valid polynomial eigenvector $l_n(x)$,
while the rate $\epsilon_n$ of Eq. \ref{ekmk} always appears in the dynamics
of the moment $m_n(t)$ with the physical meaning depending on the sign of $\epsilon_n$,
as discussed in detail in Eqs \ref{e1m1} \ref{e2m2} for $n=1,2$.

Let us now write the discrete energies $E_n$ associated to the polynomial eigenvectors $l_n(x)$ 
for the various Pearson representative examples :

$\bullet$ in the Pearson representative example (ii) where the coefficient $a$ vanishes $a=0$ while $\gamma>0$,
the spectrum of Eq. \ref{eigendiscrete} is simply linear with respect to $n$ 
and contains an infinite number of discrete energies $E_n$
associated to the infinite number of polynomial solutions $l_n(x)$ of Eq. \ref{boundstatesln} 
\begin{eqnarray}
\text{Case (ii)  } \ \ \ a=0  \ \ {\rm and } \ \ \gamma_I=\gamma>0 : \ \ \    E_n   =   \gamma  n \ \ \ \ \ \ \ { \rm with } \ \ n=0,1,2,...,+\infty
 \label{eigendiscretelinear}
\end{eqnarray}

$\bullet$  in the Pearson representative example (v) where the coefficient $a$ is negative $a=-1$ 
while $\gamma_I=\gamma+2=\alpha+\beta>0$,
the spectrum of Eq. \ref{eigendiscrete} is quadratic with respect to $n$ and contains an infinite number of discrete energies
associated to the infinite number of polynomial solutions $l_n(x)$ of Eq. \ref{boundstatesln}
\begin{eqnarray}
\text{Case (v)  }  \ \ \ a=-1  \ \ {\rm and } \ \ \gamma_I=\alpha+\beta>0 : \ \ \    E_n   = n (n-1+\alpha+\beta)  \ \ { \rm with } \ \ n=0,1,2,...,+\infty
 \label{eigendiscreteneg}
\end{eqnarray}

$\bullet$ in the Pearson representative examples (iii-iv-vi) where the coefficient $a$ is positive $a=+1$
while $\gamma_I=\gamma-2=\mu-1$,
there is only a finite number of discrete energies $E_n$
associated to the finite number of polynomial solutions $l_n(x)$ of Eq. \ref{boundstatesln}
\begin{eqnarray}
\text{Cases (iii-iv-vi) } \ \ a=+1  \ \ {\rm and } \  \gamma_I=\mu-1   : \ \ \    
E_n   =  n ( \mu- n) \ \ { \rm with } \ \ 0 \leq n <  \frac{ \mu }{2}
 \label{eigendiscretepos}
\end{eqnarray}

%%%%%%%%%%%%%%%%%%%%%%%%%%%%%%%%%%%%%%%%%%%%%%%%%%%%%%%

\subsubsection{ Summary on the energy spectrum for the various representative examples Pearson diffusions }

In summary, there are two very different possibilities for the energy spectrum of Pearson diffusions :

$ \bullet$  when the energy spectrum corresponds to an infinite series of discrete energies $E_n$ labelled by $n=0,1,..,+\infty$
as in the representative examples (ii) and (iv) of Eqs \ref{eigendiscretelinear} \ref{eigendiscreteneg},
then the energy $E_n$ of Eq. \ref{eigendiscrete}
coincides with the rate $\epsilon_n$ of Eq. \ref{ekmk} for any $n=0,1,2,..,+\infty $
\begin{eqnarray}
   E_n   =  \epsilon_n \ \ {\rm for } \ \  n=0,1,2,..,+\infty
 \label{eigendiscreteinfiniteseries}
\end{eqnarray}
Note that this means that all the rates are strictly positive $\epsilon_n >0$ for $n>0$ :
 all the moments $m_n(t)$ converge towards finite steady values $m_n^*$,
i.e. all the moments of the steady state $P_*(x)$ are finite 
\begin{eqnarray}
   m_n^* \equiv \int_{x_L}^{x_R} dx x^n P_*(x)  < + \infty \ \ {\rm for } \ \  n=0,1,2,..,+\infty
 \label{momentsallfinite}
\end{eqnarray}

$ \bullet$ when the the spectrum contains only a finite number of discrete energies $E_n$ 
as in the representative examples (iii-iv-vi) of Eq. \ref{eigendiscretepos} with $0 \leq n <  \frac{ \mu }{2} $,
the energy $E_n$ of Eq. \ref{eigendiscrete} coincides with the rate $\epsilon_n$ for $0 \leq n <  \frac{ \mu }{2} $
\begin{eqnarray}
   E_n =  n ( \mu- n)  =  \epsilon_n \ \ {\rm for } \ \  0 \leq n <  \frac{ \mu }{2}
 \label{eigendiscretefinite}
\end{eqnarray}
Note that for $k> \frac{ \mu }{2}$, the rate $\epsilon_k$ of Eq. \ref{ekmk} 
is still defined and appears in the dynamics of the moment $m_k(t)$ with the physical meaning depending on the sign of $\epsilon_k$,
as discussed in detail in Eqs \ref{e1m1} \ref{e2m2} for $k=1,2$.
 The finite number of discrete energies in Eq. \ref{eigendiscreteinfiniteseries}
 is directly related to the power-law decay of Eq. \ref{infinitypower},
 since the existence of $E_n$ requires the existence of the steady moment $m_{2n}^*$
 as a consequence of the normalization of Eq. \ref{ortholeftn2} for the left eigenvector $l_n(x)$
 that cannot be satisfied anymore for $k>  \frac{ \mu }{2}$
 \begin{eqnarray}
m_{2k}^* \equiv \int_{x_L}^{x_R} dx  x^{2k} P_*(x)=+\infty \ \ \ {\rm for } \ \  k>  \frac{ \mu }{2}
\label{m2kdv}
\end{eqnarray}
The spectral decomposition of the quantum propagator is then of the mixed form of Eq. \ref{psispectralfull}
\begin{eqnarray}
\psi_t(x \vert x_0)  \equiv  \langle x \vert e^{- H t} \vert x_0 \rangle
= \sum_{0 \leq n< \frac{\mu}{2} } e^{- t  n ( \mu- n)} \langle x \vert \phi_n \rangle \langle\phi_n \vert x_0 \rangle
+ \int_0^{+\infty}  dq e^{-t (\mu^2+q^2 )}  \langle x \vert \varphi_q \rangle \langle \varphi_q \vert x_0 \rangle
\label{psispectralfullwong}
\end{eqnarray}
where the energy continuum $E(q)=\frac{\mu^2+q^2}{4} \in ]\frac{   \mu^2   }{4  },+\infty[$ will be discussed later in Eq. 
\ref{continuum} as well as in Eqs \ref{continuumkesten} \ref{continuumfisher} \ref{continuumstudent} 
for the three Pearson examples (iii-iv-vi).
Both discrete and eigenvectors are written explicitly in terms of known families of orthogonal polynomials 
and of special functions in \cite{pearson_wong}
with explicit forms of the Fokker-Planck propagators.

%%%%%%%%%%%%%%%%%%%%%%%%%%%%%%%%

\subsubsection{ Physical meaning of the polynomial left eigenvectors $l_n(x)$ in relation with the moments $m_k(t)$ }

It is now interesting to discuss the physical meaning of the polynomial
 left eigenvector $l_n(x)$ in relation with the moments $m_k(t)$:
the associated observable $l_n^{av}[t] $ of Eq. \ref{lnav}
that displays the simple exponential dynamics of Eq. \ref{lnavdynintegrate}
corresponds here to the following linear combination of moments $m_k(t) $ using Eq. \ref{leftpolynomial}
\begin{eqnarray}
l_n^{av}[t] \equiv \int_{x_L}^{x_R} dx \left[  \sum_{k=0}^n K_{nk} x^k \right]  P_t(x)
=  \sum_{k=0}^n K_{nk} m_k(t) 
 \label{lnavpearson}
\end{eqnarray}
The coefficients $K_{nj}$ of the polynomials $l_n(x)$ of Eq. \ref{leftpolynomial}
are determined by their orthonormalization
of Eq. \ref{ortholeftalone} with respect to the steady state $P_*(x)$.
So the orthogonal polynomials $l_n(x)$ with respect to the steady state $P_*(x)$ provide the systematic construction of 
the linear combinations of the moments $m_k(t)$ that follow the simple exponential dynamics of Eq. \ref{lnavdynintegrate}.

To be more concrete, let us discuss more precisely the first left eigenvectors $l_1(x)$ 
to make the link with the dynamics of the first moments $m_1(t)$ discussed in the previous section.
For $n=1$, the polynomial $l_1(x)$ of Eq. \ref{leftpolynomial}
\begin{eqnarray}
 l_1(x) =  K_{11} x + K_{10}
 \label{leftpolynomial1}
\end{eqnarray}
should satisfy the orthonormalization of Eq. \ref{ortholeftalone}
\begin{eqnarray}
0 && =  \int_{x_L}^{x_R} dx  l_0(x) l_1(x) P_*(x) = \int_{x_L}^{x_R} dx  \left[  K_11 x + K_{10} \right] P_*(x) 
= K_{11} m_1^* + K_{10}
\nonumber \\
1 &&=  \int_{x_L}^{x_R} dx  l_1^2(x) P_*(x) = \int_{x_L}^{x_R} dx  \left[  K_11 x + K_{10} \right]^2 P_*(x)
=  K^2_{11} m_2^* + 2 K_{11} K_{10} m_1^*+K_{10}^2
\label{ortholeftl1}
\end{eqnarray}
so one obtains the two coefficients $K_{11}$ and $K_{10}$ in terms of the two first steady moments $m_1^* $ and $m_2^* $
\begin{eqnarray}
K_{10}&& = - K_{11} m_1^*
\nonumber \\
1 && =  K^2_{11} \left[ m_2^* - (m_1^*)^2 \right]
\label{ortholeftl1sol}
\end{eqnarray}
The first polynomial $l_1(x)$ can be thus 
constructed only if the two first steady moments $m_1^* $ and $m_2^* $ are finite and then reads
\begin{eqnarray}
l_1(x) = \frac{ x - m_1^*}{\sqrt{m_2^* - (m_1^*)^2 } }
 \label{l1pearson}
\end{eqnarray}
The corresponding observable of Eq. \ref{lnavpearson}
\begin{eqnarray}
l_{1}^{av}[t] =   K_{11} m_1(t) +K_{10} = \frac{ m_1(t) - m_1^*}{\sqrt{m_2^* - (m_1^*)^2 } }
 \label{lnavpearson1}
\end{eqnarray}
is then directly related to the first moment $m_1(t)$ 
discussed in Eq. \ref{dynPearsonm1integ}.

Similarly for $n=2$, the polynomial $l_2(x)$ of Eq. \ref{leftpolynomial}
\begin{eqnarray}
 l_2(x) =K_{22} x^2 +  K_{21} x + K_{20}
 \label{leftpolynomial2}
\end{eqnarray}
can satisfy the normalization of Eq. \ref{ortholeftn2} only if the fourth steady moment $m_4^*$ is finite.
Then the corresponding observable
of Eq. \ref{lnavpearson}
\begin{eqnarray}
l_{2}^{av}[t] =   K_{22} m_2(t) + K_{21} m_1(t) +K_{20}
 \label{lnavpearson2}
\end{eqnarray}
is the linear combination of the two first moments $m_1(t)$ and $m_2(t)$ that would only involve the energy $E_2$
instead of the two energies $E_1=\epsilon_1$ and $E_2=\epsilon_2$ appearing in the dynamics of the second moment $m_2(t)$ of Eq. \ref{dynPearsonm1integ}.

%%%%%%%%%%%%%%%%%%%%%%%%%%%%%%%%

\subsubsection{ Link with the next section }

Now that we have discussed the spectral properties from the point of view of the left eigenvectors $l_n(x)$,
it is interesting in the next section to analyze them from the point of view of the quantum 
eigenvectors $\phi_n(x)$ of the quantum Hamiltonian of Eq. \ref{hamiltonien}.

%%%%%%%%%%%%%%%%%%%%%%%%%%%%%%%%

\section{ Quantum supersymmetric Hamiltonians associated to Pearson diffusions}

\label{sec_susy}

In this section, we first recall the factorisation of the quantum Hamiltonian of Eq. \ref{hamiltonien}
into its supersymmetric form $H= Q^{\dagger} Q$ for a general diffusion with $F(x)$ and  $D(x)$
and then focus on the corresponding specific properties for Pearson diffusions.

\subsection{ Factorisation of the quantum Hamiltonian $H= Q^{\dagger} Q$ for a diffusion with $F(x)$ and  $D(x)$}

\subsubsection{ Factorisation of the second-order differential operator 
$H= Q^{\dagger} Q$ into a first-order operator $Q$ and its adjoint $Q^{\dagger}$}

For a general diffusion process discussed in subsection \ref{subsec_geneDiff},
the Hamiltonian of Eq. \ref{hamiltonien} can be factorized
into the well-known supersymmetric form (see the review \cite{review_susyquantum} and references therein)
\begin{eqnarray}
H =  - \frac{ \partial  }{\partial x} D(x) \frac{ \partial  }{\partial x} +V(x) =   Q^{\dagger} Q
\label{hsusy}
\end{eqnarray}
involving the first-order operator 
\begin{eqnarray}
Q   \equiv    \sqrt{ D(x) }  \left( \frac{ d }{ d x}  +\frac{ U'(x)}{2 } \right)
\label{qsusy}
\end{eqnarray}
and its adjoint
\begin{eqnarray}
Q^{\dagger}  &&\equiv  \left(   - \frac{ d }{ d x}  +\frac{ U'(x)}{2 } \right)\sqrt{ D(x) }
\label{qdaggersusy}
\end{eqnarray}

The current $j_n(x)$ of Eq. \ref{jnquantum}
 involved in the vanishing-current boundary conditions of Eq. \ref{BCjn}
can be rewritten in terms of the quantum ground-state $\phi_0(x) =\frac{ e^{ - \frac{ U(x)}{2} } }{\sqrt Z}$ of Eq. \ref{phi0}
and in terms of the action of the operator $Q$ of Eq. \ref{qsusy} on 
the quantum eigenvector $\phi_n(x)$ 
\begin{eqnarray}
  j_n(x)   = - D(x)  \frac{e^{-\frac{U(x)}{2}}}{\sqrt{Z} }  \bigg(  \frac{ d }{ d x} + \frac{U'(x)}{2}   \bigg)    \phi_n(x)
  = - \sqrt{ D(x) } \ \phi_0(x)  \ \bigg( Q \phi_n(x) \bigg)
 \label{jnquantumQ}
\end{eqnarray}

The positive quantum ground-state of Eq. \ref{phi0} is annihilated by the first-order operator $Q$ 
\begin{eqnarray}
Q  \phi_0(x)  =  \sqrt{ D(x) }  \left( \frac{ d }{ d x}  +\frac{ U'(x)}{2 } \right) \frac{ e^{ - \frac{ U(x)}{2} } }{\sqrt Z} =0
 \label{Qphi0}
\end{eqnarray}
while the energy $E_n$ of the quantum eigenstate $\phi_n(x)$ can be rewritten 
as the square of the norm of $Q \vert \phi_n \rangle $ 
\begin{eqnarray}
E_n = \langle \phi_n \vert H \vert \phi_n \rangle=   \langle \phi_n \vert  Q^{\dagger} Q \vert \phi_n \rangle
= \vert \vert Q \vert \phi_n \rangle \vert \vert^2= \int_{x_L}^{x_R} dx \bigg[ Q \phi_n(x) \bigg]^2
= \int_{x_L}^{x_R} dx D(x) \bigg[ \phi_n'(x)  +\frac{ U'(x)}{2 } \phi_n(x) \bigg]^2
\label{hsusyen}
\end{eqnarray}

%%%%%%%%%%%%%%%%%%%%%%%%%%%%%%%%%

\subsubsection{ Analysis of the supersymmetric partner $\breve{H }= Q Q^{\dagger}  $
of the Hamiltonian $H= Q^{\dagger} Q$}

The supersymmetric partner $\breve{H }  = Q Q^{\dagger}$ of the Hamiltonian $H= Q^{\dagger} Q$ of Eqs \ref{hamiltonien}
and \ref{hsusy}
\begin{eqnarray}
\breve{H } \equiv    Q Q^{\dagger} = - \frac{ \partial  }{\partial x} D(x) \frac{ \partial  }{\partial x} +\breve{V}(x)
\label{hsusypartner}
\end{eqnarray}
involves the same kinetic operator as the initial Hamiltonian $H$ of Eq. \ref{hsusy},
but the partner-potential $\breve{V}(x) $ is different from the potential $V(x)$ of Eq. \ref{vfromu}
\begin{eqnarray}
\breve{V }(x) && \equiv D(x)  \frac{ [U'(x)]^2 }{4 }  +D(x) \frac{U''(x)}{2}
 + \frac{ [D'(x)]^2 }{4 D(x) }-\frac{D''(x)}{2}
\nonumber \\
&& = \frac{ F^2(x) }{4 D(x) } - \frac{F'(x)}{2} +\frac{F(x) D'(x) }{2D(x)}+ \frac{ [D'(x)]^2 }{4 D(x) }-\frac{D''(x)}{2}
\nonumber \\
&& = \frac{ F^2_I(x) }{4 D(x) } - \frac{ F_I'(x)  }{2} 
\label{vpartner}
\end{eqnarray}
and turns out to be simpler on the last line when using the Ito force $F_I(x)= F(x)+D'(x)$ of Eq. \ref{fokkerplancklangevin}, while the initial potential $V(x)$ of Eq. \ref{vfromu}
is simpler when using the Fokker-Planck force $F(x)$.

The commutator between the two operators $ Q^{\dagger}$ and $Q$ corresponds to the difference between the two Hamiltonians $H$ and $\breve{H }  $, and thus to the difference between the two potentials $V(x)$ and $\breve{V }(x) $
\begin{eqnarray}
[ Q^{\dagger}, Q]  && \equiv    Q^{\dagger} Q -  Q Q^{\dagger} = H- \breve{H } = V(x) - \breve{V }(x) 
\nonumber \\
&& =  F'(x) -\frac{F(x) D'(x) }{2D(x)}- \frac{ [D'(x)]^2 }{4 D(x) }+\frac{D''(x)}{2}
\nonumber \\
&& =   F_S'(x) - \frac{F_S(x) D'(x) }{2D(x)} 
\label{commutateur}
\end{eqnarray}
that turns out to be simpler on the last line when using the Stratonovich force $F_S(x)  = F(x)+  \frac{D' (x)}{2} $ of Eq. \ref{fokkerplancklangevin}.

%%%%%%%%%%%%%%%%%%%%%%%%%%%%%%%%%

\subsubsection{ Relations between the discrete spectra of the Hamiltonian $H= Q^{\dagger} Q$ 
and its supersymmetric partner $\breve{H }= Q Q^{\dagger}  $}

From the excited eigenstate $\phi_n(x)$ with $n>0$ of the initial Hamiltonian $H$ of Eq. \ref{hsusy},
one can construct the state
\begin{eqnarray}
\breve{ \phi}_n(x) \equiv \frac{ Q \phi_n(x) }{\sqrt{E_n} }=  \frac{ \sqrt{ D(x) } }{ \sqrt{E_n}} \left( \phi'_n(x)  +\frac{ U'(x)}{2 }\phi_n(x) \right)
\label{phinpartner}
\end{eqnarray}
that is normalized as a consequence of Eq. \ref{hsusyen}
\begin{eqnarray}
\langle \breve{ \phi}_n \vert \breve{ \phi}_n \rangle
=  \frac{ \langle \phi_n \vert  Q^{\dagger} Q \vert \phi_n \rangle }{ E_n}
=\frac{ \langle \phi_n \vert H \vert \phi_n \rangle }{E_n}
=1 
\label{phinpartnernorm}
\end{eqnarray}
and that is an eigenstate of the supersymmetric partner $ \breve{H } $ of Eq. \ref{hsusypartner}
associated to the energy $E_n$
\begin{eqnarray}
\breve{H } \vert \breve{ \phi}_n \rangle = \frac{ 1 }{\sqrt{E_n} } Q Q^{\dagger} Q \vert  \phi_n \rangle 
= \frac{ 1 }{\sqrt{E_n} } Q H \vert  \phi_n \rangle
= E_n \frac{ Q  \vert  \phi_n \rangle}{\sqrt{E_n}}  = E_n \vert \breve{ \phi}_n \rangle
\label{phinpartnereigen}
\end{eqnarray}
while the current $j_n(x)$ of Eq. \ref{jnquantumQ}
 involved in the vanishing-current boundary conditions of Eq. \ref{BCjn}
reads in terms of $\breve{ \phi}_n(x) $
\begin{eqnarray}
  j_n(x)    = - \sqrt{ D(x) } \ \phi_0(x)  \ \bigg( Q \phi_n(x) \bigg)
    = - \sqrt{ D(x) } \ \phi_0(x)  \sqrt{E_n} \breve{ \phi}_n(x)
 \label{jnquantumQpartner}
\end{eqnarray}

So the discrete energy spectra of the initial Hamiltonian $H =Q^{\dagger} Q $ with the vanishing boundary conditions
 for $j_n(x)$ of Eq. \ref{jnquantumQ}
and of its supersymmetric-partner $ \breve{H } =Q Q^{\dagger}$
with the vanishing boundary conditions for $j_n(x)$ of Eq. \ref{jnquantumQpartner}
coincide, 
except for the ground-state $\phi_0(x) $ of the initial Hamiltonian $H$ which is annihilated by $Q$ as mentioned in Eq. \ref{phi0}
and has thus no partner via Eq. \ref{phinpartner}.

In particular, let us stress that
 at a finite boundary $x_L=0$ where the diffusion coefficient $D(x_L=0)$ and the steady state $P_*(x_L=0)$
remain finite, the vanishing-current boundary condition $j_n(x_L=0)=0$ leads to the simplified conditions
\begin{eqnarray}
\text{ If $x_L=0$ with finite $D(0)$ and $P_*(0)$ } :    j_n(0)  =0
\to  \begin{cases}
0 = \bigg( Q \phi_n(x) \bigg)\vert_{x=0}   \text{ for the eigenvectors $\phi_n(x)$ of $H$} 
\\
0 =  \breve{ \phi}_n(x) \text{for the eigenvectors $\breve{ \phi}_n(x)$ of $\breve{H }$ } 
\end{cases}
\ \  \label{nonsingularboundaryquantum}
\end{eqnarray}
i.e. vanshing boundary conditions for the eigenvectors $\breve{ \phi}_n(x)$ of 
supersymmetric-partner $ \breve{H } =Q Q^{\dagger}$
(see \cite{c_boundarydriven} for a recent discussion of these change of boundary conditions
between supersymmetric partners 
in the context of boundary-driven non-equilibrium diffusions).
For Pearson diffusions where the diffusion coefficient vanishes $D(0)=0$ 
and where the steady state $P_*(0)$ can be vanishing, diverging, or finite as stressed in Eq. \ref{zeropower},
the analysis of boundary conditions will be completely different than in Eq. \ref{nonsingularboundaryquantum}, 
as discussed in the subsection \ref{subsec_BCnormaQuantum}. 

When the energy spectrum of the Hamiltonian $H$ contains also an energy continuum,
the supersymmetric partner $ \breve{H } $ contains the same energy continuum,
and one can also analyze the relations between their continuous eigenvectors
as described around Eqs 26-31 of the review \cite{review_susyquantum}.

%%%%%%%%%%%%%%%%%%%%%%%%%%%%%%%%%%%%%%%%%%%%%%%%%%%

\subsection{ Pearson diffusions : algebraic construction of the discrete spectrum $E_n$ of the Hamiltonian $H$}

Let us now describe the specific properties
of the supersymmetric Hamiltonian $H$
for Pearson diffusions, with respect to the general analysis summarized in the previous section.

 %%%%%%%%%%%%%%%%%%%%%%%%%%%%%%%%%%%%%%%%

\subsubsection{ Boundary conditions and normalizability of the quantum eigenvectors $\phi_n(x)$ and $\breve{ \phi}_n(x)$ for Pearson diffusions} 

\label{subsec_BCnormaQuantum}

The detailed analysis of subsection \ref{subsec_BCnormaLeft} concerning
the boundary conditions and the normalizability of the left eigenvectors $l_n(x)$
can be directly translated for the quantum eigenvectors $\phi_n(x)$ 
that are in direct correspondence via Eq. \ref{rnlnphin}.
The conclusion summarized in Eq. \ref{normalizability} becomes :
if the quantum eigenvector $\phi_n(x)$ is normalizable via Eq. \ref{orthophin}
\begin{eqnarray}
1 = \langle \phi_n \vert \phi_{n} \rangle 
= \int_{x_L}^{x_R} dx \phi_n^2(x) 
\label{normalizaphin}
\end{eqnarray}
 then the boundary conditions are automatically satisfied.
 This conclusion can be further adapted using Eq. \ref{jnquantumQpartner}. 
 for the discrete eigenvectors $\breve{ \phi}_n(x)$
 of the supersymmetric-partner $ \breve{H } =Q Q^{\dagger}$
 
 In summary, for the quantum eigenvectors $\phi_n(x)$ associated to the discrete eigenvalues $E_n$ of 
 the supersymmetric Hamiltonian $H =Q^{\dagger} Q $ of Pearson diffusions, 
 as well as for the eigenvectors $\breve{ \phi}_n(x)$
 of the supersymmetric-partner $ \breve{H } =Q Q^{\dagger}$,
one only needs to focus on the normalizability of these eigenvectors via Eq. \ref{normalizaphin}
without worrying about boundary conditions anymore.

%%%%%%%%%%%%%%%%%%%%%%%%%%%%%%%%%%%%

\subsubsection{ Form of the quantum potential $V(x)$ of the supersymmetric Hamiltonian $H$ associated to a Pearson diffusion}

For a Pearson diffusion of Eq. \ref{fpearson},
the quantum potential of Eq. \ref{vfromu} displays the following dependence with respect to $x$
and with respect to the two parameters $[\lambda,\gamma]$ of the Fokker-Planck force 
$F(x)= \lambda - \gamma x $
(while the diffusion coefficient $D(x)= a x^2+b x +c$ is considered as fixed)
\begin{eqnarray}
V_{[\lambda,\gamma]}(x)  = \frac{ F^2(x) }{4 D(x) } + \frac{F'(x)}{2} 
= \frac{ \left( \lambda - \gamma x\right)^2 }{4 \left(a x^2+b x +c \right) } - \frac{ \gamma }{2}
\label{vfromupearson}
\end{eqnarray}

For later purposes, it is important to stress that its fraction-decomposition allows to rewrite it as
\begin{eqnarray}
V_{[\lambda,\gamma]}(x)  
= \Upsilon^{[0]}_{[\lambda,\gamma]} 
+\Upsilon^{[1]}_{[\lambda,\gamma]}V_1(x)+\Upsilon^{[2]}_{[\lambda,\gamma]}V_2(x)
\label{vfromupearsonfracrtion}
\end{eqnarray}
in terms of three coefficients $\Upsilon^{[i]}_{[\lambda,\gamma]}$ that depend on the two parameters $[\lambda,\gamma]$,
while the two functions $V_1(x)$ and $V_2(x)$ will turn out to play an essential role in the further sections concerning large deviations properties.

%%%%%%%%%%%%%%%%%%%%%%%%%%%%%%%%%%%%%%

\subsubsection{ Partner-potential $\breve{V }(x) $ in terms of the initial potential $V_{[.,.]}(x)   $ with other parameters }

The partner potential $\breve{V}(x) $ of Eq. \ref{vpartner}
involving the linear Ito force $F_I(x)= \lambda_I - \gamma_I x  $ of Eq. \ref{fokkerplancklangevinpearson}
\begin{eqnarray}
\breve{V }(x) && = \frac{ F_I^2(x)^2 }{4 D(x) } - \frac{ F_I'(x)  }{2} 
= \frac{ ( \lambda_I - \gamma_I x )^2 }{4 \left(a x^2+b x +c \right) } + \frac{ \gamma_I }{2} 
 =  V_{[\lambda_I,\gamma_I]}(x)  + \gamma_I
\label{vpartnerpearson}
\end{eqnarray}
can be rewritten up to the additive constant $ \gamma_I $ as the initial potential $V_{[\lambda_I,\gamma_I]}(x) $ of Eq. \ref{vfromupearson}
with the modified parameters given in Eq. \ref{fokkerplancklangevinpearson}
\begin{eqnarray}
\lambda_I && =\lambda+b
\nonumber \\
\gamma_I && =\gamma-2 a
\label{coefsvpartnerpearson}
\end{eqnarray}
This property defines the so-called shape-invariant-potentials in the field of supersymmetric quantum mechanics (see the review \cite{review_susyquantum}) and allows to obtain
the energy spectrum of the initial potential via the construction of the iterated-partner-potentials as follows.

Since the initial potential $V_{[\lambda_I,\gamma_I]}(x) $ of Eq. \ref{vfromupearson} has a ground-state at zero energy $E_0=0$, one obtains that the ground-state of the partner potential $\breve{V}(x) $ of Eq. \ref{vpartnerpearson}
is given by the remaining constant $ \gamma_I $
\begin{eqnarray}
\breve{E }_0  = \gamma_I
\label{E0partner}
\end{eqnarray}
and via the construction explained around Eqs \ref{phinpartner} \ref{phinpartnereigen}, this corresponds to the energy $E_1$ of the first excited state $\phi_1(x)$ of the initial potential $V_{[\lambda,\gamma]}(x)$
\begin{eqnarray}
E_1 = \breve{E }_0  = \gamma_I
\label{E1E0partner}
\end{eqnarray}
in agreement with Eq. \ref{eigendiscrete} for $n=1$.

%%%%%%%%%%%%%%%%%%%%%%%%%%%%%%%%%%%%%%%%%%

\subsubsection{ Recurrence to construct the full discrete spectrum $E_n$ of the initial potential $V_{[\lambda,\gamma]}(x) $}

One may now iterate the above procedure as follows.
The $n$-iterated-partner potential will have the form
\begin{eqnarray}
\breve{V }^{[n]}(x)  =  V_{[\lambda^{[n]},\gamma^{[n]}]}(x)  + \breve{E }^{[n]}_0
\label{vpartnerpearsonordern}
\end{eqnarray}
where its ground-state energy $\breve{E }^{[n]}_0 $ coincides with the energy $E_n$ 
of the $n$-th excited state of the initial potential $V_{[\lambda,\gamma]}(x)$
\begin{eqnarray}
 \breve{E }^{[n]}_0  = E_n
\label{EnE0partner}
\end{eqnarray}
 while $V_{[\lambda^{[n]},\gamma^{[n]}]}(x) $
contains the appropriate parameters $[\lambda^{[n]},\gamma^{[n]}] $.
Then the $(n+1)$-iterated-partner potential can be computed via Eq. \ref{vpartnerpearson}
\begin{eqnarray}
\breve{V }^{[n+1]}(x)  
 =  V_{[\lambda_I^{[n]},\gamma_I^{[n]}]}(x)  + \gamma_I^{[n]} +E_n
 \equiv V_{[\lambda^{[n+1]},\gamma^{[n+1]}]}(x)  + E_{n+1}
\label{vpartnerpearsoniter}
\end{eqnarray}
where the identification leads to the following recurrences using Eq. \ref{coefsvpartnerpearson}
\begin{eqnarray}
\lambda^{[n+1]} && =\lambda_I^{[n]} = \lambda^{[n]} +b
\nonumber \\
\gamma^{[n+1]} && =\gamma_I^{[n]} = \gamma^{[n]} -2 a 
 \nonumber \\
  E_{n+1} -E_n && =  \gamma_I^{[n]} =  \gamma^{[n]} -2 a
\label{Recvpartnerpearsoniter}
\end{eqnarray}
So the two parameters $[\lambda^{[n]},\gamma^{[n]}] $ are simply linear with respect to $n$
\begin{eqnarray}
\lambda^{[n]} && = \lambda + b n
\nonumber \\
\gamma^{[n]} &&  = \gamma - 2 a n
\label{Recvpartnerpearsonitersol}
\end{eqnarray}
while the energy $E_n$ is quadratic
\begin{eqnarray}
  E_n && = \sum_{k=0}^{n-1} (E_{k+1}-E_k) = \sum_{k=0}^{n-1} (\gamma - 2 a - 2a k ) 
  = (\gamma - 2 a)n -  a  n (n-1) = \gamma_I n - a n (n-1)
\label{Enquadratic}
\end{eqnarray}
in agreement with the previous computations of Eq. \ref{eigendiscrete}
as it should.

In conclusion, the formulation with the quantum supersymmetric Hamiltonian $H$
allows to construct the full discrete spectrum $E_n$ of the initial potential $V_{[\lambda,\gamma]}(x) $ algebraically, without solving any differential equation.

%%%%%%%%%%%%%%%%%%%%%%%%%%%%%%%%%

\section{ Mappings towards other diffusions with additive or multiplicative noise  }

\label{sec_mapping}

Via changes of variables for the spatial coordinate $x$,
the Pearson diffusions can be mapped onto other diffusion processes
with additive or multiplicative noise, that will inherit the spectra and other properties of the Pearson diffusions,
even if they are not Pearson diffusions by themselves.
In this section, we describe two important examples.

%%%%%%%%%%%%%%%%%%%%%%%%%%%%%%%%%%

\subsection{ Change of variables $x \to z$ towards a diffusion process $z(t)$ with constant diffusion coefficient $d(z)=1$} 

Whenever the diffusion coefficient $D(x)$ depends on the spatial coordinate $x$,
 it is interesting to consider the change of variables from $x$ to the new space-coordinate $z$ 
\begin{eqnarray}
dz \equiv \frac{dx}{\sqrt{ D(x)} }
\label{xtoz}
\end{eqnarray}
that will produce the constant diffusion coefficient $d(z)=1 $ for the process $z(t)$.
The Stratonovich interpretation of Eq. \ref{langevin} leads to the following Langevin dynamics for $z(t)$
\begin{eqnarray}
dz(t) \equiv \frac{dx(t)}{\sqrt{ D(x(t))} } = \frac{ F_S( x (t) ) }{ \sqrt{ D(x(t))}} dt + \sqrt{ 2  }  \ dB(t)
\equiv f(z) dt + \sqrt{ 2  }  \ dB(t)
 \label{stratozz}
\end{eqnarray}
where the three forces, namely the Fokker-Planck force and the Ito force and the Stratonovich force 
coincide (since the diffusion coefficient $d(z)=1$ does not depend on $z$ in Eq. \ref{fokkerplancklangevin})
\begin{eqnarray}
 f(z) = f_S(z)=f_I(z) = \frac{F_S(x) }{\sqrt{ D(x)}} \bigg\vert_{x=x(z)}
= \frac{F(x) + \frac{D'(x)}{2} }{\sqrt{ D(x)}} \bigg\vert_{x=x(z)}
\label{fokkerplanckzforce}
\end{eqnarray}
As a consequence, the Fokker-Planck Eq. \ref{fokkerplanck} becomes for the probability density $p_t(z)$
to be at coordinate $z$ at time $t$
\begin{eqnarray}
 \partial_t p_t(z)    && =   -  \partial_z   j_t(z)
  \nonumber \\
 j_t(z) && \equiv f(z)   p_t(z) -   \partial_z p_t(z)
\label{fokkerplanckz}
\end{eqnarray}
 where the current $j_t(z) $ associated to $p_t(z) $
that involves the force $f(z) $ and the diffusion coefficient $d(z)=1$
satisfies the vanishing-current boundary conditions translated from Eq. \ref{jboundaries}
at the left boundary $z_L=z(x_L)$ and at the right boundary $z_R=z(x_R)$ obtained via the mapping $x \to z(x)$
\begin{eqnarray}
j_t(z_L) && =0
\nonumber \\
j_t(z_R) && =0
\label{jboundariesz}
\end{eqnarray}

The corresponding potential $u(z) $ obtained from Eq. \ref{Ux} with $ d(z)=1$
\begin{eqnarray}
u'(z) = - f(z) 
\label{Uz}
\end{eqnarray}
governs the steady state of Eq. \ref{steadyeq} for the process $z(t)$
\begin{eqnarray}
  p_*(z)  = \frac{ e^{ - u(z)} }{ \int_{z_L}^{z_R} dz' e^{ - u(z')}}
 \label{steadyeqz}
\end{eqnarray}
that can also be obtained from the initial steady state $P_*(x)$ 
via the application of the change of variables $x \to z(x)$ for the probability densities $p_*(z) dz = P_*(x) dx$.

The quantum supersymmetric Hamiltonian of Eq. \ref{hsusy} \ref{qsusy}
\begin{eqnarray}
 h = h^{\dagger} && =\left(   - \frac{ d }{ d z}  +\frac{ u'(z)}{2 } \right)
 \left( \frac{ d }{ d z}  +\frac{ u'(z)}{2 } \right) =  - \frac{d^2}{dz^2}  +v(z)
\label{hamiltonienz}
\end{eqnarray}
involves the standard Laplacian $\frac{d^2}{dz^2} $  associated to a constant mass 
(in contrast to Eq. \ref{hamiltonien} with an effective space-dependent mass)
and the quantum potential
\begin{eqnarray}
 v(z) && =  \frac{ [u'(z)]^2 }{4 }  - \frac{u''(z)}{2} 
 = \frac{ f^2(z) }{4  } + \frac{f'(z)}{2}
\label{quantumvz}
\end{eqnarray}
while the partner potential of Eq. \ref{vpartner} reads
\begin{eqnarray}
\breve{v }(z)  = \frac{ f^2(z) }{4  } - \frac{ f'(z)  }{2} 
\label{vpartnerz}
\end{eqnarray}

For Pearson diffusions, the change of variables of Eq. \ref{xtoz} involves elementary functions
and the corresponding potential $v(z)$ will correspond to solvable potentials
that are standard in the field of supersymmetric quantum mechanics 
(see the review \cite{review_susyquantum} and references therein).
In addition, the quantum Hamiltonian $h$ of Eq. \ref{hamiltonienz}
allows to see directly if there is a continuous spectrum of the form 
\begin{eqnarray}
 \text{ Continuous spectrum :} \ \ \ \ E \in ]v_{\infty},+\infty[
\label{continuum}
\end{eqnarray}
via the computation of the minimum of the two limiting values of the quantum potential $v(z)$ of Eq. \ref{quantumvz} 
as $z \to \pm \infty$
\begin{eqnarray}
 v_{\infty} = \min[ v(z \to +\infty) ; v(z \to - \infty) ]
\label{vinfty}
\end{eqnarray}
If $v_{\infty}<+\infty$ is finite, then the continuous spectrum is given by Eq. \ref{continuum},
while if $v_{\infty}=+\infty$ is infinite, then there is no continuous spectrum but only discrete energies. 

It will be also interesting to
translate the moments $m_k(t)$ of Eq. \ref{mktk} of the initial Pearson process $x(t)$ 
\begin{eqnarray}
m_k(t) \equiv  \int_{x_L}^{x_R} dx P_t(x) x^k  = \int_{z_L}^{z_R} dz  p_t(z) \left[ x(z) \right]^k
 \label{mktkz}
\end{eqnarray}
that involve the observables $\left[ x(z) \right]^k $ for the process $z(t)$.

%%%%%%%%%%%%%%%%%%%%%%%%%%%%%%%%%%%%%%%%%%

\subsection{ Change of variables $x \to  y= x^{- \frac{1}{q} }$ for positive Pearson diffusions $x(t) \in ]0,+\infty[$}

For some positive Pearson diffusion processes $x(t) \in ]0,+\infty[$, another interesting 
 change of variable involves the parameter $q>0$
\begin{eqnarray}
y= x^{- \frac{1}{q} } \in ]0,+\infty[
\label{xtoy}
\end{eqnarray}
The Stratonovich dynamics for $y(t)$ is obtained from the Stratonovich SDE for $x(t)$ of Eq. \ref{langevin}
with the force $F_S(x) = \lambda_S - \gamma_S x $ of Eq. \ref{fokkerplancklangevinpearson}
\begin{eqnarray}
dy(t)   && =  - \frac{1}{q} y^{1+q}  dx(t) 
= - \frac{1}{q} y^{1+q}(t) \left[  ( \lambda_S - \gamma_S y^{-q} (t) )dt + \sqrt{ 2 D(y^{-q}(t)) } dB_t \right] 
\nonumber \\
&& =  \frac{\gamma_S y (t)- \lambda_S y^{1+q}(t) }{q}   dt
 -     \sqrt{ 2 \frac{y^{2+2q}(t)}{q^2} D(y^{-q}(t)) } dB_t
 \equiv  {\cal F}_S( y (t) ) \ dt - \sqrt{ 2 {\cal D} ( y (t) ) }  \ dB(t)
\label{ystrato}
\end{eqnarray}
where the diffusion coefficient 
\begin{eqnarray}
{\cal D} ( y  )= \frac{y^{2+2q}}{q^2} D(y^{-q})
=  \frac{y^{2+2q}}{q^2} ( a y^{-2q}+by^{-q}+c)
= \frac{y^{2}}{q^2} ( a +by^{q}+c y^{2q})
\label{dy}
\end{eqnarray}
will be a polynomial in $y$ when $q$ is an integer,
while the Stratonovich force $ {\cal F}_S( y  )$ contains a linear contribution in $y$
and a non-linear contribution in $ y^{1+q}$ 
\begin{eqnarray}
{\cal F}_S ( y  )= \frac{\gamma_S  }{q} y - \frac{ \lambda_S  }{q} y^{1+q}
\label{fsy}
\end{eqnarray}
The corresponding Fokker-Planck force ${\cal F} ( y  ) $ and the Ito force ${\cal F}_I ( y  ) $ obtained from Eq. \ref{fokkerplancklangevin} 
\begin{eqnarray}
  {\cal F} ( y  ) = {\cal F}_S ( y  ) - \frac{{\cal D}' ( y  )}{2} 
  = \left( \frac{\gamma_S  }{q} -  \frac{a}{q^2} \right) y 
  - \left[ \frac{ \lambda_S  }{q} + b \left( \frac{1}{q^2}+\frac{1}{2 q} \right)\right] y^{1+q}
  -c \left( \frac{1}{q^2}+\frac{1}{q} \right) y^{1+2q}
\nonumber \\
  {\cal F}_I ( y  ) = {\cal F}_S ( y  ) + \frac{{\cal D}' ( y  )}{2} 
  = \left( \frac{\gamma_S  }{q} +  \frac{a}{q^2} \right) y 
   - \left[ \frac{ \lambda_S  }{q} - b \left( \frac{1}{q^2}+\frac{1}{2 q} \right)\right] y^{1+q}
  +c \left( \frac{1}{q^2}+\frac{1}{q} \right) y^{1+2q}
\label{ffiy}
\end{eqnarray}
may contain a supplementary non-linear contribution in $ y^{1+2q}$ when $c \ne 0$
with respect to the two other types of contributions involving $y$ and $y^{1+q}$ 
already present  the Stratonovich force $ {\cal F}_S( y  )$ of Eq. \ref{fsy}.
The Fokker-Planck Eq. \ref{fokkerplanck} reads for the probability density ${\cal P}_t(y)$
to be at coordinate $y$ at time $t$
\begin{eqnarray}
 \partial_t {\cal P}_t(y)    && =   -  \partial_z  {\cal J}_t(y)
  \nonumber \\
{\cal J}_t(y) && \equiv   {\cal F} ( y  )   {\cal P}_t(y) -  {\cal D} ( y  ) \partial_y {\cal P}_t(y)
\label{fokkerplancky}
\end{eqnarray}
 where the current ${\cal J}_t(y) $ associated to ${\cal P}_t(y)$
that involves the force ${\cal F} ( y  ) $ and the diffusion coefficient ${\cal D} ( y  )$
satisfies the vanishing-current boundary conditions translated from Eq. \ref{jboundaries}
at the left boundary $y_L=y(x_L)$ and at the right boundary $y_R=y(x_R)$ obtained via the mapping $x \to y(x)$
\begin{eqnarray}
{\cal J}_t(y_L) && =0
\nonumber \\
{\cal J}_t(y_R) && =0
\label{jboundariesy}
\end{eqnarray}
The moments $m_k(t)$ of Eq. \ref{mktk} of the initial Pearson process translate into
\begin{eqnarray}
m_k(t) \equiv  \int_{x_L}^{x_R} dx P_t(x) x^k  = \int_{y_L}^{y_R} dy  {\cal P}_t(y) y^{-q k} 
 \label{mktky}
\end{eqnarray}

These processes have attracted a lot of interest in the field of multiplicative stochastic processes \cite{review_multiplicative,carleman_multiplicative} :
for physical applications, the most relevant cases seem to be 
the first integer values $q=1$ and $q=2$ 
corresponding to quadratic and cubic non-linearities in the Stratonovich force $ {\cal F}_S( y  )$ 
of Eq. \ref{fsy}, while the diffusion coefficient ${\cal D} ( y  ) $ of Eq. \ref{dy} is a polynomial of low degree,
that will depend on the values of $(a,b,c)$ of the initial Pearson process,
so that it will be more appropriate to continue the discussion in the sections concerning 
the Pearson representative examples.

%%%%%%%%%%%%%%%%%%%%%%%%%%%%%%%%%%%%%%%%%

%%%%%%%%%%%%%%%%%%%%%%%%%%%%%%%%%%%

\section{ Large deviations at level 1 for time-averaged observables over $[0,T]$ } 

\label{sec_level1}

In this section, the goal is to study the statistical properties of 
time-averaged observables over the time-window $[0,T]$ 
and to analyze their large deviations properties in the limit of large time $T \to + \infty$.
We will stress the specific properties for Pearson diffusions with respect to the 
general diffusion process discussed in subsection \ref{subsec_geneDiff}.

%%%%%%%%%%%%%%%%%%%%%%%%%%%%%%%%%%%%%%%%%%%%%%

\subsection{ Time-averaged observables 
over the time-window $[0,T]$ for each given trajectory $x(0 \leq t \leq T)$}

For an observable $w(x)$, besides the averaged value $w^{av}[t] $ at time $t$ discussed in Eq. \ref{Oav}, 
it is interesting to consider the time-average over the time-window $t \in [0,T]$
for each given trajectory $x(0 \leq t \leq T)$
\begin{eqnarray}
W [x(0 \leq t \leq T) ]  \equiv  \frac{1}{T}   \int_0^T  dt w(x(t))
 \label{empi}
\end{eqnarray}

For a general diffusion discussed in subsection \ref{subsec_geneDiff}, 
we will consider in particular the cases 
where the observable $w(x)$ is a left eigenvector $l_n(x)$ with $n>0$, 
\begin{eqnarray}
L_n [x(0 \leq t \leq T) ] \equiv \frac{1}{T}   \int_0^T  dt  \ l_n(x(t))
 \label{Lnempi}
\end{eqnarray}
since the corresponding averaged value $l_n^{av}[t] $ at time $t$ of Eq. \ref{lnav}
follows the simple dynamics of Eq. \ref{lnavdynintegrate}.

For Pearson diffusions, it is also natural to study 
the time-averaged moment of order $k$ corresponding to $w(x)=x^k$ 
\begin{eqnarray}
M_k [x(0 \leq t \leq T) ]=   \frac{1}{T}   \int_0^T  dt \ x^k(t) 
 \label{Mkempi}
\end{eqnarray}
in order to compare with the moment $m_k(t)$ of order $k$ at time $t$ of Eq. \ref{mktk}.
We will also identify the specific observables $w(x)$
whose rate function can be explicitly computed.

%%%%%%%%%%%%%%%%%%%%%%%%%%%%%%%%%%%%%%%%%%%%%%

\subsection{ Averaged value $W^{av}[T]$ of $W [x(0 \leq t \leq T) ] $ over the trajectories $x(0 \leq t \leq T) $ as a function of $T$}

The averaged value of Eq. \ref{empi} over the trajectories $x(0 \leq t \leq T)$
reduces to the time-average over the time-window $t \in [0,T]$
of the averaged value $w^{av}[t] $ at time $t$ of Eq. \ref{Oav}
\begin{eqnarray}
W^{av} (T) \equiv \overline{ W [x(0 \leq t \leq T) ] } 
=  \frac{1}{T}   \int_0^T  dt  \overline{ w(x(t)) } 
=  \frac{1}{T}   \int_0^T  dt \int_{x_L}^{x_R} dx w(x) P_t(x)
= \frac{1}{T}   \int_0^T  dt w^{av}[t]
 \label{empiav}
\end{eqnarray}
and will converge towards the steady value $w_*$ for large $T$
\begin{eqnarray}
W^{av} (T) =  \frac{1}{T}   \int_0^T  dt \int_{x_L}^{x_R} dx w(x) P_t(x)
\opsimeq_{T \to + \infty} \int_{x_L}^{x_R} dx w(x) P_*(x) \equiv w_*
 \label{empiavSteady}
\end{eqnarray}

For a general diffusion with $F(x)$ and $D(x)$, 
the averaged value of $L_n [x(0 \leq t \leq T) ]$ of Eq. \ref{Lnempi} with $n>0$
over the trajectories $x(0 \leq t \leq T)$
can be obtained from the solution of Eq. \ref{lnavdynintegrate}
for $l_n^{av}[t] $
\begin{eqnarray}
L_n^{av} [x(0 \leq t \leq T) ] = \frac{1}{T}   \int_0^T  dt  \ l_n^{av}[t]
=  \frac{1}{T}   \int_0^T  dt l_n^{av}[0] e^{- t E_n} 
= l_n^{av}[0]  \frac{1- e^{-T E_n} }{T E_n }  
 \label{Lnempiav}
\end{eqnarray}
so that it converges as $\frac{1}{T}$ towards its vanishing steady value $l_n^*=0$
as a consequence of Eq. \ref{ortholeftalone} with $l_0(x)=1$ 
\begin{eqnarray}
l_n^* \equiv \int_{x_L}^{x_R} dx l_n(x) P_*(x) =\int_{x_L}^{x_R} dx l_0(x) l_n(x) P_*(x) = 0
 \label{LnempiavSteady}
\end{eqnarray}

For Pearson diffusions, 
 the results for the averaged valued $m_k(t)$ at time $t$ of Eq. \ref{mktk}
allow to obtain the corresponding behaviors of the time-averages over the time-window $t \in [0,T]$
\begin{eqnarray}
M^{av}_k (T)= \overline{  \frac{1}{T}   \int_0^T  dt x^k(t) } 
= \frac{1}{T}   \int_0^T  dt m_k(t)
 \label{empiavMk}
\end{eqnarray}
For $k=1$, the explicit result of Eq. \ref{dynPearsonm1integ} for $m_1(t)$ yields
\begin{eqnarray}
M_1^{av} (T) = \frac{1}{T}   \int_0^T  dt   m_1(t)    
=  \frac{\lambda_I}{\gamma_I} 
+ \left(  m_1(0)  - \frac{\lambda_I}{\gamma_I} \right) \frac{ 1- e^{-  T E_1}  }{ T E_1 }
\ i.e.   \begin{cases}
\text{ convergence towards $m_1^*=\frac{\lambda_I}{\gamma_I}  $
as $\frac{1}{T}$ for $E_1 >0$ } \\
 \text{ exponential growth towards $m_1^*=+\infty$ for $E_1 <0$}
\end{cases}
 \label{M1empi}
\end{eqnarray}
and one can similarly use the explicit result of Eq. \ref{OavdynPearsonkm2integ} for $m_2(t)$ to obtain 
\begin{eqnarray}
M_2^{av} (T) = \frac{1}{T}   \int_0^T  dt   m_2(t)   
\ \ \ {\rm i.e. } \ \   \begin{cases}
\text{ convergence towards $m_2^*=\frac{c +  ( b   + \lambda_I  )   \frac{\lambda_I}{\gamma_I}}{ (  \gamma_I -a ) }$ 
 for $E_1 >0$ and $E_2 >0 $} \\
 \text{ exponential growth towards $m_2^*=+\infty$ otherwise}
\end{cases}
 \label{M2empi}
\end{eqnarray}

%%%%%%%%%%%%%%%%%%%%%%%%%%%%%%%%%%%%%%%%%%%%%%

\subsection{ Rescaled variance of $W [x(0 \leq t \leq T) ] $ over the trajectories $x(0 \leq t \leq T) $ for large $T$}

The second moment of $W [x(0 \leq t \leq T) ] $ of Eq. \ref{empi}
involves the two-time-correlation between $w(x(t)) $ and $w(x(t+\tau)) $
\begin{eqnarray}
  \overline{ W^2 [x(0 \leq t \leq T) ] } 
&& =   \frac{2}{T^2}   \int_0^T  dt   \int_0^{T-t}   d\tau   \overline{ w(x(t+\tau)) w(x(t)) }
\nonumber \\
&& = \int dx w(x) \int dy  w(y)  \frac{2}{T^2}   \int_0^T  dt   \int_0^{T-t}   d\tau P_{\tau}(x \vert y) P_t(y \vert x_0)
 \label{empi2}
\end{eqnarray}
while the variance involves the corresponding two-time-connected-correlation
\begin{eqnarray}
 && \overline{ W^2 [x(0 \leq t \leq T) ] } - \left( \overline{ W [x(0 \leq t \leq T) ] }\right)^2
 =   \frac{2}{T^2}   \int_0^T  dt   \int_0^{T-t}   d\tau  
\left[  \overline{ w(x(t+\tau)) w(x(t)) } - \overline{ w(x(t+\tau)) } \times \overline { w(x(t)) }  \right] 
\nonumber \\
&& =   \int dx w(x) \int dy  w(y)  \frac{2}{T^2}   \int_0^T  dt   \int_0^{T-t}   d\tau 
\left[ P_{\tau}(x \vert y) - P_{t+\tau}(x \vert x_0) \right] P_t(y \vert x_0) 
 \label{empivar}
\end{eqnarray}

The multiplication of Eq \ref{empivar} by $T$
allows to obtain the following finite limit as $T \to + \infty$ for the rescaled variance 
\begin{eqnarray}
T \left[  \overline{ W^2 [x(0 \leq t \leq T) ] } - \left( \overline{ W [x(0 \leq t \leq T) ] }\right)^2 \right]
&& \opsimeq_{T \to + \infty} 
2   \int dx w(x)    \int dy  w(y) \int_0^{+\infty}   d\tau
\left[ P_{\tau}(x \vert y) - P_*(x ) \right] P_*(y )
\nonumber \\
&& \equiv 2  \int dx  \int dy  w(x) G(x,y) w(y) P_*(y )
 \label{empivargreen}
\end{eqnarray}
where the Green function $G(x,y)$ characterizes the convergence of the propagator $P_{\tau}(x \vert y) $
with its spectral decomposition of Eq. \ref{FPspectralq}
towards the steady state $P_*(x ) $
\begin{eqnarray}
G(x,y) \equiv  \int_0^{+\infty}   d\tau \left[ P_{\tau}(x \vert y) - P_*(x ) \right]
=  \int_0^{+\infty}   d\tau \left[ \sum_{n>0} r_n(x) l_n(y) e^{- E_n \tau}\right]
=\sum_{n>0} \frac{ r_n(x) l_n(y) }{ E_n }
 \label{greeninteg}
\end{eqnarray}
The properties of this Green function are discussed in detail in \cite{c_ruelle,c_SkewDB,us_kemeny}.

Alternatively, the limit of the rescaled variance of Eq. \ref{empivargreen}
can be written
\begin{eqnarray}
T \left[  \overline{ W^2 [x(0 \leq t \leq T) ] } - \left( \overline{ W [x(0 \leq t \leq T) ] }\right)^2 \right]
&& \opsimeq_{T \to + \infty} 
2  \int dy  w(y) P_*(y )
\int_0^{+\infty}   d\tau 
\left[  \int dx w(x)    P_{\tau}(x \vert y) -  \int dx w(x)    P_*(x ) \right] 
\nonumber \\
&& \equiv 
2  \int dy  w(y) P_*(y )
\int_0^{+\infty}   d\tau \left(  w^{av}[\tau \vert y ] - w_* \right)
 \label{rescaledvar}
\end{eqnarray}
in terms of the average $w^{av}[\tau \vert y ]$ of $w(x(\tau))$ when the starting point is $x(\tau=0)=y$
\begin{eqnarray}
w^{av}[\tau \vert y ] =  \int dx w(x)   P_{\tau}(x \vert y) 
 \label{conditionalav}
\end{eqnarray}
that has for initial value at $\tau=0$
\begin{eqnarray}
w^{av}[\tau=0 \vert y ] =  \int dx w(x)   P_{\tau=0}(x \vert y) = w(y)
 \label{conditionalavzero}
\end{eqnarray}
and that converges for $\tau \to +\infty$ towards the steady value $w_*$
\begin{eqnarray}
w^{av}[\tau \vert y ] \opsimeq_{\tau \to + \infty}  \int dx w(x)   P_*(x ) = w_*
 \label{conditionalavinfinity}
\end{eqnarray}

%%%%%%%%%%%%%%%%%%%%%%%%%%%%%%%%%%%%%%%%%%%%%%

\subsubsection{ Rescaled variance of $L_n [x(0 \leq t \leq T) ] $ corresponding to $w(x)=l_n(x)$ for a general diffusion with $F(x)$ and $D(x)$}

The application of Eq. \ref{empivargreen} to $L_n [x(0 \leq t \leq T) ] $ of Eq. \ref{Lnempi} corresponding to $w(x)=l_n(x)$ with $n>0$ yields that its rescaled variance
reads using Eqs \ref{ortholr} and \ref{ortholeftalone}
\begin{eqnarray}
&& T \left[  \overline{ L_n^2 [x(0 \leq t \leq T) ] } - \left( \overline{ L_n [x(0 \leq t \leq T) ] }\right)^2 \right]
 \opsimeq_{T \to + \infty} 
 2  \int dx  \int dy  l_n(x) G(x,y) l_n(y) P_*(y )
 \nonumber \\
 && =  2  \int dx  \int dy  l_n(x) \left[\sum_{n'>0} \frac{ r_{n'}(x) l_{n'}(y) }{ E_{n'} } \right]  l_n(y) P_*(y )
   =  \sum_{n'>0} \frac{2 }{ E_{n'} }  \left[ \int dx    l_n(x)  r_{n'}(x) \right]
  \left[ \int dy l_{n'}(y)  l_n(y) P_*(y )  \right] 
  \nonumber \\
 && 
   =  \sum_{n'>0} \frac{2 }{ E_{n'} }  \delta_{n,n'} \delta_{n,n'} = \frac{2}{E_n} 
    \label{empivargreenLn}
\end{eqnarray}
so that it only involves the inverse of the energy $E_n$ associated to the left eigenvector $l_n(x)$.

As an example of application of the alternative formula of Eq. \ref{rescaledvar},
we can use the simple dynamics of Eq. \ref{lnavdynintegrate} for $l_n^{av}[t] $
to obtain the conditional moment of Eq. \ref{conditionalav}
\begin{eqnarray}
l_n^{av}[\tau \vert y ] =  \int dx l_n(x)   P_{\tau}(x \vert y) = l_n(y) e^{- \tau E_n} 
 \label{lnconditionalav}
\end{eqnarray}
so that the asymptotic rescaled variance of Eq. \ref{rescaledvar} reads using Eq. \ref{ortholeftalone}
\begin{eqnarray}
&& T \left[  \overline{ L_n^2 [x(0 \leq t \leq T) ] } - \left( \overline{ L_n [x(0 \leq t \leq T) ] }\right)^2 \right]
 \opsimeq_{T \to + \infty} 
2  \int dy  l_n(y) P_*(y )\int_0^{+\infty}   d\tau l_n^{av}[\tau \vert y ] 
\nonumber \\
&& = 2  \int dy  l_n(y) P_*(y )\int_0^{+\infty}   d\tau  l_n(y) e^{- \tau E_n} 
= \frac{2}{E_n} \int dy  l^2_n(y) P_*(y ) = \frac{2}{E_n}
 \label{Lnrescaledvar}
\end{eqnarray}
in agreement with the previous analysis of Eq. \ref{empivargreenLn}.

The computation of Eq. \ref{empivargreenLn} suggests to consider the observables $w(x)$
that can be decomposed onto the series of the left eigenvectors $l_n(x)$
\begin{eqnarray}
w(x)= \langle w \vert x \rangle =\langle w \vert \bigg( \sum_{n=0}^{+\infty} \vert r_n \rangle \langle l_n \vert \bigg) \vert x \rangle =  \sum_{n=0}^{+\infty} w_n  l_n(x) 
 \label{wxln}
\end{eqnarray}
with the coefficients
\begin{eqnarray}
w_n= \langle w \vert  r_n \rangle = \int dx w(x) r_n(x) = \int dx w(x) l_n(x) P_*(x)
 \label{wxlncoefs}
\end{eqnarray}
Then the application of Eq. \ref{empivargreen} 
reads using Eqs \ref{ortholr} and \ref{ortholeftalone}
\begin{eqnarray}
&& T \left[  \overline{ W^2 [x(0 \leq t \leq T) ] } - \left( \overline{ W [x(0 \leq t \leq T) ] }\right)^2 \right]
 \opsimeq_{T \to + \infty} 
 2  \int dx  \int dy  w(x) G(x,y) w(y) P_*(y )
 \nonumber \\
 && =  2  \int dx  \int dy \left[\sum_{n'=0}^{+\infty} w_{n'}  l_{n'}(x)  \right]  \left[\sum_{n>0} \frac{ r_{n}(x) l_{n}(y) }{ E_n } \right] \left[\sum_{n''=0}^{+\infty} w_{n''}  l_{n''}(y)  \right] P_*(y )
   \nonumber \\
 &&  =  \sum_{n>0} \frac{2 }{ E_n } \sum_{n'=0}^{+\infty} w_{n'}\sum_{n''=0}^{+\infty} w_{n''}
  \left[ \int dx    l_{n'}(x)  r_{n}(x) \right]
  \left[ \int dy l_{n}(y)  l_{n''}(y) P_*(y )  \right] 
  \nonumber \\
 && 
   =  \sum_{n>0} \frac{2 }{ E_n } \sum_{n'=0}^{+\infty} w_{n'}\sum_{n''=0}^{+\infty} w_{n''}
    \delta_{n',n} \delta_{n'',n} = 2 \sum_{n>0} \frac{ w_n^2}{E_n} 
    \label{empivargreenwxlnn}
\end{eqnarray}
so that it involves the energies $E_n$ and the squares of the coefficients $w_n$ of Eq. \ref{wxlncoefs}.
 
 %%%%%%%%%%%%%%%%%%%%%%%%%%%%%%%%%%%%%%%

\subsubsection{ Rescaled variance of $M_k [x(0 \leq t \leq T) ] $ associated to $w(x)=x^k$ for Pearson diffusions}

For Pearson diffusions, let us first consider $M_1 [x(0 \leq t \leq T) ] $ 
of Eq. \ref{Mkempi}
associated to $w(x)=x$.
One can adapt the explicit result of Eq. \ref{dynPearsonm1integ} for $m_1(t)$
to obtain the moment of Eq. \ref{mkconditionalav}
when $E_1=\gamma_I>0$ 
with the finite steady value $m_1^* =\frac{\lambda_I}{\gamma_I} $
of Eq. \ref{conditionalav}
\begin{eqnarray}
m_1[\tau \vert y ] =  \int dx x   P_{\tau}(x \vert y) = 
m_1^* + \left(  y  - m_1^* \right) e^{-  \tau E_1}
 \label{mkconditionalav}
\end{eqnarray}
so that the asymptotic rescaled variance of Eq. \ref{rescaledvar} for $M_1 [x(0 \leq t \leq T) ] $ of Eq. \ref{Mkempi}
reads 
using Eq. \ref{ortholeftalone}
\begin{eqnarray}
T \left[  \overline{ M_1^2 [x(0 \leq t \leq T) ] } - \left( \overline{ M_1 [x(0 \leq t \leq T) ] }\right)^2 \right]
&& \opsimeq_{T \to + \infty} 
2  \int dy  y P_*(y )
\int_0^{+\infty}   d\tau \left( m_1[\tau \vert y ]  - m_1^* \right)
\nonumber \\
&& = 2  \int dy  y P_*(y )
\int_0^{+\infty}   d\tau  \left(  y  - m_1^* \right) e^{-  \tau E_1}
= \frac{2}{E_1}  \int dy  \left(  y^2  - m_1^* y\right) P_*(y )
\nonumber \\
&&= \frac{2}{E_1} [ m_2^* - (m_1^*)^2 ]
 \label{M1rescaledvar}
\end{eqnarray}
i.e. the numerator involves the variance $[ m_2^* - (m_1^*)^2 ]$ in the steady state,
while the denominator involves the energy $E_1$.
This result can be recovered from the inversion of Eq. \ref{l1pearson}
\begin{eqnarray}
  x = m_1^* + \sqrt{m_2^* - (m_1^*)^2 } l_1(x)
 \label{l1pearsoninversion}
\end{eqnarray}
that corresponds to the decomposition of Eq. \ref{wxln}
that only involves the two terms $n=0$ and $n=1$ with the two coefficients $w_0=m_1^*$ 
and $w_1=\sqrt{m_2^* - (m_1^*)^2 }$, so that Eq. \ref{M1rescaledvar} is in agreement with Eq. \ref{empivargreenwxlnn}.

The application of Eq. \ref{empivargreenwxlnn}
to $M_k [x(0 \leq t \leq T) ] $ of Eq. \ref{Mkempi} associated to $w(x)=x^k$
\begin{eqnarray}
&& T \left[  \overline{ M_k^2 [x(0 \leq t \leq T) ] } - \left( \overline{ M_k [x(0 \leq t \leq T) ] }\right)^2 \right]
 \opsimeq_{T \to + \infty} 
  2 \sum_{n=1}^k \frac{ w_n^2}{E_n} 
    \label{empivargreenwxlnnMk}
\end{eqnarray}
involves the $k$ coefficients of Eq. \ref{wxlncoefs} with $n=1,..,k$
\begin{eqnarray}
w_n=   \int dx x^k l_n(x) P_*(x)
 \label{wxlncoefsMk}
\end{eqnarray}
that appear in the decomposition of $x^k$ onto the left eigenvectors $l_n(x)$
\begin{eqnarray}
x^k = \sum_{n=0}^{k} w_n  l_n(x) 
 \label{wxlnMk}
\end{eqnarray}
that corresponds to the inversion of Eq. \ref{leftpolynomial}.

%%%%%%%%%%%%%%%%%%%%%%%%%%%%%%%%%%%%%

\subsection{ Large deviations of time-averaged observables for large $T$ for arbitrary $F(x)$ and $D(x)$  }

\subsubsection{ Generating function $Z^{[p]}_T (x \vert x_0) $ of $W [x(0 \leq t \leq T) ] $ over the trajectories $x(0 \leq t \leq T) $  }

To analyze the statistical properties of $W [x(0 \leq t \leq T) ] $ of Eq. \ref{empi}
beyond its averaged value and its rescaled variance described above,
it is convenient to consider the generating function of parameter $p$ 
over the trajectories 
starting at position $x_0$ at time $t=0$ and ending at position $x$ at time $T$
 \begin{eqnarray}
Z^{[p]}_T (x \vert x_0) && \equiv \overline{ e^{-p T W [x(0 \leq t \leq T) ] } }  
=  \overline{ e^{   -p \int_0^T dt w(x(t))   } }
\label{multifzpath}
\end{eqnarray}
whose dynamics is governed by the following p-deformed generator 
${\cal L}^{[p]}_x $ with respect to the initial Fokker-Planck generator ${\cal L}_x={\cal L}^{[p=0]}_x $ 
\begin{eqnarray}
\partial_T  Z^{[p]}_T (x \vert x_0)  =  {\cal L}^{[p]}_x Z^{[p]}_T (x \vert x_0)
&& =   -  \partial_{x}  \bigg( F(x)  - D (x)  \partial_{x}  \bigg) Z^{[p]}_T (x \vert x_0)  -p w (x) Z^{[p]}_T (x \vert x_0) 
\nonumber \\
&& =   -  \partial_{x}  J^{[p]}_T (x \vert x_0)  -p w (x) Z^{[p]}_T (x \vert x_0)   
\label{fokkerplancktilte}
\end{eqnarray}
The first contribution corresponds to the space derivative of the current $J^{[p]}_T (x \vert x_0) $ associated to $Z^{[p]}_T (x \vert x_0) $
\begin{eqnarray}
J^{[p]}_T (x \vert x_0)  \equiv  F(x)Z^{[p]}_T (x \vert x_0)  - D (x)  \partial_{x}   Z^{[p]}_T (x \vert x_0)
\label{currentZp}
\end{eqnarray}
while $p w (x)$ plays the role of a killing rate if $p w(x) >0$ or of a reproducing rate if $p w(x)<0$.
In particular, the integral of $Z^{[p]}_T (x \vert x_0)$ is not conserved
\begin{eqnarray}
\partial_T  \int_{x_L}^{x_R} dx Z^{[p]}_T (x \vert x_0)  
= - \left[J^{[p]}_T (x \vert x_0) \right]_{x_L}^{x_R} - p  \int_{x_L}^{x_R} dx w (x) Z^{[p]}_T (x \vert x_0) 
\label{fokkerplancktiltenotconserved}
\end{eqnarray}
as a consequence of the killing or reproducing rate $p w(x)$ in the bulk $x \in ]x_L,x_R[$,
but not as a consequence of currents flowing through the boundaries
(the physical picture is that a varying number of diffusive particles can be reproduced or killed in the bulk $x \in ]x_L,x_R[$,
but they cannot escape the interval, and exterior particles cannot enter either).
So the current $ J^{[p]}_T (x \vert x_0)$ of Eq. \ref{currentZp}
should vanish at the two boundaries as in Eq. \ref{jboundaries} for the initial problem corresponding to $p=0$
(see more detailed discussions in \cite{touchette-reflected} and in the PhD Thesis \cite{duBuisson_thesis}) : 
\begin{eqnarray}
J^{[p]}_T(x_L\vert x_0)=0=J^{[p]}_T(x_R \vert x_0)
\label{jboundariespZ}
\end{eqnarray}

 %%%%%%%%%%%%%%%%%%%%%%%%%%%%%%%%%%%%%%%%%%%%%%
 
 \subsubsection { Spectral decomposition of the generating function $Z^{[p]}_T (x \vert x_0) $ }

If one performs the same change of variables as in Eq. \ref{ppsi} for the generating function 
\begin{eqnarray}
Z^{[p]}_T (x \vert x_0)  = \sqrt{ \frac{P_*(x) }{P_*(x_0) }} \  \psi^{[p]}_T(x \vert x_0) 
\label{ppsitilt}
\end{eqnarray}
the $p$-deformed forward dynamics
of Eq. \ref{fokkerplancktilte} for the generating function $Z^{[p]}_T (x \vert x_0) $
translates into the euclidean Schr\"odinger equation 
\begin{eqnarray}
-\partial_T \psi^{[p]}_T(x \vert x_0) = H_p \psi^{[p]}_T(x \vert x_0)
\label{schropsitilte}
\end{eqnarray}
involving the $p$-deformed
Hermitian Hamiltonian with respect to Eq. \ref{hamiltonien}
\begin{eqnarray}
H_p  && = H_p^{\dagger} =  H +p w(x)  =  - \frac{ \partial  }{\partial x} D(x) \frac{ \partial  }{\partial x} +V_p(x)
\label{hamiltonianp}
\end{eqnarray}
where the scalar potential $V_p(x)$ involves the additional contribution $[p w(x)]$ with respect to the initial potential $V(x)$ of Eq. \ref{vfromu}
\begin{eqnarray}
V_p(x) = V(x) +p w(x)&&  = \frac{ F^2(x) }{4 D(x) } + \frac{F'(x)}{2}+p w(x)
\label{vfromup}
\end{eqnarray}

The spectral decomposition analogous to Eq. \ref{psispectral}
for the quantum propagator $\psi^{[p]}_T(x \vert x_0) $ 
\begin{eqnarray}
\psi^{[p]}_T(x \vert x_0)  \equiv  \langle x \vert e^{- t H_p } \vert x_0 \rangle
= \sum_{n=0}^{+\infty} e^{- t E_n(p) } \phi_n^{[p]}(x) \phi_n^{[p]}(x_0)
\label{psispectralp}
\end{eqnarray}
involving the quantum eigenvectors $\phi_n^{[p]}(x) $ of $H_p$ associated to the eigenvalues $E_n(p)$
\begin{eqnarray}
E_n(p) \phi_n^{[p]}(x)  = H_p \phi_n^{[p]}(x) = \bigg( - \frac{ \partial  }{\partial x} D(x) \frac{ \partial  }{\partial x} +V(x) +p w(x)\bigg)  \phi_n^{[p]}(x)
\label{htilteigen}
\end{eqnarray}
leads to the spectral decomposition analog to Eq. \ref{FPspectralq} for the generating function $Z^{[p]}_T (x \vert x_0)  $
\begin{eqnarray}
Z^{[p]}_T (x \vert x_0)   = \sum_{n=0}^{+\infty} e^{- t E_n(p)}  r^{[p]}_n(x) l^{[p]}_n(x_0)
\label{ZPspectral}
\end{eqnarray}
where
\begin{eqnarray}
r_n^{[p]}(x) && =  \sqrt{ P^*(x) } \phi_n^{[p]}(x) =  \frac{ e^{ - \frac{ U(x)}{2} } }{\sqrt Z} \phi_n^{[p]}(x)
\nonumber \\
l_n^{[p]}(x_0) && = \frac{1}{ \sqrt{ P^*(x_0) }}  \phi_n^{[p]}(x_0) = {\sqrt Z} e^{ + \frac{ U(x_0)}{2} } \phi_n^{[p]}(x_0)
\label{changeleftright}
\end{eqnarray}
are the right and the left eigenvectors of the p-deformed generator $ {\cal L}^{[p]}_x $
\begin{eqnarray}
-E_n(p) r^{[p]}_n(x)  && =  {\cal L}^{[p]}_x r^{[p]}_n(x) 
 =   -  \partial_{x}  \bigg( F(x) r^{[p]}_n(x) \bigg )  + \partial_{x}  \bigg(  D (x)  \partial_{x} r^{[p]}_n(x) \bigg) 
 -p w (x) r^{[p]}_n(x)
\nonumber \\
-E_n(p) l^{[p]}_n(x) && =  \left({\cal L}^{[p]}_x\right)^{\dagger} l^{[p]}_n(x) 
 =  F(x) \partial_x l^{[p]}_n(x)    +  \partial_x \bigg( D(x)  \partial_x l^{[p]}_n(x) \bigg)
-p w(x) l^{[p]}_n(x) 
\label{zpeigen}
\end{eqnarray}
satisfying the orthonormalization as in Eqs \ref{orthophin}  \ref{ortholr} \ref{ortholeftalone}
\begin{eqnarray}
\delta_{nm} && =  \langle \phi_n^{[p]} \vert \phi_m^{[p]} \rangle = \int_{x_L}^{x_R} dx  \phi_n^{[p]}(x)  \phi_m^{[p]}(x) 
\nonumber \\
&&  = \langle l_n^{[p]} \vert r_m^{[p]} \rangle = \int_{x_L}^{x_R} dx  l_n^{[p]}(x)  r_m^{[p]}(x) 
 = \int_{x_L}^{x_R} dx l_n^{[p]}(x)  l_m^{[p]}(x)  P_*(x)
 \label{Zportho}
\end{eqnarray}
The vanishing-current boundary conditions inherited from Eq. \ref{jboundariespZ} 
\begin{eqnarray}
  j^{[p]}_n(x_L)=0=j^{[p]}_n(x_R)
 \label{BCjnpZ}
\end{eqnarray}
involve the current $j^{[p]}_n(x)$ associated to the right eigenvector $r^{[p]}_n(x)$
\begin{eqnarray}
  j^{[p]}_n(x) && \equiv \bigg( F(x)  - D (x)  \partial_{x} \bigg) r_n^{[p]}(x) 
  = - D(x) \bigg(  U'(x)  +  \partial_{x} \bigg)  r_n^{[p]}(x) 
 \label{jnrightpZ}
\end{eqnarray}
that can be translated for the quantum eigenvector $\phi^{[p]}_n(x)$ via Eq. \ref{changeleftright}
\begin{eqnarray}
  j^{[p]}_n(x)  && = - D(x) \bigg(  U'(x)  +  \partial_{x} \bigg)  \phi^{[p]}_n(x)   \frac{e^{-\frac{U(x)}{2}}}{\sqrt{Z} } 
 = - D(x)  \frac{e^{-\frac{U(x)}{2}}}{\sqrt{Z} }  \bigg(  \frac{U'(x)}{2} + \partial_{x} \bigg)    \phi^{[p]}_n(x) 
 \label{jnquantumpZ}
\end{eqnarray}
and for the left eigenvectors $ l_n^{[p]}(x) $ into
\begin{eqnarray}
  j^{[p]}_n(x)  = - D(x) \bigg(  U'(x)  +  \partial_{x} \bigg)  l^{[p]}_n(x) \frac{e^{-U(x)}}{Z}
  = -   D(x) \frac{e^{-U(x)}}{Z} \partial_x  l_n^{[p]}(x) 
  = - D(x) P_*(x) \partial_x l_n^{[p]}(x)
 \label{jnleftpZ}
\end{eqnarray}
i.e. the boundary conditions are exactly the same as in Eqs \ref{jnright} \ref{jnquantum} \ref{jnleft}.
concerning the initial process $p=0$.

As for $p=0$, it is simpler in practice to focus 
either on the quantum eigenvectors $\phi^{[p]}_n(x)$ 
or on the left eigenvectors $l^{[p]}_n(x)$.
In the next subsection, we summarize the spectral problem for 
the left eigenvector $l^{[p]}_0(x) $ associated to the lowest energy $E_0(p)$
that dominates the generating function of Eq. \ref{ZPspectral} for large time $T$
\begin{eqnarray}
Z^{[p]}_T (x \vert x_0) 
\opsimeq_{T \to + \infty} e^{ - T E_0(p) } r^{[p]}_0(x) l^{[p]}_0(x_0)
\label{ZpspectrallargeT}
\end{eqnarray}

%%%%%%%%%%%%%%%%%%%%%%%%%%%%%%%%%%%%%%%%

\subsubsection{ Spectral problem for the positive left eigenvector $l^{[p]}_0(x) $ associated to the lowest energy $E_0(p)$  }

\label{subsec_spectralLeftp}

The computation of the lowest eigenvalue $E_0(p)$
requires to solve the following spectral problem for the associated positive left eigenvector $l^{[p]}_0(x) $
corresponding to the $p$-deformation of the trivial unperturbed left eigenvector $l_0^{[p=0]}(x) =1$ : 

(1) the left eigenvector $l_0^{[p]}(x) $  
should satisfy the eigenvalue Eq. \ref{zpeigen} that can also be rewritten
 in terms of the Ito force $ $ of Eq. \ref{fokkerplancklangevin}
using Eq. \ref{adjointIto}
\begin{eqnarray}
-E_0(p) l^{[p]}_0(x)  =  \left({\cal L}^{[p]}_x\right)^{\dagger} l^{[p]}_0(x) 
&&  =  F(x) \partial_x l^{[p]}_0(x)    +  \partial_x \bigg( D(x)  \partial_x l^{[p]}_0(x) \bigg)-p w(x) l^{[p]}_0(x) 
 \nonumber \\
 && = F_I(x) \frac{d l^{[p]}_0(x)}{dx}     +   D(x)  \frac{d^2 l^{[p]}_0(x)}{dx^2} -p w(x) l^{[p]}_0(x)
\label{zpeigenzero}
\end{eqnarray}

(2) the left eigenvector $l_0^{[p]}(x) $ should satisfy the normalization of Eq. \ref{Zportho}
\begin{eqnarray}
1 =\langle l_0^{[p]} \vert r_0^{[p]} \rangle
= \int_{x_L}^{x_R} dx  l_0^{[p]}(x)  r_0^{[p]}(x) 
 = \int_{x_L}^{x_R} dx  \left[ l_0^{[p]}(x) \right]^2 P_*(x)
\label{orthophinzero}
\end{eqnarray}

(3) the left eigenvector $l_0^{[p]}(x) $ should satisfy the vanishing-current boundary conditions at $x_L$ and $x_R$
\begin{eqnarray}
  j^{[p]}_n(x_L) && = 0 
  \nonumber \\
    j^{[p]}_n(x_R) && = 0 
 \label{jnleftpZzeroxlxr}
\end{eqnarray}
for the current of Eq. \ref{jnleftpZ}
\begin{eqnarray}
  j^{[p]}_n(x)    = - D(x) P_*(x) \frac{d l^{[p]}_0(x)}{dx} 
 \label{jnleftpZzero}
\end{eqnarray}

%%%%%%%%%%%%%%%%%%%%%%%%%%%%%%%%%%%%%

\subsubsection{ Eigenvalue $E_0(p)$ as the generating function of the rescaled cumulants of $W [x(0 \leq t \leq T) ] $}

\label{subsec_E0cumulants}

Let us now recall why the lowest eigenvalue $E_0(p)$ dominating Eq. \ref{ZpspectrallargeT}
represents the generating function of the rescaled cumulants of $W [x(0 \leq t \leq T) ] $.
Indeed, the cumulants $\kappa_k(T)$ of $W [x(0 \leq t \leq T) ] $ are defined by the series expansion 
 \begin{eqnarray}
 \ln \left[  \overline{ e^{-p T W [x(0 \leq t \leq T) ] } }  \right]
&& =  \sum_{k=1}^{+\infty}  \frac{ (-p T)^k}{k!} \kappa_k(T)
\nonumber \\
&& = - p T \kappa_1 (T)
+ \frac{  p^2 T^2}{2} \kappa_2 (T)
- \frac{ p^3 T^3}{3!} \kappa_3 (T)
+  \frac{ p^4 T^4}{4!} \kappa_4 (T)+...
\label{defcumulants}
\end{eqnarray}
The large-time behavior of Eq. \ref{ZpspectrallargeT}
yields
\begin{eqnarray}
 E_0(p)  && = \lim_{T \to + \infty} \left[ - \frac{\ln \left[  \overline{ e^{-p T W [x(0 \leq t \leq T) ] } }  \right]}{T} \right]
 =   \sum_{k=1}^{+\infty} (-1)^{k+1} p^k  \lim_{T \to + \infty} \left[\frac{  T^{k-1}  \kappa_k(T) }{k!} \right]
\nonumber \\
&& =  p \lim_{T \to + \infty} \left[  \kappa_1 (T) \right]
- p^2  \lim_{T \to + \infty} \left[ \frac{   T \kappa_2 (T) }{2} \right] 
+ p^3  \lim_{T \to + \infty} \left[\frac{  T^2 \kappa_3 (T) }{3!} \right]
- p^4  \lim_{T \to + \infty} \left[\frac{  T^3 \kappa_4 (T) }{4!} \right]+...
\label{E0cumulants}
\end{eqnarray}
So the power expansion in $p$ 
of the eigenvalue $E_0(p)$ involving coefficients $e_k$ that do not depend on $T$
 \begin{eqnarray}
E_0(p)  = p e_1 - p^2 e_2 + p^3 e_3 - p^4 e_4 +O(p^5)
  \label{eseriesp}
\end{eqnarray}
allows to obtain the first cumulants via
the identification with Eq. \ref{E0cumulants} as follows :

(1) the first cumulant $\kappa_1(T)$, i.e. the average of $W [x(0 \leq t \leq T) ] $,
 converges to the finite limit $e_1$ for $T \to +\infty$ 
\begin{eqnarray}
   \kappa_1 (T) \equiv \overline{ W [x(0 \leq t \leq T) ] } \opsimeq_{T \to + \infty}  e_1
\label{cumulant1}
\end{eqnarray}

(2) the second cumulant $\kappa_2(T)$, i.e. the variance of $W [x(0 \leq t \leq T) ] $,
 scales as $\frac{1}{T}$ with an amplitude $[2e_2]$
\begin{eqnarray}
   \kappa_2(T)  \equiv \overline{ \left( W [x(0 \leq t \leq T) ]  -  \overline{ W [x(0 \leq t \leq T) ] }\right)^2 }  
    \opsimeq_{T \to + \infty} \frac{2 e_2}{T}
    \label{cumulant2}
\end{eqnarray}

(3) the third cumulant $\kappa_3(T)$
 scales as $\frac{1}{T^2}$ with an amplitude $[6 e_3]$
\begin{eqnarray}
   \kappa_3(T)  \equiv \overline{ \left( W [x(0 \leq t \leq T) ]  -  \overline{ W [x(0 \leq t \leq T) ] }\right)^3 }  
    \opsimeq_{T \to + \infty} \frac{6 e_3}{T^2}
    \label{cumulant3}
\end{eqnarray}

As a consequence, to obtain these first cumulants, 
one just needs to use the perturbation theory in $p$ \cite{c_ruelle,c_SkewDB}
to compute the first coefficients of the expansion of $E_0(p)$ of Eq. \ref{eseriesp}
and one obtains in terms of the unperturbed left and right eigenvectors 
$l_0(x)=1$ and $r_0(x)=P_*(x)$ of Eq. \ref{r0l0} 
the following results \cite{c_ruelle,c_SkewDB} :

(1) The first-order correction $e_1$
corresponds to the averaged value of the perturbation $w$
 computed in the unperturbed zero-eigenvalue subspace 
\begin{eqnarray}
e_1 = \langle  l_0 \vert  w  \vert r_0 \rangle =  \int_{x_L}^{x_R} dx  w(x) P_*(x) = w_*
  \label{energy1}
\end{eqnarray}
so that the first cumulant of Eq. \ref{cumulant1} is in agreement with the previous direct computation of Eq. \ref{empiavSteady}.

(2) The second-order correction 
\begin{eqnarray}
  e_2  =   \langle  l_0 \vert  w  G w  \vert r_0 \rangle
  =   \int_{x_L}^{x_R} dx   \int_{x_L}^{x_R} dy  w(x) G(x,y) w(y) P_*(y)
  \label{energy2}
\end{eqnarray}
involves the Green function of Eq. \ref{greeninteg},
so that the second cumulant of Eq. \ref{cumulant2} is in agreement with 
the previous direct computation of Eq. \ref{empivargreen}.

(3) The Green function $G$ also governs all the higher orders of perturbation theory.
For instance the third-order correction $e_3$ reads
\begin{eqnarray}
e_3 && =  \langle  l_0 \vert  w  G w G w \vert r_0 \rangle
  - e_1 \langle  l_0 \vert  w  G^2 w  \vert r_0 \rangle
  \nonumber \\
  && =  \int_{x_L}^{x_R} dx   \int_{x_L}^{x_R} dy \int_{x_L}^{x_R} dz w(x) G(x,y) w(y) G(y,z) w(z) P_*(z)
  - e_1  \int_{x_L}^{x_R} dx   \int_{x_L}^{x_R} dy  w(x) [G^2](x,y) w(y) P_*(y)
  \ \ 
  \label{energy3}
\end{eqnarray}

When the observable $w(x)$ is the left eigenvector $l_n(x)$ with $n>0$,
we have already discussed the averaged value in Eq. \ref{LnempiavSteady}
and the rescaled variance in Eq. \ref{empivargreenLn},
so that it is interesting to apply now Eq. \ref{energy3}
to obtain its rescaled third cumulant via Eq. \ref{cumulant3}.
Using Eqs \ref{ortholr} and \ref{ortholeftalone}
 and Eq. \ref{LnempiavSteady}, Eq. \ref{energy3} reduces for $w(x)=l_n(x)$ to
\begin{eqnarray}
e_3 && = 
 \int dx   \int dy \int dz 
 l_n(x) 
 \left[\sum_{n'>0} \frac{  r_{n'} (x) l_{n'}(y) }{ E_{n'} } \right] 
 l_n(y) 
 \left[\sum_{n''>0} \frac{  r_{n''} (y) l_{n''}(z) }{ E_{n''} } \right] 
 l_n(z) P_*(z)
 \nonumber \\
&& = \sum_{n'>0} \frac{1}{E_{n'}}\sum_{n''>0} \frac{1}{E_{n''}}
 \left[ \int dx    l_n(x)    r_{n'} (x) \right]
\left[ \int dy   l_{n'} (y)  l_n(y)  r_{n''} (y) \right]
\left[ \int dz  l_{n''}(z)  l_n(z) P_*(z) \right]
 \nonumber \\
&& = \sum_{n'>0} \frac{1}{E_{n'}}\sum_{n''>0} \frac{1}{E_{n''}}
 \delta_{n,n'}
\left[ \int dy   l_{n'} (y)  l_n(y)  r_{n''} (y) \right]
\delta_{n'',n}
 \nonumber \\
&& =  \frac{1}{E_n^2} \left[ \int dy    l_n^2(y)   r_n (y) \right]
=  \frac{1}{E_n^2}  \int dy    l_n^3(y)   P_* (y) 
  \label{energy3ln}
\end{eqnarray}

For the Pearson diffusions where the first left eigenvector $l_1(x)$ is given by Eq. \ref{l1pearson},
Eq. \ref{energy3ln} becomes
\begin{eqnarray}
e_3 && 
=  \frac{1}{E_1^2} \int dy    l_1^3(y)   P_* (y) 
=  \frac{1}{E_1^2} \int dy   \left[  \frac{ y - m_1^*}{\sqrt{m_2^* - (m_1^*)^2 } }\right]^3   P_* (y) 
=  \frac{ m_3^* - 3 m_1^* m_2^* + 2 m_1^3 }{E_1^2 [ \sqrt{m_2^* - (m_1^*)^2 }]^3 }
  \label{energy3l1}
\end{eqnarray}
i.e. the numerator involves the third cumulant of the steady state $P_*(x)$, 
while the denominator involves its variance.

%%%%%%%%%%%%%%%%%%%%%%%%%%%%%%%%%%%%%

\subsubsection{ Legendre transform of the eigenvalue $E_0(p)$ to obtain the rate function $I(W)$ governing the large deviations}

The link between the eigenvalue $E_0(p)$ discussed above
and the rate function $I(W)$ that governs the asymptotic behavior 
for large $T$ of the probability $P_T(W)$ to see a given value $W$ 
of $W [x(0 \leq t \leq T) ] $ of Eq. \ref{empi}
\begin{eqnarray}
P_T(W)  \opsimeq_{T \to +\infty}  e^{- \displaystyle T  I(W)  }
\label{level1}
\end{eqnarray}
is based on the saddle-point evaluation for large $T$ of the generating function $Z^{[p]}_T $
from the probability distribution $P_T[ W] $ of Eq. \ref{level1} 
\begin{eqnarray}
Z^{[p]}_T \equiv  \int d W \ P_T( W) \ e^{  - p T W  }
\opsimeq_{ T\to + \infty} \int d W \ e^{ - T \left[      I(W) +p W\right] } 
\opsimeq_{ T \to + \infty} e^{ - T E_0(p) }
\label{multifz}
\end{eqnarray}
The energy $E_0(p)  $ 
 thus corresponds to the Legendre transform of the rate function $I(W)$ 
\begin{eqnarray}
 I(W) +p W  && = E_0(p)
\nonumber \\
  I'(W) +p && =0
\label{legendre}
\end{eqnarray}
with the reciprocal Legendre transform
\begin{eqnarray}
I(W)   && =  E_0(p)-pW 
\nonumber \\
0 && = E_0'(p) - W 
\label{legendrereci}
\end{eqnarray}
In particular, 
the rate function $ I(W) $ vanishes 
at the steady value $W_*=w_*$ of Eq. \ref{empiavSteady}
where it is minimum
\begin{eqnarray}
I(W_* ) = 0 = I'(W_*)
 \label{Iminastar}
\end{eqnarray}
This corresponds to $p=0$ in Eq. \ref{legendre} \ref{legendrereci}, so that the steady value $W^*$
corresponds to the first derivative of $E_0(p)$ at $p=0$ in agreement with Eq. \ref{energy1}
\begin{eqnarray}
 W_*= E_0'(p=0) =e_1
\label{firstderivative}
\end{eqnarray}

%%%%%%%%%%%%%%%%%%%%%%%%%%%%%%%%%%%%%%%%%%

\subsubsection { Canonical conditioned process of parameter $p$ }

\label{sub_condi}

From the generating function $Z^{[p]}_T (x \vert x_0) $ of Eq. \ref{ZPspectral},
it is interesting to construct the propagator ${\mathring P}_{t} (x \vert x_0) $ of the canonical conditioned process of parameter $p$ via
\begin{eqnarray}
{\mathring P}^{[p]}_{t} (x \vert x_0)
\equiv e^{ t E_0(p)} \frac{ l^{[p]}_0(x) } { l^{[p]}_0(x_0) } Z^{[p]}_t (x \vert x_0)   
=  l^{[p]}_0(x)r^{[p]}_0(x) + \sum_{n=1}^{+\infty} e^{- t [E_n(p)-E_0(p) ] } l^{[p]}_0(x)r^{[p]}_n(x) \frac{  l^{[p]}_n(x_0) } { l^{[p]}_0(x_0) } 
\label{Ringpropagator}
\end{eqnarray}
that converges for $t \to +\infty$ towards the conditioned steady state ${\mathring P}_*^{[p]} (x ) $
given by the product of $l_0^{[p]}(x) $ and $r_0^{[p]}(x) $ or by the 
various other equivalent expressions using Eq. \ref{changeleftright}
\begin{eqnarray}
{\mathring P}_*^{[p]} (x ) && = l_0^{[p]}(x)  r_0^{[p]}(x) = \left[ l^{[p]}_0(x)\right]^2 P_*(x)
\nonumber \\
&& = \left[ \phi^{[p]}_0(x)\right]^2
\label{RingSteady}
\end{eqnarray}

The corresponding probability-preserving Fokker-Planck generator $  {\mathring {\cal L}}_x^{[p]}$ 
\begin{eqnarray}
{\mathring {\cal L}}_x^{[p]}   =   l_0^{[p]}(x) {\cal L}^{[p]}_x   \frac{ 1}{ l_0^{[p]}(x) }  +E_0(p)  
    =  - \partial_x \left[ {\mathring F}^{[p]}(x)   - D(x)   \partial_x  \right]
\label{RingGenerator}
\end{eqnarray}
involves the same diffusion coefficient $D(x)$ as the initial process,
while the conditioned force ${\mathring F}^{[p]}(x)$ 
contains the initial force $F(x)$ and the logarithmic derivative of the positive left eigenvector $l_0^{[p]}(x) $  
\begin{eqnarray}
{\mathring F}^{[p]}(x) \equiv F(x) + 2 D(x) \frac{d}{dx} \ln \left( l_0^{[p]}(x)\right)
 \label{Ringforce}
\end{eqnarray}
or by the following expressions using Eq. \ref{changeleftright} and $U'(x)= - \frac{F(x)}{D(x)}$ of Eq. \ref{Ux}
\begin{eqnarray}
{\mathring F}^{[p]}(x)  && = F(x)  + 2 D(x) \frac{ d }{dx} \ln \left( l_0^{[p]}(x)\right)
= -D(x) U'(x)   + 2 D(x) \frac{ d }{dx}  \left[  \ln({\sqrt Z}) + \frac{ U(x)}{2} + \ln \left( \phi^{[p]}_0(x) \right) \right]
\nonumber \\
&& =  2 D(x)  \frac{ d  }{dx}   \ln \left( \phi^{[p]}_0(x) \right)
= D(x)  \frac{ d  }{dx}   \ln {\mathring P}_*^{[p]} (x)
 \label{RingforceQ}
\end{eqnarray}
The last expression in terms of the conditioned steady state ${\mathring P}_*^{[p]} (x ) $ of Eq. \ref{RingSteady}
means that the conditioned steady current ${\mathring J}_*^{[p]} (x ) $ identically vanishes
\begin{eqnarray}
 {\mathring J}_*^{[p]} (x ) \equiv {\mathring F}^{[p]}(x) {\mathring P}_* (x ) - D (x)  \partial_{x} {\mathring P}_* (x ) =0 
 \ \ \ {\rm for } \ \ x \in ]x_L,x_R[
\label{jringsteady}
\end{eqnarray}
i.e. the conditioned process satisfies detailed-balance.

The quantum supersymmetric Hamiltonian of Eq. \ref{hamiltonien}
associated to the conditioned process reads 
\begin{eqnarray}
 {\mathring H}^{[p]}  =  - \frac{ \partial  }{\partial x} D(x) \frac{ \partial  }{\partial x} 
 +{\mathring V}^{[p]} (x)
\label{Ringhamiltonien}
\end{eqnarray}
where the potential ${\mathring V}^{[p]} (x) $ involves the conditioned force ${\mathring F}^{[p]}(x) $
\begin{eqnarray}
{\mathring V}^{[p]} (x) \equiv \frac{ \left( {\mathring F}^{[p]}(x)\right)^2 }{4 D(x) }
 + \frac{1}{2} \frac{ d {\mathring F}^{[p]}(x)}{dx}
\label{Ringv}
\end{eqnarray}

The physical interpretation 
is that this canonical conditioned process of parameter $p$
is equivalent for large $T$ to the microcanonical conditioned process
producing the value $W=E'_0(p)$ of the Legendre transform of Eq. \ref{legendrereci}
(see the very detailed papers \cite{chetrite_conditioned,chetrite_optimal,derrida-conditioned} and references therein) :
as a consequence, it allows to understand and to generate the stochastic trajectories
that will dominate a given rare fluctuation of the observable $W$ for the initial process.
Recently, the deformed-Markov-generator approach has been applied to analyze the 
large deviations properties of interesting time-averaged observables for very many different Markov processes
\cite{peliti,derrida-lecture,sollich_review,lazarescu_companion,lazarescu_generic,jack_review,vivien_thesis,lecomte_chaotic,lecomte_thermo,lecomte_formalism,lecomte_glass,kristina1,kristina2,jack_ensemble,simon1,simon2,tailleur,simon3,Gunter1,Gunter2,Gunter3,Gunter4,chetrite_canonical,chetrite_conditioned,chetrite_optimal,chetrite_HDR,touchette_circle,touchette_langevin,touchette_occ,touchette_occupation,garrahan_lecture,Vivo,c_ring,c_detailed,chemical,derrida-conditioned,derrida-ring,bertin-conditioned,touchette-reflected,touchette-reflectedbis,c_lyapunov,previousquantum2.5doob,quantum2.5doob,quantum2.5dooblong,c_ruelle,lapolla,c_east,chabane,us_gyrator,duBuisson_gyrator}
where the corresponding conditioned process is often also constructed.

Besides this physical interest, the conditioned process is actually also useful at the technical level
as explained in the next subsection.

%%%%%%%%%%%%%%%%%%%%%%%%%%%%%%%%%%%%%%%%%

\subsubsection{ Spectral problem for the lowest energy $E_0(p)$ reformulated in terms of the properties of conditioned process}

The spectral problem for the positive left eigenvector $l^{[p]}_0(x) $ associated to the lowest energy $E_0(p)$
as summarized in subsection \ref{subsec_spectralLeftp}
can be reformulated in terms of the properties of conditioned process :

(1) The linear second-order differential equation for left eigenvector $ l_0^{[p]}(x)$ of Eq. \ref{zpeigenzero}
translates into the following non-linear Riccati first-order differential equation  
for the conditioned force ${\mathring F}^{[p]}(x) $ of Eq. \ref{Ringforce}
 \begin{eqnarray}
\frac{ \left( {\mathring F}^{[p]}(x)\right)^2 }{4 D(x) }
 + \frac{1}{2} \frac{ d {\mathring F}^{[p]}(x)}{dx} 
  = V(x) +p w(x) - E_0(p) = \frac{ F^2(x) }{4 D(x) } + \frac{F'(x)}{2}+p w(x) - E_0(p)
\label{riccati}
\end{eqnarray}
i.e. the quantum potential ${\mathring V}^{[p]} (x) $ of Eq. \ref{Ringv}
should be equal to the $p$-deformed potential of Eq. \ref{vfromup} minus its ground-state energy $E_0(p)$
 \begin{eqnarray}
{\mathring V}^{[p]} (x) =  V(x) +p w(x) - E_0(p) = V_p(x) - E_0(p)
\label{susyrefactor}
\end{eqnarray}

(2) the normalization of Eq. \ref{orthophinzero} for
left eigenvector $l_0^{[p]}(x) $ with respect to $P_*(x)$ can be rewritten as the normalization 
of the conditioned steady state ${\mathring P}_*^{[p]} (x ) $ of Eq. \ref{RingSteady}
\begin{eqnarray}
1  = \int_{x_L}^{x_R} dx  \left[ l_0^{[p]}(x) \right]^2 P_*(x)
=  \int_{x_L}^{x_R} dx {\mathring P}_*^{[p]} (x ) 
\label{orthophinzeroRing}
\end{eqnarray}

(3)  the current of Eq. \ref{jnleftpZ}
involved in the vanishing-current boundary conditions of Eq. \ref{jnleftpZzeroxlxr}
 can be rewritten in terms the conditioned force ${\mathring F}^{[p]}(x) $ of Eq. \ref{Ringforce} 
 and in terms of the conditioned steady state ${\mathring P}_*^{[p]} (x ) $ of Eq. \ref{RingSteady}
 as
\begin{eqnarray}
  j^{[p]}_n(x)  &&  = - D(x) P_*(x) \frac{d l^{[p]}_0(x)}{dx} = - \frac{P_*(x) l^{[p]}_0(x)}{2} \left[{\mathring F}^{[p]}(x) - F(x) \right] 
\nonumber \\
&&   = - \frac{1}{2} \sqrt{ P_*(x) {\mathring P}_*^{[p]} (x ) }  \left[{\mathring F}^{[p]}(x) - F(x) \right]
 \label{jnleftpZzeroforce}
\end{eqnarray}

For an arbitrary observable $w(x)$ of an arbitrary diffusion process with $F(x)$ and $D(x)$
discussed in subsection \ref{subsec_geneDiff},
the solution for $E_0(p)$ as a function of $p$ is unfortunately not explicit,
and the only possibility is then to use the perturbation theory in $p$
to obtain the first rescaled cumulants,
as explained around Eq. \ref{eseriesp}.
However one can also consider the problem the other way around 
and determine whether there are some specific observables $w(x)$ 
for a given diffusion process 
where $E_0(p) $ can be written explicitly for any $p$,
as discussed in the next subsection for Pearson diffusions.

%%%%%%%%%%%%%%%%%%%%%%%%%%%%%%%%%%%%

\subsection{ Simplifications for Pearson diffusions with explicit large deviations for specific observables $w(x)$  }

\label{subsec_pearsonw}

For Pearson diffusions with quadratic diffusion coefficient $D(x)=a x^2+b x +c$ 
and linear force $F(x)=\lambda-\gamma x$ of Eq. \ref{fpearson},
let us analyze whether explicit solutions
can be found when the conditioned force ${\mathring F}^{[p]}(x) $ is also linear in $x$ with two coefficients 
${\mathring \lambda}_p$ and ${\mathring \gamma}_p$
\begin{eqnarray}
{\mathring F}^{[p]}(x)   = {\mathring \lambda}_p - {\mathring \gamma}_p x 
\label{fpearsonring}
\end{eqnarray}
by considering the three items of the previous subsection :

(1) the Riccati Eq. \ref{riccati} for the conditioned force ${\mathring F}^{[p]}(x) $ of Eq. \ref{fpearsonring}
becomes
 \begin{eqnarray}
p w(x) - E_0(p) && = \frac{ \left( {\mathring F}^{[p]}(x)\right)^2 - F^2(x)}{4 D(x) }
 + \frac{1}{2} \frac{ d {\mathring F}^{[p]}(x)}{dx} -  \frac{1}{2} \frac{ d F(x)}{dx} 
\nonumber \\
&& = \frac{ \left( {\mathring \lambda}_p - {\mathring \gamma}_p x\right)^2 - ( \lambda - \gamma x)^2}{4 (a x^2+b x +c) }+ \frac{\gamma - {\mathring \gamma}_p  }{2}
\label{riccatiPearsonlinear}
\end{eqnarray}
The fraction decomposition of the right handside
yields that for the observables $w(x)$ corresponding to linear combinations
of the two functions $V_1(x)$ and $V_2(x)$ introduced in Eq. \ref{vfromupearsonfracrtion}
with two coefficients $c_1$ and $c_2$
\begin{eqnarray}
w(x)=c_1V_1(x)+c_2 V_2(x)
\label{wexplicit}
\end{eqnarray}
one can satisfy Eq. \ref{riccatiPearsonlinear} and obtain three equations that will determine
together the energy $E_0(p)$ and 
the two parameters ${\mathring \lambda}_p$ and ${\mathring \gamma}_p $ as a function of $p$. 

(2) the normalization of Eq. \ref{orthophinzeroRing}
simply means that the conditioned steady state ${\mathring P}_*^{[p]} (x ) $
associated to the linear conditioned force of Eq. \ref{fpearsonring}
and to the same diffusion coefficient $D(x)$
should remain a normalizable Pearson steady state  
in the same family as the initial process (see the table \ref{tablePearson} for the normalizability domain of 
each Pearson representative example).

(3)  the current of Eq. \ref{jnleftpZzeroforce}
 reads using the initial linear force $F(x)=\lambda-\gamma x$
 and the linear conditioned force ${\mathring F}^{[p]}(x) $ of Eq. \ref{fpearsonring}
\begin{eqnarray}
  j^{[p]}_n(x)    && =  \frac{1}{2} \sqrt{ P_*(x) {\mathring P}_*^{[p]} (x ) }  \left[  F(x) - {\mathring F}^{[p]}(x)\right]
  \nonumber \\
  && = \frac{1}{2} \sqrt{ P_*(x) {\mathring P}_*^{[p]} (x ) }  \left[ (\lambda- {\mathring \lambda}_p)
   + ({\mathring \gamma}_p- \gamma) x\right]
 \label{jnleftpZzeroforcepearson}
\end{eqnarray}
Let us analyze 
the vanishing-current boundary conditions of Eq. \ref{jnleftpZzeroxlxr}
for infinite and finite boundaries :

$ \bullet$ When the boundary is infinite, as $x_R=+\infty$ in the 
four representative examples (ii-iii-iv-vi) of the table \ref{tablePearson},
both the initial Pearson steady state $P_*(x) $ and the conditioned Pearson steady state ${\mathring P}_*^{[p]} (x ) $
 have to be normalizable for $x \to +\infty $,
so that the vanishing of the current of Eq. \ref{jnleftpZzeroforcepearson} is always satisfied
\begin{eqnarray}
 \lim_{x \to x_R=+\infty}  j^{[p]}_n(x)    =0 \ \text { : the boundary condition 
at $x_R=+\infty$ is always satisfied } 
 \label{jnleftpZzeroforceinfinity}
\end{eqnarray}

$ \bullet$ When the boundary is finite, as $x_L=0$ in the 
four representative examples (ii-iii-iv-v) of the table \ref{tablePearson},
both the initial Pearson steady state $P_*(x) $ and the conditioned Pearson steady state ${\mathring P}_*^{[p]} (x ) $
 have to be normalizable for $x \to 0$, so the contribution involving $[ ({\mathring \gamma}_p- \gamma) x]$
  in Eq. \ref{jnleftpZzeroforcepearson}
 cannot survive in the limit $x \to 0$ and the remaining boundary condition 
 reduces to the other contribution involving $(\lambda- {\mathring \lambda}_p) $
\begin{eqnarray}
0= \lim_{x \to x_L=0}  j^{[p]}_n(x)    = (\lambda- {\mathring \lambda}_p) \lim_{x \to x_L=0} 
 \sqrt{ P_*(x) {\mathring P}_*^{[p]} (x ) }
\to  \begin{cases}
\text{always satisfied if} \displaystyle  \lim_{x \to x_L=0}  \sqrt{ P_*(x) {\mathring P}_*^{[p]} (x ) } =0
\\
{\mathring \lambda}_p =\lambda  \ \ \text{ if } \ \ \displaystyle   \lim_{x \to x_L=0}  \sqrt{ P_*(x) {\mathring P}_*^{[p]} (x ) }  \ne 0
\end{cases}
 \label{jnleftpZzeroforcefinite}
\end{eqnarray}
The second possibility that imposes ${\mathring \lambda}_p =\lambda $
requires the non-vanishing of $\sqrt{ P_*(x) {\mathring P}_*^{[p]} (x ) } $ as $x \to 0$ 
that can occur in the Pearson representative examples (ii) (iv) (v) 
with the power-laws at the origin of Eq. \ref{zeropower}
\begin{eqnarray}
 P_*(x) && \oppropto_{x \to 0^+} x^{\alpha-1}  
 \nonumber \\
{\mathring P}_*^{[p]} (x ) && \oppropto_{x \to 0^+} x^{{\mathring \alpha}_p-1}  
\nonumber \\
 \sqrt{ P_*(x) {\mathring P}_*^{[p]} (x ) } && \oppropto_{x \to 0^+} x^{ \frac{ \alpha +{\mathring \alpha}_p}{2} -1}  
\ \ \  \text{ non vanishing in the region }  \ \ \ 0< \frac{ \alpha +{\mathring \alpha}_p}{2} \leq 1
\label{zeropowerConditioned}
\end{eqnarray}
while $\lambda=\alpha-1$ in these cases.
Since our goal is to consider the coefficient ${\mathring \alpha}_p $ as a function of $p$
that coincides for $p=0$ with the initial coefficient ${\mathring \alpha}_{p=0}=\alpha $,
the only boundary condition that needs to be taken into account in the present analysis
can be summarized as the following restriction : 
\begin{eqnarray}
\text{ Cases (ii) (iv) (v) : the possibility ${\mathring \alpha}_p \ne \alpha  $ can be considered only in the region $\alpha>1$
but not for $ 0< \alpha \leq 1$ } \ \ \ \ 
 \label{restriction}
\end{eqnarray}
that will be stressed again in Eqs \ref{gammarestriction} \ref{fisherrestriction} \ref{jacobirestriction} in the respective sections devoted
to the cases (ii) (iv) (v),
and that will be adapted in Eq. \ref{jacobirestrictionbeta} to the other finite boundary $x_R=1$ of the case (v).

In conclusion, for each representative example of Pearson diffusions considered in the further sections, 
we will analyze the specific observables $w(x)$ of the form of Eq. \ref{wexplicit}
whose large deviations properties can be written explicitly.

%%%%%%%%%%%%%%%%%%%%%%%%%%%%%%%%%%%%%%%%%

\section{ Explicit large deviations at level 2 with application to inference }

\label{sec_level2}

In this section, we focus on the large deviations at level 2 for the empirical density $ \rho_T(x)$
seen during the time-window $[0,T]$
and we discuss the application to the statistical inference of the two parameters of the Pearson linear force
from the data of a long stochastic trajectory $x(0 \leq t \leq T)$.

\subsection{ Explicit large deviations at level 2 for the empirical density $ \rho_T(x)$ 
for a diffusion with $F(x)$ and $D(x)$}

For a general diffusion process discussed in subsection \ref{subsec_geneDiff},
the empirical density $ \rho_T(.)$ 
represents the histogram of the position $x$ seen in a stochastic trajectory $x(0 \leq t \leq T) $ during the time window $[0, T]$
\begin{eqnarray}
 \rho_T(x)  \equiv \frac{1}{T} \int_0^T dt \  \delta( x(t)-x)  
\label{rhoempi}
\end{eqnarray}
that satisfies the normalization
\begin{eqnarray}
  \int_{x_L}^{x_R} dx \rho_T(x)  = 1
\label{rhonorma}
\end{eqnarray}

The empirical density of Eq. \ref{rhoempi} allows to reconstruct 
all the time-averaged observables of Eq. \ref{empi} discussed in the previous section
\begin{eqnarray}
W [x(0 \leq t \leq T) ]  \equiv  \frac{1}{T}   \int_0^T  dt w(x(t))
=  \int_{x_L}^{x_R} dx w(x) \rho_T(x) 
 \label{empifromrho}
\end{eqnarray}

The probability $P_T^{[2]}[ \rho(.)]  $ to see the empirical density $\rho_T(.)=\rho(.)$ 
satisfies the large deviation form at level 2 for large $T$
\begin{eqnarray}
 P_T^{[2]}[ \rho(.) ]   \opsimeq_{T \to +\infty}  \delta \left( \int_{x_L}^{x_R} dx \rho(x)  -1  \right) 
 e^{- \displaystyle T  I_{2}[ \rho(.)]  }
\label{level2}
\end{eqnarray}
where the prefactor corresponds to the normalization constraint of Eq. \ref{rhonorma},
while the rate function $I_2 [ \rho(.)]   $ 
at level 2 is explicit 
as a consequence of detailed-balance
\begin{eqnarray}
I_2 [ \rho(.)]  
=  \int_{x_L}^{x_R} \frac{ dx } { 4 D(x) \rho(x) } \left[   \rho(x) F(x) - D(x)  \rho'(x)     \right]^2
\label{rate2}
\end{eqnarray}

Using $F(x)=-D(x) U'(x)$ of Eq. \ref{Ux} and $U'(x)= - \frac{d}{dx} \ln (P_*(x))$, 
the rate function at level 2 can be rewritten as
\begin{eqnarray}
I_2 [ \rho(.)]  
&& = \frac{1}{4}  \int_{x_L}^{x_R} dx  D(x) \rho(x)  \left[   U'(x) +  \frac{ \rho'(x) }{\rho(x) }    \right]^2  
=  \frac{1}{4}  \int_{x_L}^{x_R} dx  D(x) \rho(x)  \left[    \frac{ \rho'(x) }{\rho(x) }  - \frac{P_*'(x) }{P_*(x) }  \right]^2  
\nonumber \\
 && = \frac{1}{4}  \int_{x_L}^{x_R} dx  D(x) \rho(x)  \left[   \frac{d}{dx} \ln \left(  \frac{ \rho(x) }{P_*(x) } \right)   \right]^2
\label{rate2bis}
\end{eqnarray}
in order to make obvious that it vanishes only when the normalized empirical density $\rho(x)$ coincides with the steady state $P_*(x)$.

%%%%%%%%%%%%%%%%%%%%%%%%%%%%%%%%%%%%%%%%%%%%%%%%%%%

\subsection{ Rephrasing as the probability to infer the steady state $ {\hat P}_* (x) $ instead of the true steady state $P_*(x)$}

Another point of view on the large deviations at level 2 is based 
on the inverse problem of statistical inference of the model parameters \cite{c_inference} :
from the data of a long dynamical trajectory $x(0 \leq t \leq T)$, 
one measures the empirical density $\rho(x)$ of Eq. \ref{rhoempi}
that gives directly
the best steady state that one can infer from the data
\begin{eqnarray}
 {\hat P}_* (x) = \rho(x)
\label{inferPstar}
\end{eqnarray}

As a consequence, the large deviations at level 2 of Eq. \ref{level2}
can be rephrased for the probability to infer a given steady state ${\hat P}_*(.) $ for large $T$
\begin{eqnarray}
 P_T^{[2]}[ {\hat P}_*(.) ]   \opsimeq_{T \to +\infty}  \delta \left( \int_{x_L}^{x_R} dx {\hat P}_* (x)  -1  \right) 
 e^{- \displaystyle T  I_{2}[ {\hat P}_* (.)]  }
\label{level2Infersteady}
\end{eqnarray}
with the rate function of Eq. \ref{rate2bis}
\begin{eqnarray}
I_2 [ {\hat P}_*(.)]   = \frac{1}{4}  \int_{x_L}^{x_R} dx  D(x) {\hat P}_*(x)  
\left[   \frac{d}{dx} \ln \left(  \frac{ {\hat P}_*(x) }{P_*(x) } \right)   \right]^2
\label{rateInfer}
\end{eqnarray}
As explained in details in \cite{c_inference},
the diffusion coefficient cannot fluctuate $\hat D(x) \equiv D(x)$ in the strict continuous-time limit,
so that the inference of the steady state ${\hat P}_*(x) $ via Eq. \ref{inferPstar}
can be further reformulated as the inference of the force ${\hat F} (x) $ 
that would produce the steady state ${\hat P}_* (x) $
\begin{eqnarray}
 {\hat F} (x) = D(x) \frac{d}{dx} \ln {\hat P}_* (x)
\label{inferForce}
\end{eqnarray}
with the corresponding rate function translated from Eq. \ref{rateInfer}
\begin{eqnarray}
I_2^{Force} [ {\hat F}(.)]   =   \int_{x_L}^{x_R} dx   \frac{ {\hat P}_*(x)  }{ 4 D(x) }
\left[   {\hat F}(x) - F(x)  \right]^2
\label{rateInferForce}
\end{eqnarray}
where the inferred steady state$ {\hat P}_*(x)$ should be computed in terms of the 
inferred force $ {\hat F}(x) $ via Eqs \ref{steadyeq} \ref{Ux}
\begin{eqnarray}
  {\hat P}_*(x)  && = \frac{ e^{ - {\hat U}(x)} }{\int dz e^{ -  {\hat U}(z) } }
  \nonumber \\
  {\hat U}(x) && = - \int_{x_{ref}}^x dy \frac{{\hat F}(y)}{D(y)}
 \label{steadyinfer}
\end{eqnarray}

%%%%%%%%%%%%%%%%%%%%%%%%%%%%%%%%%%%%%%%%%%%%

\subsection{ Application to the inference of the two parameters of the Pearson linear force }

For a Pearson diffusion process with the quadratic diffusion coefficient $D(x)$ and the linear force 
$F(x)= \lambda - \gamma x$ of Eq. \ref{fpearson},
one obtains that the probability $P_T^{[Infer]}(\hat \lambda , \hat \gamma) $ to infer 
the two coefficients $(\hat \lambda , \hat \gamma)$
of the linear force
\begin{eqnarray}
{\hat F}(x)  =\hat \lambda - \hat \gamma x 
\label{fpearsonInfer}
\end{eqnarray}
 displays the large deviation form for large $T$
\begin{eqnarray}
 P_T^{[Infer]}(\hat \lambda , \hat \gamma)   \opsimeq_{T \to +\infty}   
 e^{- \displaystyle T  I^{Infer} (\hat \lambda , \hat \gamma) }
\label{level2Infer}
\end{eqnarray}
where the rate function $I^{Infer} (\hat \lambda , \hat \gamma) $ obtained from Eq. \ref{rateInferForce}
involves the inferred steady state ${\hat P}_*(x) = P_*^{[\hat \lambda , \hat \gamma]}(x) $ corresponding to the Pearson steady state $P_*^{[\hat \lambda , \hat \gamma]}(x) $ with the inferred parameters $(\hat \lambda , \hat \gamma) $
\begin{eqnarray}
I^{Infer} (\hat \lambda , \hat \gamma)   
&& =  \int_{x_L}^{x_R} dx P_*^{[\hat \lambda , \hat \gamma]} (x) 
\frac{\left[ (\hat \lambda- \lambda)  - (\hat \gamma - \gamma) x    \right]^2 }{ 4  (a x^2+b x +c ) }
\nonumber \\
&& = \int_{x_L}^{x_R} dx P_*^{[\hat \lambda , \hat \gamma]} (x) 
\frac{(\hat \lambda- \lambda)^2 + (\hat \gamma - \gamma)^2 x^2
- 2  (\hat \lambda- \lambda) (\hat \gamma - \gamma) x  }{ 4  (a x^2+b x +c ) }
\label{rateInferpara}
\end{eqnarray}
The fraction decomposition of the function besides $P_*^{[\hat \lambda , \hat \gamma]} (x)  $ 
will again produce the two functions $V_1(x)$ and $V_2(x)$
introduced in \ref{vfromupearsonfracrtion} and encountered again in Eq. \ref{wexplicit}
during the analysis of large deviations at level 1.
In each representative example of Pearson diffusions discussed in the following sections,
we will thus explicitly compute the rate function of Eq. \ref{rateInferpara}
for the inference of the two parameters $(\hat \lambda , \hat \gamma) $
of the Pearson linear force.

%%%%%%%%%%%%%%%%%%%%%%%%%%%%%%%%%%%

\section{ Case $D(x)=x$ and Gamma-distribution
for the steady state $P_*(x)$ on $ ]0,+\infty[ $      }

\label{sec_gamma}

In this section, we focus on the Pearson diffusion with the linear diffusion coefficient $D(x)=x$ 
while the steady state $P_*(x)$ is the $\Gamma$-distribution 
\begin{eqnarray}
% D(x)  =x   \ \ \ {\rm and} \ \ 
P_*(x) = \frac{\gamma^{\alpha}}{ \Gamma(\alpha)} x^{\alpha-1}e^{- \gamma x} \ \ \ \ \ {\rm for } \ \ x \in ]0,+\infty[
\ \ \ {\rm with } \ \ \alpha>0 \ \ \ \ { \rm and } \ \ \ \ \gamma>0
\label{gamma}
\end{eqnarray}
where the two parameters $\alpha>0$ and $\gamma>0$
parametrize the two coefficients of the corresponding linear forces of Eq. \ref{fpearson} and \ref{fokkerplancklangevinpearson}
\begin{eqnarray}
 F(x) && =D (x)  \frac{d \ln P_*( x) }{dx} = (\alpha-1) - \gamma x
 \nonumber \\
    F_I(x) && = F(x)+ D' (x) = \alpha- \gamma x 
\nonumber \\
   F_S(x) && = F(x)+  \frac{D' (x)}{2} = \left(\alpha-\frac{1}{2} \right) - \gamma x
\label{forcesgamma}
\end{eqnarray}

The moments $m_k^*$ 
of the steady state $P_*(x)$ of Eq. \ref{gamma}
can be computed even for non-integer $k \in ]-\alpha,+\infty[ $
\begin{eqnarray}
 m_k^* =    \int_0^{+\infty} dx x^k P_*(x) 
 =  \int_0^{+\infty} dx   \frac{\gamma^{\alpha}}{ \Gamma(\alpha)} x^{k+\alpha-1}e^{- \gamma x}
 = \frac{\Gamma(\alpha+k)}{ \gamma^k \Gamma(\alpha)} \ \ \ {\rm for } \ \ \ k \in ]-\alpha,+\infty[
 \label{mkgammasteady}
\end{eqnarray}
so that here all the integer moments are finite for $k=1,2,..,+\infty$.

The corresponding diffusion process is
called either the Square-Root process \cite{dufresne} or the Cox-Ingersoll-Ross process \cite{CIR},
while the spectral decomposition of the propagator involves the Laguerre polynomials 
\cite{pearson_wong,pearson_class,pearson2012,PearsonHeavyTailed,pearson2018}.

%%%%%%%%%%%%%%%%%%%%%%%%%%%%%%%%%%%%%%%%

\subsection{ Dynamical equations for the moments $m_k(t)$ and for the Laplace transform ${\hat P}_t(s) $ }

The dynamical equation of Eq. \ref{dynPearsonmk} for the moment $m_k(t)$ of order $k$
only involves the previous moment $m_{k-1}(t)$ of order $(k-1)$
\begin{eqnarray}
\partial_t m_k(t)  = - k  \gamma m_k(t)+ k (   k+  \alpha-1)  m_{k-1}(t)
 \label{dynPearsonmkgamma}
\end{eqnarray}
The convergence of the moment $m_k(t)$ towards its finite steady value of Eq. \ref{mkgammasteady}
involves the $k$ relaxation rates $(\epsilon_1,..,epsilon_k)$ of Eq. \ref{ekmk} that are simply linear with respect to $k$
\begin{eqnarray}
 \epsilon_k = k  \gamma 
  \label{ekmkgamma}
\end{eqnarray}

The Laplace transform ${\hat P}_t(s) $ of Eq. \ref{laplace}
follows the closed dynamics of Eq. \ref{laplacedyn} 
\begin{eqnarray}
\partial_t {\hat P}_t(s) = - s ( s +\gamma) \partial_s {\hat P}_t(s) - \alpha s  {\hat P}_t(s)
 \label{laplacedyngamma}
\end{eqnarray}
and converges for $t \to +\infty$ towards the Laplace transform of the steady state $P_*(x)$
\begin{eqnarray}
 {\hat P}_*(s)  =  \int_0^{+\infty} dx e^{-s x} P_*(x) =  \int_0^{+\infty} dx   
  \frac{\gamma^{\alpha}}{ \Gamma(\alpha)} x^{\alpha-1}e^{- (\gamma+s) x}
  = \frac{\gamma^{\alpha}}{(\gamma+s)^{\alpha}}
 \label{laplacegammasteady}
\end{eqnarray}

%%%%%%%%%%%%%%%%%%%%%%%%%%%%%%%%%%%%%%%%%%%%

\subsection{ Observables $w(x)$ with explicit large deviations for the time-average $W [x(0 \leq t \leq T) ]  \equiv  \frac{1}{T}   \int_0^T  dt w(x(t)) $  }

 The quantum supersymmetric Hamiltonian of Eq. \ref{hamiltonien}
\begin{eqnarray}
 H = H^{\dagger} =  - \frac{ \partial  }{\partial x} x \frac{ \partial  }{\partial x} +V(x)
\label{gammahamiltonien}
\end{eqnarray}
involves the potential of Eq \ref{vfromu}
\begin{eqnarray}
V(x)   = \frac{ F^2(x) }{4 D(x) } + \frac{F'(x)}{2}
=\frac{ [(\alpha-1) - \gamma x]^2 }{4 x } - \frac{  \gamma }{2}
 =  \frac{\gamma^2}{4} x + \frac{ (\alpha-1)^2  }{4 x } - \frac{  \gamma \alpha }{2} 
\label{gammavfromu}
\end{eqnarray}
so that the two functions introduced in Eq. \ref{vfromupearsonfracrtion}
are simply
\begin{eqnarray}
V_1(x)  && = x 
\nonumber \\
V_2(x) && =\frac{1}{x}
\label{gammav1v2}
\end{eqnarray}

As discussed in detail in section \ref{subsec_pearsonw},
it is interesting to consider the linear conditioned force
\begin{eqnarray}
{\mathring F}^{[p]}(x)   = ({\mathring \alpha}_p-1) - {\mathring \gamma}_p x
 \ \ \ \ \ {\rm with} \ \ {\mathring \alpha}_p>0 \ \ \ \ {\rm and} \ \  {\mathring \gamma}_p >0
\label{gammafpearsonring}
\end{eqnarray}
with the restriction of Eq. \ref{restriction} that we repeat here for clarity :
\begin{eqnarray}
\text {the possibility ${\mathring \alpha}_p \ne \alpha  $ can be considered only in the region $\alpha>1$
but not for $ 0< \alpha \leq 1$ }
\label{gammarestriction}
\end{eqnarray}

The associated potential ${\mathring V}^{[p]} (x) $
has the same form as Eq. \ref{gammavfromu} with different coefficients
\begin{eqnarray}
{\mathring V}^{[p]} (x) = \frac{ \left( {\mathring F}^{[p]}(x)\right)^2 }{4 D(x) }
 + \frac{1}{2} \frac{ d {\mathring F}^{[p]}(x)}{dx}
=  \frac{{\mathring \gamma}_p^2}{4} x 
+ \frac{ ({\mathring \alpha}_p-1)^2  }{4 x } 
- \frac{  {\mathring \gamma}_p {\mathring \alpha}_p}{2} 
\label{gammaRingv}
\end{eqnarray}
so that Eq. \ref{susyrefactor} reads
\begin{eqnarray}
p w(x) - E_0(p) = {\mathring V}^{[p]} (x) - V(x)  
= \frac{{\mathring \gamma}_p^2- \gamma^2}{4} x 
+ \frac{ ({\mathring \alpha}_p-1)^2  - (\alpha-1)^2}{4 x } 
+ \frac{\gamma \alpha  -  {\mathring \gamma}_p {\mathring \alpha}_p}{2} 
\label{gammasusyrefactor}
\end{eqnarray}
This equation can be satisfied 
for the observables $w(x)$ corresponding to linear combinations of Eq. \ref{wexplicit}
with the two functions of Eq. \ref{gammav1v2}
\begin{eqnarray}
 w(x) = c_1V_1(x)+c_2 V_2(x) = c_1 x + \frac{ c_2 }{x}
\label{gammawx}
\end{eqnarray}
that leads to the system
\begin{eqnarray}
p c_1 && = \frac{{\mathring \gamma}_p^2- \gamma^2}{4}  
\nonumber \\
p c_2 && = \frac{ ({\mathring \alpha}_p-1)^2  - (\alpha-1)^2}{4  } 
\nonumber \\
E_0(p) && = \frac{  {\mathring \gamma}_p {\mathring \alpha}_p- \gamma \alpha  }{2} 
\label{gammasusyrefactorsystem}
\end{eqnarray}
The two first equations allow to compute the two parameters ${\mathring \gamma}_p $ 
and ${\mathring \alpha}_p $ of the conditioned force of Eq. \ref{gammafpearsonring} as a function of $p$
with ${\mathring \gamma}_{p=0} =\gamma$ and ${\mathring \alpha}_{p=0} =\alpha$
\begin{eqnarray}
{\mathring \gamma}_p && = \gamma \sqrt{  1 + \frac{4 p c_1}{\gamma^2} }
\nonumber \\
  {\mathring \alpha}_p  && = (\alpha-1)\sqrt{ 1 + \frac{4 p c_2}{(\alpha-1)^2} } +1
\label{gammaparametersring}
\end{eqnarray}
that can be plugged into the last equation of the system \ref{gammasusyrefactorsystem} 
to obtain the energy $E_0(p)$ as a function of $p$
\begin{eqnarray}
E_0(p) && = \frac{  {\mathring \gamma}_p {\mathring \alpha}_p- \gamma \alpha  }{2} 
= \frac{1}{2} \left[  \gamma \sqrt{  1 + \frac{4 p c_1}{\gamma^2} }
\left(  (\alpha-1)\sqrt{ 1 + \frac{4 p c_2}{(\alpha-1)^2} } +1 \right) - \gamma \alpha\right]
\label{gammaE0p}
\end{eqnarray}

Let us now focus on the two interesting special cases $[c_1=1,c_2=0]$ and $[c_1=0,c_2=1]$.

%%%%%%%%%%%%%%%%%%%%%%%%%%%%%%%%%%%%%%%%%%%%

\subsubsection{ Case $c_1=1$ and $c_2=0$ : explicit large deviations for the time-average $W [x(0 \leq t \leq T) ]  \equiv  \frac{1}{T}   \int_0^T  dt x(t) $ }

For the special case $c_1=1$ and $c_2=0$, the scaled cumulant generating function $E_0(p)$ of Eq. \ref{gammaE0p}
reduces to
\begin{eqnarray}
E_0(p) && 
= \frac{\alpha}{2} \left[  \gamma \sqrt{  1 + \frac{4 p }{\gamma^2} } - \gamma \right]
\label{gammaE0pw1}
\end{eqnarray}

For the Legendre transform of Eq. \ref{legendrereci}, one needs to invert
\begin{eqnarray}
W = E_0'(p) = \frac{\alpha}{\sqrt{\gamma^2+4 p}}
\label{gammaw1legendrereci}
\end{eqnarray}
into
\begin{eqnarray}
p= \frac{1}{4} \left( \frac{\alpha^2}{W^2} - \gamma^2 \right)
\label{gammaw1legendrereciinvert}
\end{eqnarray}
that can be plugged into Eq. \ref{legendrereci} to obtain the rate function $I(W)$
\begin{eqnarray}
I(W)   && =  E_0(p)-W p 
=  \frac{\alpha}{2} \left( \frac{\alpha}{W}  - \gamma \right) 
- \frac{W}{4} \left( \frac{\alpha}{W}  - \gamma \right) \left( \frac{\alpha}{W} + \gamma \right)
% \nonumber \\ && 
= \frac{W}{4} \left( \frac{\alpha}{W}  - \gamma \right)^2 \ \ \ {\rm for } \ \ W \in ]0,+\infty[
\label{gammaw1Rate}
\end{eqnarray}
that vanishes only for the steady value
\begin{eqnarray}
W_*=  \frac{\alpha}{ \gamma } = m_1^*
\label{gammaw1Ratevanish}
\end{eqnarray}
and that diverges near the two boundaries as
\begin{eqnarray}
I(W)   &&  \opsimeq_{W \to 0^+} \frac{\alpha^2}{4 W}  
\nonumber \\
I(W)   &&  \opsimeq_{W \to +\infty} \frac{\gamma^2}{4}   W
\label{gammaw1RateBoundaries}
\end{eqnarray}

%%%%%%%%%%%%%%%%%%%%%%%%%%%%%%%%%%%

\subsubsection{ Case $c_1=0$ and $c_2=1$ : explicit large deviations for the time-average $W [x(0 \leq t \leq T) ]  \equiv  \frac{1}{T}   \int_0^T  dt \frac{1}{x(t) }$ when $\alpha>1$}

For the special case $c_1=0$ and $c_2=1$, 
the scaled cumulant generating function $E_0(p)$ of Eq. \ref{gammaE0p}
reduces to
\begin{eqnarray}
E_0(p) 
= \frac{\gamma (\alpha-1)}{2} \left[  
  \sqrt{ 1 + \frac{4 p }{(\alpha-1)^2} } -1 \right]
\label{gammaE0pw2}
\end{eqnarray}

For the Legendre transform of Eq. \ref{legendrereci}, one needs to invert
\begin{eqnarray}
W = E_0'(p) =\frac{\gamma}{ (\alpha-1)\sqrt{ 1 + \frac{4 p }{(\alpha-1)^2} } }
\label{gammaw2legendrereci}
\end{eqnarray}
into
\begin{eqnarray}
p= \frac{1}{4} \left( \frac{\gamma^2}{W^2} - (\alpha-1)^2 \right)
\label{gammaw2legendrereciinvert}
\end{eqnarray}
that can be plugged into Eq. \ref{legendrereci} to obtain the rate function $I(W)$
\begin{eqnarray}
I(W)   && =  E_0(p)-W p 
=  \frac{\gamma}{2} \left( \frac{\gamma}{W} - (\alpha-1) \right) 
-  \frac{W}{4} \left( \frac{\gamma}{W} - (\alpha-1) \right) \left( \frac{\gamma}{W} + (\alpha-1) \right)
\nonumber \\
&& =  \frac{W}{4} \left( \frac{\gamma}{W} - (\alpha-1) \right)^2 \ \ \ {\rm for } \ \ W \in ]0,+\infty[
\label{gammaw2Rate}
\end{eqnarray}
that vanishes only for the steady value
\begin{eqnarray}
W_*= \frac{\gamma}{\alpha -1} = m_{k=-1}^*
\label{gammaw2Ratevanish}
\end{eqnarray}
and that diverges near the two boundaries as
\begin{eqnarray}
I(W)   &&  \opsimeq_{W \to 0^+} \frac{\gamma^2}{4 W} 
\nonumber \\
I(W)   &&  \opsimeq_{W \to +\infty} \frac{(\alpha-1)^2}{4} W
\label{gammaw2RateBoundaries}
\end{eqnarray}

%%%%%%%%%%%%%%%%%%%%%%%%%%%%%%%%%%%%%%%%%%%

\subsection{ Change of variables $x \to z$ towards the diffusion process $z(t)$ with constant diffusion coefficient $d(z)=1$ }

The change of variables of Eq. \ref{xtoz} 
towards a diffusion process $z(t)$ with constant diffusion coefficient $d(z)=1$ reads
\begin{eqnarray}
z && = \int_0^x \frac{dy}{\sqrt{ D(y)} }= \int_0^x \frac{dy}{ \sqrt{y} } = 2 \sqrt{ x } \in ]0,+ \infty[
\nonumber \\
x && = \frac{z^2}{4}
\label{xtozgamma}
\end{eqnarray}
The force of Eq. \ref{fokkerplanckzforce} is given by
\begin{eqnarray}
 f(z) = \frac{F_S(x) }{\sqrt{ D(x)}} \bigg\vert_{x=x(z)} = \frac{ 2 \alpha-1}{z}-  \frac{\gamma}{2} z = - u'(z)
\label{fokkerplanckzforcegamma}
\end{eqnarray}
while the steady state of Eq. \ref{steadyeqz} reads
\begin{eqnarray}
 p^*(z) = \frac{e^{-u(z)} }{ \int_{-\infty}^{+\infty} dz' e^{- u(z') }} 
= \frac{\gamma^{\alpha}}{ \Gamma(\alpha)} \left( \frac{z}{2} \right) ^{2\alpha-1}e^{- \gamma \frac{z^2}{4}} \ \ \ \ \ {\rm for } \ \ z \in ]0,+\infty[
\label{zsteadygamma}
\end{eqnarray}

The moments $m_k(t)$ of Eq. \ref{mktk} of the initial Pearson process translate into Eq. \ref{mktkz}
\begin{eqnarray}
m_k(t) \equiv  \int_0^{+\infty} dx P_t(x) x^k  = \int_0^{+\infty} dz  p_t(z) \left[  \frac{z^2}{4} \right]^k
 \label{mktkzgamma}
\end{eqnarray}

The quantum supersymmetric Hamiltonian of Eq. \ref{hamiltonienz}
\begin{eqnarray}
 h = h^{\dagger} && =\left(   - \frac{ d }{ d z} +  \frac{\gamma}{4} z- \frac{ 2 \alpha-1}{2z} \right)
 \left( \frac{ d }{ d z}  +  \frac{\gamma}{4} z- \frac{ 2 \alpha-1}{2z} \right) =  - \frac{d^2}{dz^2}  +v(z)
\label{hamiltonienzgamma}
\end{eqnarray}
involves the potential of Eq. \ref{quantumvz}
\begin{eqnarray}
v(z) 
&& = \frac{ f^2(z) }{4  } + \frac{f'(z)}{2}
 = \frac{1}{4} \left[ \frac{ (2 \alpha-1)^2}{z^2}+ \frac{\gamma^2}{4} z^2 - \gamma (2 \alpha-1)
 - 2 \frac{ 2 \alpha-1}{z^2}- \gamma 
 \right]
 \nonumber \\
 && = \frac{1}{4} \left[ \frac{\gamma^2}{4} z^2
+ \frac{ (2 \alpha-1)(2 \alpha-3) }{z^2} - 2 \alpha \gamma
 \right]
\label{quantumvzgamma}
\end{eqnarray}
This is the effective one-dimensional potential for the radial part
of the Schr\"odinger equation concerning
 the three-dimensional quantum harmonic oscillator with angular momentum (see the review \cite{review_susyquantum} and references therein).

Finally, the observables that have explicit large deviations for their time-averages
can be translated from Eq. \ref{gammawx} via the change of variables of Eq. \ref{xtozgamma}
\begin{eqnarray}
 w(x) = c_1 x + \frac{ c_2 }{x} = \frac{c_1}{4} z^2 +  \frac{4 c_2 }{z^2}
\label{wxzgamma}
\end{eqnarray}
or equivalently, they correspond to the two functions of $z$ appearing
in the quantum potential $v(z)$ of Eq. \ref{quantumvzgamma}.

%%%%%%%%%%%%%%%%%%%%%%%%%%%%%%%%%%%%%%%%%%

\subsection{ Change of variables $x \to  y= x^{- \frac{1}{q} }$ involving the parameter $q>0$} 

Via the change of variables $ y= x^{- \frac{1}{q} }$ of Eq. \ref{xtoy},
the diffusion coefficient of Eq. \ref{dy} involves only the power $y^{2+q} $
\begin{eqnarray}
{\cal D} ( y  ) =  \frac{y^{2+q}}{q^2} 
\label{dygamma}
\end{eqnarray}
while the three forces of Eqs \ref{fsy} \ref{ffiy} 
involve a linear contribution in $y$
and a non-linear contribution in $ y^{1+q}$ parametrized by $q$
\begin{eqnarray}
{\cal F}_S ( y  ) && = \frac{\gamma  }{q} y - \frac{ \left(\alpha-\frac{1}{2} \right)  }{q} y^{1+q}
\nonumber \\
  {\cal F} ( y  ) && = \frac{\gamma  }{q}  y 
  - \left[ \frac{ \left(\alpha-\frac{1}{2} \right)  }{q} +  \left( \frac{1}{q^2}+\frac{1}{2 q} \right)\right] y^{1+q}
\nonumber \\
  {\cal F}_I ( y  ) && = {\cal F}_S ( y  ) + \frac{{\cal D}' ( y  )}{2} 
  =  \frac{\gamma  }{q}  y 
   - \left[ \frac{ \left(\alpha-\frac{1}{2} \right)  }{q} -  \left( \frac{1}{q^2}+\frac{1}{2 q} \right)\right] y^{1+q}
\label{fsygamma}
\end{eqnarray}
The steady state
\begin{eqnarray}
{\cal P}_*(y) = P_*(x) \bigg\vert \frac{dx}{dy} \bigg\vert 
= \frac{ q \gamma^{\alpha}}{ \Gamma(\alpha) y^{1+ q \alpha} } e^{- \frac{\gamma }{y^q} } 
\ \ \ {\rm for } \ \ y \in ]0,+ \infty[
\label{ysteadygamma}
\end{eqnarray}
reduces to the Inverse-Gamma-distribution for $q=1$, 
but the process is different from the case that will be studied in
the next section as a Pearson diffusion, since the diffusion coefficient ${\cal D} ( y  ) $ 
is not quadratic and the forces are not linear in $y$ here.

The observables that have explicit large deviations for their time-averages
can be translated from Eq. \ref{gammawx} via the change of variable $x=y^{-q}$
\begin{eqnarray}
 w(x) =  c_1 x + \frac{ c_2 }{x} = c_1 y^{-q} + c_2 y^{q}
\label{wxygamma}
\end{eqnarray}

It is thus interesting to mention the two special cases :

(1) for $q=1$, the diffusion coefficient is cubic and the force is quadratic
\begin{eqnarray}
{\cal D} ( y  ) && =   y^3 
\nonumber \\
{\cal F}_S ( y  ) && = \gamma   y -  \left(\alpha-\frac{1}{2} \right)   y^2
\label{q2ygamma}
\end{eqnarray}
while the observables of Eq. \ref{wxygamma} involve $\frac{1}{y}$ and $y$
\begin{eqnarray}
 w(x) =  c_1 x + \frac{ c_2 }{x} = \frac{c_1}{y} + c_2 y
\label{q1wxygamma}
\end{eqnarray}

(2) for $q=2$,
the diffusion coefficient is quartic and the force is cubic
\begin{eqnarray}
{\cal D} ( y  ) && = \frac{y^4}{4} 
\nonumber \\
{\cal F}_S ( y  ) && =  \frac{\gamma  }{2} y - \frac{ \left(\alpha-\frac{1}{2} \right)  }{2} y^3
\label{q1ygamma}
\end{eqnarray}
while the observables of Eq. \ref{wxygamma} involve $\frac{1}{y^2}$ and $y^2$
\begin{eqnarray}
 w(x) =  c_1 x + \frac{ c_2 }{x} = \frac{c_1}{y^2} + c_2 y^2
\label{q2wxygamma}
\end{eqnarray}

%%%%%%%%%%%%%%%%%%%%%%%%%%%%%%%%%%%%%%%%%%%%

\subsection{ Explicit rate function for the inference of the two parameters $(\alpha,\gamma) $ of the Pearson linear force}

The application of Eq. \ref{level2Infer} \ref{rateInferpara}
to the model of Eq. \ref{gamma}
yields that the rate function $I^{Infer} (\hat \alpha , \hat \gamma)  $ 
associated to the probability $ P_T^{[Infer]}(\hat \alpha , \hat \gamma)  $ 
of the two inferred parameters $(\hat \alpha, \hat\gamma) $ reads
\begin{eqnarray}
I^{Infer} (\hat \alpha , \hat \gamma)   
&& = \int_0^{+\infty} dx P_*^{[\hat \alpha , \hat \gamma]} (x) 
\frac{\left[ (\hat \alpha- \alpha)  - (\hat \gamma - \gamma) x    \right]^2 }{ 4  x }
\nonumber \\
&&
= \int_0^{+\infty} dx  \frac{{\hat \gamma}^{\hat \alpha}}{ \Gamma(\hat \alpha)} x^{\hat \alpha-1}e^{- \hat \gamma x} 
\left[ \frac{(\hat \alpha- \alpha)^2   }{ 4  x }
+ \frac{ (\hat \gamma - \gamma)^2  }{ 4   } x
- \frac{   (\hat \alpha- \alpha) (\hat \gamma - \gamma)   }{ 2 }
\right]
\nonumber \\
&& = \frac{1}{4} \left[ (\hat \alpha- \alpha)^2 \frac{ \hat \gamma }{\hat \alpha -1}
+ (\hat \gamma - \gamma)^2 \frac{\hat \alpha}{ \hat \gamma }
- 2  (\hat \alpha- \alpha) (\hat \gamma - \gamma)  \right] \ \ \ {\rm for } \ \ \ \hat \alpha>1 
\label{rateInferparagamma}
\end{eqnarray}
while for $0 < \hat \alpha \leq 1$, the rate function $I^{Infer} (\hat \alpha , \hat \gamma) $ is infinite unless $\hat \alpha = \alpha $.

%%%%%%%%%%%%%%%%%%%%%%%%%%%%%%%%%%%%%%%%%%

\section{ Case $D(x)=x^2$ and Inverse-Gamma-distribution
for $P_*(x)$ on $ ]0,+\infty[ $      }

\label{sec_kesten}

In this section, we focus on the Pearson diffusion with the quadratic diffusion coefficient $D(x)=x^2$ 
while the steady state $P_*(x)$ is the Inverse-Gamma-distribution
\begin{eqnarray}
% D(x)  =  x^2  \ \ \ {\rm and} \ \ 
P_*(x) = \frac{\lambda^{\mu}}{\Gamma(\mu) x^{1+\mu} } e^{- \frac{\lambda}{x}} \ \ \ {\rm } \ \ {\rm for } \ \ x \in ]0,+\infty[
\ \ \ {\rm with } \ \ \lambda>0 \ \ {\rm and } \ \ \ \mu>0
\label{kesten}
\end{eqnarray}
where the two parameters $\lambda>0$ and $\mu>0$
parametrize the two coefficients of the corresponding linear forces of Eq. \ref{fpearson} and \ref{fokkerplancklangevinpearson}
\begin{eqnarray}
 F(x) && =D (x)  \frac{d \ln P_*( x) }{dx} =  \lambda - (\mu+1)  x
 \nonumber \\
    F_I(x) && = F(x)+ D' (x) =  \lambda - (\mu-1)  x
\nonumber \\
   F_S(x) && = F(x)+  \frac{D' (x)}{2} = \lambda - \mu x
\label{forceskesten}
\end{eqnarray}

The moments $m_k^*$ 
of the steady state $P_*(x)$ of Eq. \ref{kesten}
can be computed even for non-integer $k \in ]-\infty,\mu[ $
\begin{eqnarray}
 m_k^*  =  \int_0^{+\infty} dx x^k P_*(x) 
 = \int_0^{+\infty} dx  \frac{\lambda^{\mu}}{\Gamma(\mu)  } x^{k-\mu-1}e^{- \frac{\lambda}{x}}
 = \lambda^k  \frac{\Gamma(\mu-k)}{\Gamma(\mu)  }
 \ \ {\rm for } \ \ k \in ]-\infty,\mu[
 \label{mksteadykesten}
\end{eqnarray}
so here the integer moments are finite only for $k<\mu$.

This Pearson process is often called the reciprocal gamma process
(see  \cite{pearson_wong,pearson_class,pearson2012,PearsonHeavyTailed,pearson2018}
and references therein).
Independently of the Pearson family,
this process has been also much studied
as the following exponential functional of the Brownian motion $B_t$
\cite{c_flux,c_these,yor}
\begin{eqnarray}
x(t)  \equiv \lambda \int_0^t ds e^{-\mu (t-s) + \sqrt{2} (B_t -B_s) }  
\label{xexpfunctionalBrownian}
\end{eqnarray}
since its time derivative leads to the following Stratonovich Stochastic differential Equation
\begin{eqnarray}
dx(t)   = \left[ \lambda - \mu x(t) \right] dt + x(t) \sqrt{2} dB_t
 \ \ \ \ \ \ \ \ \ \ \ \ \ \ [{\rm Stratonovich \ Interpretation}]
\label{zstrato}
\end{eqnarray}
The exponential functional of Eq. \ref{xexpfunctionalBrownian} 
has attracted a lot of interest since it corresponds to the continuous limit 
of the Kesten random variables 
that appears in many disordered systems
\cite{Kesten,Solomon,sinai,jpb_review,annphys90,Der_Hil,Cal,strong_review,c_microcano,c_watermelon,c_mblcayley,c_reset}.

%%%%%%%%%%%%%%%%%%%%%%%%%%%%%%%%%%%%%%

\subsection{ Dynamical equations for the moments $m_k(t)$ and for the Laplace transform ${\hat P}_t(s) $ }

The characteristic rate $\epsilon_k$ of Eq. \ref{ekmk} is positive only for $ k < \mu$
\begin{eqnarray}
 \epsilon_k = k \bigg( \mu -  k \bigg)
\ \ \ i.e. \ \  \begin{cases}
\text{ exponential relaxation as $e^{ - t k (\mu-k) } $  for $k<\mu$ } \\
 \text{ exponential growth as $e^{  t k (k-\mu) }  $  for $k>\mu$ }
\end{cases}
 \label{ekkesten}
\end{eqnarray}
In particular, it is interesting to write the explicit dynamics for the first moment $m_1(t)$ of Eq. \ref{dynPearsonm1integ}
\begin{eqnarray}
  m_1(t)    =  \frac{\lambda}{\mu -1} + \left(  m_1(0)  -  \frac{\lambda}{\mu -1} \right) e^{-  t (\mu-1)}  
\ i.e.   \begin{cases}
\text{ exponential relaxation as $e^{-  t (\mu-1)}  $ towards $m_1^*= \frac{\lambda}{\mu -1}  $ for $\mu >1$ } \\
 \text{ exponential growth as $e^{  t (1-\mu)}  $ towards $m_1^*=+\infty$ for $0<\mu <1$}
\end{cases}
 \label{kestenm1integ}
\end{eqnarray}
and for the second moment $m_2(t)$ of Eq. \ref{dynPearsonm2integ}
\begin{eqnarray}
 m_2(t)  && =  e^{- 2 (  \mu-2 ) t}  m_2(0)
 + \lambda^2\frac{ 1- e^{ -2 (\mu-2 ) t} }{ ( \mu -2) (\mu-1) } 
 +  \frac{ e^{- (\mu-1) t } - e^{- 2 (  \mu-2 ) t}}{ \mu -3} 2  \lambda  \left(  m_1(0)  - \frac{\lambda}{\mu-1} \right)   
 \nonumber \\
&&   i.e.
  \begin{cases}
  \text{ exponential relaxation towards $m_2^*= \frac{\lambda^2}{( \mu -2) (\mu-1)}$ for $\mu >2$ } 
  \\
 \text{ exponential growth as $e^{ 2 (2-\mu) t} $ towards $m_2^*=+\infty  $ for $0<\mu <2$}
\end{cases}
 \label{kestenm2integ}
\end{eqnarray}

The dynamical equation of Eq. \ref{dynPearsonmk} for the moment $m_k(t)$ of order $k$
only involves the previous moment $m_{k-1}(t)$ of order $(k-1)$
\begin{eqnarray}
\partial_t m_k(t)  = k (  k- \mu ) m_k(t)+ k  \lambda  m_{k-1}(t)
 \label{dynPearsonmkkesten}
\end{eqnarray}

The Laplace transform ${\hat P}_t(s) $
follows the closed dynamics of Eq. \ref{laplacedyn} 
\begin{eqnarray}
\partial_t {\hat P}_t(s) =  s^2 \partial_s^2  {\hat P}_t(s)+ (1-\mu) s \partial_s  {\hat P}_t(s) -\lambda s  {\hat P}_t(s)
 \label{laplacedynkesten}
\end{eqnarray}
and converges for $t \to +\infty$ towards the Laplace transform of the steady state $P_*(x)$
\begin{eqnarray}
 {\hat P}_*(s)  =  \int_0^{+\infty} dx e^{-s x} P_*(x) =  \int_0^{+\infty} dx   
 \frac{\lambda^{\mu}}{\Gamma(\mu) x^{1+\mu} } e^{- s x - \frac{\lambda}{x}} 
 = \frac{2}{\Gamma(\mu)  }  (\lambda s)^{\frac{\mu}{2} } K_{\mu} (2 \sqrt{\lambda s})
 \label{laplaceKestensteady}
\end{eqnarray}

For the special case $\mu=\frac{1}{2}$, the steady state corresponds to the L\'evy stable law
\begin{eqnarray}
P_*(x) && = \frac{ \sqrt{\lambda}}{\sqrt{\pi} x^{\frac{3}{2}} } e^{- \frac{\lambda}{x}} 
\nonumber \\
 {\hat P}_*(s)  &&  
 = \frac{2}{\sqrt{\pi}  }  (\lambda s)^{\frac{1}{4} } K_{\frac{1}{2}} (2 \sqrt{\lambda s})
 = e^{ - 2 \sqrt{\lambda s } }
\label{Dxdemi}
\end{eqnarray}

%%%%%%%%%%%%%%%%%%%%%%%%%%%%%%%%%%%%%%%%%%%

\subsection{ Observables $w(x)$ with explicit large deviations for the time-average $W [x(0 \leq t \leq T) ]  \equiv  \frac{1}{T}   \int_0^T  dt w(x(t)) $  }

The quantum supersymmetric Hamiltonian of Eq. \ref{hamiltonien}
\begin{eqnarray}
 H = H^{\dagger} =  - \frac{ \partial  }{\partial x} x^2 \frac{ \partial  }{\partial x} +V(x)
\label{kestenhamiltonien}
\end{eqnarray}
involves the potential of Eq \ref{vfromu}
\begin{eqnarray}
V(x) && = \frac{ F^2(x) }{4 D(x) } + \frac{F'(x)}{2}
= \frac{  \lambda^2+ (\mu+1)^2  x^2 -2 \lambda (\mu+1)  x}{4 x^2} - \frac{  (\mu+1)  }{2}
\nonumber \\
&& = \frac{  \lambda^2 }{4 x^2} - \frac{    \lambda (\mu+1)  }{2 x}+ \frac{  \mu^2-1  }{4}
\label{kestenvfromu}
\end{eqnarray}
so that the two functions introduced in Eq. \ref{vfromupearsonfracrtion}
are 
\begin{eqnarray}
V_1(x)  && = \frac{1}{x} 
\nonumber \\
V_2(x) && =\frac{1}{x^2}
\label{kestenv1v2}
\end{eqnarray}

As discussed in detail in section \ref{subsec_pearsonw},
it is interesting to consider the linear conditioned force
\begin{eqnarray}
{\mathring F}^{[p]}(x) =
  {\mathring \lambda}_p - ({\mathring \mu}_p+1)  x
   \ \ \ \ \ {\rm with} \ \ {\mathring \lambda}_p>0 \ \ \ \ {\rm and} \ \  {\mathring \mu}_p >0
\label{kestenfpearsonring}
\end{eqnarray}

The corresponding potential ${\mathring V}^{[p]} (x) $ 
has the same form as Eq. \ref{kestenvfromu} with different coefficients
\begin{eqnarray}
{\mathring V}^{[p]} (x) = \frac{ \left( {\mathring F}^{[p]}(x)\right)^2 }{4 D(x) }
 + \frac{1}{2} \frac{ d {\mathring F}^{[p]}(x)}{dx}
=  \frac{  {\mathring \lambda}_p^2 }{4 x^2} - \frac{    {\mathring \lambda}_p ({\mathring \mu}_p+1)  }{2 x}+ \frac{  {\mathring \mu}_p^2-1  }{4}
\label{kestenRingv}
\end{eqnarray}
so that Eq. \ref{susyrefactor} reads
\begin{eqnarray}
p w(x) - E_0(p) = {\mathring V}^{[p]} (x) - V(x)  
= \frac{  {\mathring \lambda}_p^2 -  \lambda^2 }{4 x^2} 
+ \frac{  \lambda (\mu+1)    - {\mathring \lambda}_p ({\mathring \mu}_p+1)  }{2 x}
+ \frac{  {\mathring \mu}_p^2-\mu^2  }{4}
\label{kestensusyrefactor}
\end{eqnarray}
This equation can be satisfied 
for the observables $w(x)$ corresponding to linear combinations of Eq. \ref{wexplicit}
with the two functions of Eq. \ref{kestenv1v2}
\begin{eqnarray}
 w(x) = c_1V_1(x)+c_2 V_2(x)= \frac{ c_1 }{x} + \frac{ c_2 }{x^2}
\label{kestenwx}
\end{eqnarray}
that leads to the system
\begin{eqnarray}
p c_2 && =  \frac{  {\mathring \lambda}_p^2 -  \lambda^2 }{4 } 
\nonumber \\
p c_1 && = \frac{  \lambda (\mu+1)    - {\mathring \lambda}_p ({\mathring \mu}_p+1)  }{2 }
\nonumber \\
E_0(p) && =\frac{ \mu^2 - {\mathring \mu}_p^2  }{4}
\label{kestensusyrefactorsystem}
\end{eqnarray}
The two first equations allow to compute the two parameters ${\mathring \lambda}_p $ 
and ${\mathring \mu}_p $ of the conditioned force of Eq. \ref{kestenfpearsonring} as a function of $p$
with ${\mathring \lambda}_{p=0} =\lambda$ and ${\mathring \mu}_{p=0} =\mu$
\begin{eqnarray}
{\mathring \lambda}_p && = \lambda \sqrt{  1 + \frac{4 p c_2}{\lambda^2} }
\nonumber \\
  {\mathring \mu}_p && = \frac{ \lambda (\mu+1)    -  2p c_1 }{ {\mathring \lambda}_p} -1
  = \frac{  (\mu+1)    -  \frac{ 2p c_1}{ \lambda }  }{   \sqrt{  1 + \frac{4 p c_2}{\lambda^2} }} -1
\label{kestenparametersring}
\end{eqnarray}
that can be plugged into the last equation of the system \ref{kestensusyrefactorsystem} 
to obtain the energy $E_0(p)$ as a function of $p$
\begin{eqnarray}
E_0(p) && = \frac{ \mu^2 - {\mathring \mu}_p^2  }{4}
= \frac{1}{4} \left[ \mu^2- \left( \frac{  (\mu+1)    -  \frac{ 2p c_1}{ \lambda }  }{   \sqrt{  1 + \frac{4 p c_2}{\lambda^2} }} -1 \right)^2\right] 
\label{kestenE0p}
\end{eqnarray}

Let us now focus on the two interesting special cases $[c_1=1,c_2=0]$ and $[c_1=0,c_2=1]$.

%%%%%%%%%%%%%%%%%%%%%%%%%%%%%%%%%%%%%%%%%%%%

\subsubsection{ Case $c_1=1$ and $c_2=0$ : explicit large deviations for the time-average $W [x(0 \leq t \leq T) ]  \equiv  \frac{1}{T}   \int_0^T  dt \frac{1}{x(t) } $ }

For the special case $c_1=1$ and $c_2=0$, the scaled cumulant generating function $E_0(p)$ of Eq. \ref{kestenE0p}
reduces to
\begin{eqnarray}
E_0(p) && 
= \frac{1}{4} \left[ \mu^2
- \left( \mu    -  \frac{ 2p }{ \lambda }   \right)^2\right] 
=  \frac{ \mu p }{\lambda}    -  \frac{p^2 }{\lambda^2}  
\label{kestenE0pw1}
\end{eqnarray}
For the Legendre transform of Eq. \ref{legendrereci}, one needs to invert
\begin{eqnarray}
W = E_0'(p) = \frac{ \mu}{\lambda}    -  \frac{ 2 }{\lambda^2}  p
\label{kestenw1legendrereci}
\end{eqnarray}
into
\begin{eqnarray}
p= \frac{\lambda^2}{ 2 } \left( \frac{ \mu}{\lambda} - W\right)
\label{kestenw1legendrereciinvert}
\end{eqnarray}
that can be plugged into Eq. \ref{legendrereci} to obtain the rate function $I(W)$
\begin{eqnarray}
I(W)    =  E_0(p)-W p
=  \frac{\lambda^2}{ 4 } \left( \frac{ \mu}{\lambda} - W\right)^2 \ \ \ {\rm for } \ \ W \in ]0,+\infty[
\label{kestenw1Rate}
\end{eqnarray}
that vanishes only for the steady value
\begin{eqnarray}
W_*=  \frac{ \mu}{\lambda} = m_{k=-1}^*
\label{kestenw1Ratevanish}
\end{eqnarray}
and that diverges at $(+\infty)$
\begin{eqnarray}
I(W)   &&  \opsimeq_{W \to +\infty}  \frac{\lambda^2}{ 4 }  W^2
\label{kestenw1RateBoundaries}
\end{eqnarray}
while it remains finite at the boundary $W =0$
\begin{eqnarray}
I(W=0)   
=  \frac{ \mu^2 }{ 4 \lambda} 
\label{kestenw1RateZero}
\end{eqnarray}

%%%%%%%%%%%%%%%%%%%%%%%%%%%%%%%%%%%

\subsubsection{ Case $c_1=0$ and $c_2=1$ : explicit large deviations for the time-average $W [x(0 \leq t \leq T) ]  \equiv  \frac{1}{T}   \int_0^T  dt \frac{1}{x^2(t) }$ }

For the special case $c_1=0$ and $c_2=1$, 
the scaled cumulant generating function $E_0(p)$ of Eq. \ref{kestenE0p}
reduces to
\begin{eqnarray}
E_0(p) && = \frac{1}{4} \left[ \mu^2- \left( \frac{  \mu+1  }{   \sqrt{  1 + \frac{4 p }{\lambda^2} }} -1 \right)^2\right] 
 = \frac{\mu+1}{4} \left[ \mu- 1 - \frac{\mu+1}{ 1 + \frac{4 p }{\lambda^2}}
+ \frac{  2 }{   \sqrt{  1 + \frac{4 p }{\lambda^2} }} \right] 
\label{kestenE0pw2}
\end{eqnarray}

For the Legendre transform of Eq. \ref{legendrereci}, the inversion of
\begin{eqnarray}
W = E_0'(p) =  
\frac{  \mu+1   }{ \lambda^2  \left[  1 + \frac{4 p }{\lambda^2}  \right]^{\frac{3}{2}}}
  \left( \frac{  \mu+1  }{   \sqrt{  1 + \frac{4 p }{\lambda^2} }} -1 \right) 
\label{kestenw2legendrereci}
\end{eqnarray}
 requires the solution of a quartic equation, which is somewhat lengthy,
 so that here it is simpler to write the rate function $I(W)$ in the parametric form
 as a function of $p$
 \begin{eqnarray}
I(W(p))    =  E_0(p)-W p =
 \frac{\mu+1}{4} \left(  \frac{ 1  }{   \sqrt{  1 + \frac{4 p }{\lambda^2} }} -1  \right)^2
 \left( \frac{\mu+1}{ 1 + \frac{4 p }{\lambda^2}} +\frac{ 2\mu+ 1  }{   \sqrt{  1 + \frac{4 p }{\lambda^2} }} +\mu -1  \right)
\label{kestenw2rateparametric}
\end{eqnarray}
while $W$ as a function of $p$ is given by Eq. \ref{kestenw2legendrereci}

%%%%%%%%%%%%%%%%%%%%%%%%%%%%%%%%%%%%%%%%%%%

\subsection{ Change of variables $x \to z$ towards the diffusion process $z(t)$ with constant diffusion coefficient $d(z)=1$ }

The change of variables of Eq. \ref{xtoz} 
towards a diffusion process $z(t)$ with constant diffusion coefficient $d(z)=1$ reads
\begin{eqnarray}
z && = \int_1^x \frac{dy}{\sqrt{ D(y)} }=\int_1^x \frac{ dy}{ y} = \ln x  \in ]-\infty,+\infty[
\nonumber \\
x && = e^z
\label{xtozkesten}
\end{eqnarray}
The force of Eq. \ref{fokkerplanckzforce} is then exponential with respect to $z$
\begin{eqnarray}
 f(z) =\frac{F_S(x) }{\sqrt{ D(x)}} \bigg\vert_{x=x(z)} = \lambda e^{-z}- \mu = - u'(z)\label{fokkerplanckzforcekesten}
\end{eqnarray}
while the steady state of Eq. \ref{steadyeqz}
\begin{eqnarray}
 p^*(z) = \frac{e^{-u(z)} }{ \int_{-\infty}^{+\infty} dz' e^{- u(z') }} = 
 \frac{\lambda^{\mu} }{\Gamma(\mu) } e^{- \mu z - \lambda e^{-z} } \ \ {\rm for } \ \ z \in ]-\infty,+\infty[
\label{zsteadykesten}
\end{eqnarray}
corresponds for $\mu=1=\lambda$ to the Gumbel distribution of the field of Extreme-Value-Statistics.

The moments $m_k(t)$ of Eq. \ref{mktk} of the initial Pearson process translate into Eq. \ref{mktkz}
\begin{eqnarray}
m_k(t)   = \int_{-\infty}^{+\infty} dz  p_t(z) e^{k z}
 \label{mktkzkesten}
\end{eqnarray}

The quantum supersymmetric Hamiltonian of Eq. \ref{hamiltonienz}
\begin{eqnarray}
 h = h^{\dagger} && =\left(   - \frac{ d }{ d z}  +\frac{ \mu- \lambda e^{-z}}{2 } \right)
 \left( \frac{ d }{ d z}  +\frac{ \mu- \lambda e^{-z} }{2 } \right) =  - \frac{d^2}{dz^2}  +v(z)
\label{hamiltonienzkesten}
\end{eqnarray}
involves the potential of Eq. \ref{quantumvz}
\begin{eqnarray}
v(z)  = \frac{ f^2(z) }{4  } + \frac{f'(z)}{2}
 = \frac{  \lambda^2 e^{- 2 z} + \mu^2  -2 (\mu+1) \lambda e^{-z} }{4  } 
  \label{quantumvzkesten}
\end{eqnarray}
known as the Morse potential in the field of exactly solvable quantum supersymmetric potentials (see the review \cite{review_susyquantum} and references therein).
Its asymptotic behaviors for $z \to \pm \infty$
\begin{eqnarray}
 v(z)  && \opsimeq_{z \to - \infty} \frac{  \lambda^2 e^{ 2 (-z) }  }{4  } 
 \nonumber \\
  v(z)  && \opsimeq_{z \to + \infty} \frac{   \mu^2   }{4  } 
  \label{largezquantumvzkesten}
\end{eqnarray}
yields that there is a continuous spectrum of the form of Eq \ref{continuum} with 
the lower boundary $v_{\infty}=\frac{   \mu^2   }{4  } $
\begin{eqnarray}
 \text{ Continuous spectrum :} \ \ \ \ E \in ] v_{\infty} = \frac{   \mu^2   }{4  },+\infty[
\label{continuumkesten}
\end{eqnarray}
besides the finite number of discrete levels of Eq. \ref{eigendiscretepos}
\begin{eqnarray}
 E_n = n \bigg( \mu -  n \bigg)  \ \ \ {\rm for } \ \  0 \leq n < \frac{\mu}{2}
  \label{enkesten}
\end{eqnarray}

Finally, the observables that have explicit large deviations for their time-averages
can be translated from Eq. \ref{kestenwx} via the change of variables of Eq. \ref{xtozkesten}
\begin{eqnarray}
 w(x) = \frac{ c_1 }{x} + \frac{ c_2 }{x^2} = c_1 e^{-z} + c_2 e^{-2 z}
\label{wxzkesten}
\end{eqnarray}
or equivalently, they correspond to the two functions of $z$ appearing
in the quantum potential $v(z)$ of Eq. \ref{quantumvzkesten}.

%%%%%%%%%%%%%%%%%%%%%%%%%%%%%%%%%%%%%%%%

%%%%%%%%%%%%%%%%%%%%%%%%%%%%%%%%%%%%%%%%%%

\subsection{ Change of variables $x \to  y= x^{- \frac{1}{q} }$ involving the parameter $q>0$} 

Via the change of variables $ y= x^{- \frac{1}{q} }$ of Eq. \ref{xtoy},
the diffusion coefficient of Eq. \ref{dy} is always quadratic in $y$
\begin{eqnarray}
{\cal D} ( y  ) = \frac{y^2}{q^2} 
\label{dykesten}
\end{eqnarray}
while the three forces of Eqs \ref{fsy} \ref{ffiy} 
involve a linear contribution in $y$
and a non-linear contribution in $ y^{1+q}$ parametrized by $q$
\begin{eqnarray}
{\cal F}_S ( y  ) && = \frac{\mu  }{q} y - \frac{ \lambda  }{q} y^{1+q}
\nonumber \\
  {\cal F} ( y  ) && = {\cal F}_S ( y  ) - \frac{{\cal D}' ( y  )}{2} 
  = \left( \frac{\mu  }{q} -  \frac{1}{q^2} \right) y   -  \frac{ \lambda  }{q} y^{1+q}
\nonumber \\
  {\cal F}_I ( y  ) && = {\cal F}_S ( y  ) + \frac{{\cal D}' ( y  )}{2} 
  = \left( \frac{\mu  }{q} +  \frac{1}{q^2} \right) y 
   -  \frac{ \lambda  }{q} y^{1+q}
\label{fsykesten}
\end{eqnarray}
These processes play a major role in the field of multiplicative stochastic processes \cite{review_multiplicative,carleman_multiplicative}.

The steady state
\begin{eqnarray}
{\cal P}_*(y) = P_*(x) \bigg\vert \frac{dx}{dy} \bigg\vert = \frac{ q \lambda^{\mu}}{\Gamma(\mu)  } y^{q \mu-1} e^{- \lambda y^q}  
\ \ \ {\rm for } \ \ y \in ]0,+ \infty[
\label{ysteadykesten}
\end{eqnarray}
reduces to the Gamma-distribution for $q=1$, but the process is different from the case studied in
the previous section, since the diffusion coefficient ${\cal D} ( y  ) $ and the forces are quadratic in $y$.

The observables that have explicit large deviations for their time-averages
can be translated from Eq. \ref{kestenwx} via the change of variable $x=y^{-q}$
\begin{eqnarray}
 w(x) = \frac{ c_1 }{x} + \frac{ c_2 }{x^2} = c_1 y^q + c_2 y^{2q}
\label{wxykesten}
\end{eqnarray}

It is thus interesting to discuss the two special cases :

(1) for $q=1$, the forces of Eq. \ref{fsykesten} involve a quadratic non-linearity
\begin{eqnarray}
{\cal D} ( y  ) && = y^2
\nonumber \\
{\cal F}_S ( y  ) && = \mu y -  \lambda  y^2
\label{q1ykesten}
\end{eqnarray}
and the observables of Eq. \ref{wxykesten} involve linear and quadratic contributions in $y$
\begin{eqnarray}
 w(x) =  \frac{ c_1 }{x} + \frac{ c_2 }{x^2} =  c_1 y + c_2 y^2
\label{q1wxykesten}
\end{eqnarray}

(2) for $q=2$
the forces of Eq. \ref{fsykesten} involve a cubic non-linearity
\begin{eqnarray}
{\cal D} ( y  ) && =  \frac{y^2}{4} 
\nonumber \\
{\cal F}_S ( y  ) && = \frac{\mu  }{2} y - \frac{ \lambda  }{2} y^3
\label{q2ykesten}
\end{eqnarray}
and the observables of Eq. \ref{wxykesten} involve quadratic and quartic contributions in $y$
\begin{eqnarray}
 w(x) =  \frac{ c_1 }{x} + \frac{ c_2 }{x^2} =  c_1 y^2 + c_2 y^4
\label{q2wxykesten}
\end{eqnarray}

%%%%%%%%%%%%%%%%%%%%%%%%%%%%%%%%%%%%%%%%%%%

\subsection{ Explicit rate function for the inference of the two parameters $(\lambda,\mu) $ of the Pearson linear force}

The application of Eq. \ref{level2Infer} \ref{rateInferpara}
to the model of Eq. \ref{kesten}
yields that the rate function $I^{Infer} (\hat \lambda , \hat \mu)  $ 
associated to the probability $ P_T^{[Infer]}(\hat \lambda , \hat \mu)  $ 
of the two inferred parameters $(\hat \lambda , \hat \mu) $ reads
\begin{eqnarray}
I^{Infer} (\hat \lambda , \hat \mu) 
&&= \int_0^{+\infty} dx P_*^{[\hat \lambda , \hat \mu]} (x) 
\frac{\left[ (\hat \lambda- \lambda)  - (\hat \mu - \mu) x    \right]^2 }{ 4  x^2 }
\nonumber \\
&& 
=  \int_0^{+\infty} dx \frac{\hat \lambda^{\hat \mu}}{\Gamma(\hat \mu) x^{1+\hat \mu} } e^{- \frac{\hat \lambda}{x}}
\left[ \frac{(\hat \lambda- \lambda)^2  }{ 4  x^2 }
+\frac{ (\hat \mu - \mu)^2   }{ 4   }
- \frac{ 2  (\hat \lambda- \lambda) (\hat \mu - \mu)   }{ 4  x }
\right]
\nonumber \\
&& =  \frac{1}{4} \left[
(\hat \lambda- \lambda)^2   \frac{\hat \mu(\hat \mu+1)}{\hat \lambda^2  }
 + (\hat \mu - \mu)^2 
- 2  (\hat \lambda- \lambda) (\hat \mu - \mu)   \frac{\hat \mu}{\hat \lambda  }
\right]
\label{rateInferparakesten}
\end{eqnarray}

%%%%%%%%%%%%%%%%%%%%%%%%%%%%%%%%%%%%%%%%%%

\section{ Case $D(x)=x (x+1)$ and Fisher-Snedecor-distribution
for $P_*(x)$ on $ ]0,+\infty[ $      }

\label{sec_fisher}

In this section, we focus on the Pearson diffusion with the quadratic diffusion coefficient $D(x)=x (x+1)$ 
while the steady state $P_*(x)$ is the Fisher-Snedecor-distribution 
\begin{eqnarray}
% D(x)  =  x (x+1) \ \ \ {\rm and} \ \ 
P_*(x) = \frac{\Gamma(\alpha+\mu)}{ \Gamma(\alpha) \Gamma(\mu) } \frac{ x^{\alpha-1} }{(1+x)^{\alpha+\mu} }
 \ \ \ {\rm for } \ x \in ]0,+\infty[
 \ \ \ {\rm with } \ \ \alpha>0 \ \ {\rm and } \ \ \mu>0
\label{fisher}
\end{eqnarray}
where the two parameters $\alpha>0$ and $\mu$
parametrize the two coefficients of the corresponding linear forces of Eq. \ref{fpearson} and \ref{fokkerplancklangevinpearson}
\begin{eqnarray}
 F(x) && =D (x)  \frac{d \ln P_*( x) }{dx} = (\alpha-1) - (\mu+1) x
 \nonumber \\
    F_I(x) && = F(x)+ D' (x) = \alpha-(\mu-1) x
\nonumber \\
   F_S(x) && = F(x)+  \frac{D' (x)}{2} =  \left(\alpha- \frac{1}{2}\right) - \mu x
\label{forcesfisher}
\end{eqnarray}

The moments $m_k^*$ 
of the steady state $P_*(x)$ of Eq. \ref{fisher}
can be computed even for non-integer $k \in ]-\alpha,\mu[ $
\begin{eqnarray}
 m_k^*  =  \int_0^{+\infty} dx x^k P_*(x) = 
  \int_0^{+\infty} dx
   \frac{\Gamma(\alpha+\mu)}{ \Gamma(\alpha) \Gamma(\mu) } \frac{ x^{k+\alpha-1} }{(1+x)^{\alpha+\mu} }
  =\frac{\Gamma(\alpha+k)\Gamma(\mu-k)}{\Gamma(\alpha)\Gamma(\mu) }
 \ \ {\rm for } \ \ k \in ]-\alpha,\mu[ 
 \label{mksteadyfisher}
\end{eqnarray}
so here the integer moments are finite only for $k<\mu$.

The corresponding process is called the Fisher-Snedecor diffusion 
(see  \cite{pearson_wong,pearson_class,pearson2012,PearsonHeavyTailed,pearson2018,pearson_fisher,pearson_fisherSnedecor}
and references therein).

%%%%%%%%%%%%%%%%%%%%%%%%%%%%%%%%%%%

\subsection{ Dynamical equations for the moments $m_k(t)$ and for the Laplace transform ${\hat P}_t(s) $ }

The characteristic rate $\epsilon_k$ of Eq. \ref{ekmk} is positive only for $ k < \mu$
\begin{eqnarray}
 \epsilon_k =  k \bigg( \mu- k  \bigg)
\ \ \ i.e. \ \  \begin{cases}
\text{ exponential relaxation as $e^{ - t k (\mu-k) } $  for $k<\mu$ } \\
 \text{ exponential growth as $e^{  t k (k-\mu) }  $  for $k>\mu$ }
\end{cases}
 \label{ekfisher}
\end{eqnarray}
In particular, it is interesting to write the explicit dynamics for the first moment $m_1(t)$ of Eq. \ref{dynPearsonm1integ}
\begin{eqnarray}
  m_1(t)    =   \frac{\alpha}{\mu-1} + \left(  m_1(0)  - \frac{\alpha}{\mu-1} \right) e^{-  t (\mu-1)} 
\ i.e.   \begin{cases}
\text{ exponential relaxation as $e^{-  t (\mu-1)}  $ towards $m_1^*=  \frac{\alpha}{\mu-1}  $ for $\mu >1$ } \\
 \text{ exponential growth as $e^{  t (1-\mu)}  $ towards $m_1^*=+\infty$ for $0<\mu <1$}
\end{cases}
 \label{fisherm1integ}
\end{eqnarray}
and for the second moment $m_2(t)$ of Eq. \ref{dynPearsonm2integ}
\begin{eqnarray}
 m_2(t)  && =   e^{- 2 (  \mu-2 ) t}  m_2(0)
 + [ 1-e^{ -2 ( \mu-2 ) t} ] 
     \frac{\alpha ( 1   + \alpha  ) }{(\mu-1) ( \mu -2 )} 
 +  \frac{ e^{- (\mu-1) t } - e^{- 2 (   \mu-2 ) t}}{ \mu-3} 2 ( 1   + \alpha  )\left(  m_1(0)  - \frac{\alpha}{\mu-1} \right) 
 \nonumber \\
&&   i.e.
  \begin{cases}
  \text{ exponential relaxation towards $m_2^*=  \frac{\alpha ( 1   + \alpha  ) }{(\mu-1) ( \mu -2 )} $ for $\mu >2$ } 
  \\
 \text{ exponential growth as $e^{ 2 (2-\mu) t} $ towards $m_2^*=+\infty  $ for $0<\mu <2$}
\end{cases}
 \label{fisherm2integ}
\end{eqnarray}

The dynamical equation of Eq. \ref{dynPearsonmk} for the moment $m_k(t)$ of order $k$
only involves the previous moment $m_{k-1}(t)$ of order $(k-1)$
\begin{eqnarray}
\partial_t m_{k}(t)  =  k(k- \mu)  m_k(t)
+ k (  k + \alpha -1 ) m_{k-1}(t)
 \label{OavdynPearsonkfisher}
\end{eqnarray}
while the Laplace transform follows the dynamics of Eq. \ref{laplacedyn}
\begin{eqnarray}
\partial_t {\hat P}_t(s)
 =   s^2 \partial_s^2  {\hat P}_t(s) + s [ (\mu-1) -s ]   \partial_s  {\hat P}_t(s) - \alpha s   {\hat P}_t(s)
 \label{laplacedynfisher}
\end{eqnarray}
%and converges towards its steady state value 
%that involves the hypergeometric F11

%%%%%%%%%%%%%%%%%%%%%%%%%%%%%%%%%%%%

\subsection{ Observables $w(x)$ with explicit large deviations for the time-average $W [x(0 \leq t \leq T) ]  \equiv  \frac{1}{T}   \int_0^T  dt w(x(t)) $  }

 The quantum supersymmetric Hamiltonian of Eq. \ref{hamiltonien}
\begin{eqnarray}
 H = H^{\dagger} =  - \frac{ \partial  }{\partial x} x (x+1) \frac{ \partial  }{\partial x} +V(x)
\label{fisherhamiltonien}
\end{eqnarray}
involves the potential of Eq \ref{vfromu}
\begin{eqnarray}
V(x) && = \frac{ F^2(x) }{4 D(x) } + \frac{F'(x)}{2}
=  \frac{  (\mu+1)^2 x^2 - 2 (\alpha-1)  (\mu+1) x + (\alpha-1)^2 +}{4 x (x+1) } - \frac{ (\mu+1) }{2}
\nonumber \\
&& = \frac{\mu^2-1}{4}+ \frac{(\alpha-1)^2}{4x}- \frac{(\alpha+\mu)^2}{4(x+1)}
\label{fishervfromu}
\end{eqnarray}
so that the two functions introduced in Eq. \ref{vfromupearsonfracrtion}
are simply
\begin{eqnarray}
V_1(x)  && = \frac{1}{x} 
\nonumber \\
V_2(x) && =\frac{1}{x+1}
\label{fisherv1v2}
\end{eqnarray}

As discussed in detail in section \ref{subsec_pearsonw},
it is interesting to consider the linear conditioned force
\begin{eqnarray}
{\mathring F}^{[p]}(x)   = ({\mathring \alpha}_p-1) - ({\mathring \mu}_p +1)x 
 \ \ \ \ \ {\rm with} \ \ {\mathring \alpha}_p>0 \ \ \ \ {\rm and} \ \  {\mathring \mu}_p >0
\label{fisherfpearsonring}
\end{eqnarray}
with the restriction of Eq. \ref{restriction} that we repeat here for clarity :
\begin{eqnarray}
\text {the possibility ${\mathring \alpha}_p \ne \alpha  $ can be considered only in the region $\alpha>1$
but not for $ 0< \alpha \leq 1$ }
\label{fisherrestriction}
\end{eqnarray}

The potential ${\mathring V}^{[p]} (x) $ has the same form as Eq. \ref{fishervfromu} with different coefficients
\begin{eqnarray}
{\mathring V}^{[p]} (x) = \frac{ \left( {\mathring F}^{[p]}(x)\right)^2 }{4 D(x) }
 + \frac{1}{2} \frac{ d {\mathring F}^{[p]}(x)}{dx}
=  \frac{{\mathring \mu}_p^2-1}{4}+ \frac{({\mathring \alpha}_p-1)^2}{4x}- \frac{({\mathring \alpha}_p+{\mathring \mu}_p)^2}{4(x+1)}
\label{fisherRingv}
\end{eqnarray}
so that Eq. \ref{susyrefactor} reads
\begin{eqnarray}
p w(x) - E_0(p) = {\mathring V}^{[p]} (x) - V(x)  
=   \frac{{\mathring \mu}_p^2-\mu^2}{4}+ \frac{({\mathring \alpha}_p-1)^2- (\alpha-1)^2}{4x}
+ \frac{(\alpha+\mu)^2 - ({\mathring \alpha}_p+{\mathring \mu}_p)^2}{4(x+1)}
\label{fishersusyrefactor}
\end{eqnarray}
This equation can be satisfied 
for the observables $w(x)$ corresponding to linear combinations of Eq. \ref{wexplicit}
with the two functions of Eq. \ref{fisherv1v2}
\begin{eqnarray}
 w(x) = c_1V_1(x)+c_2 V_2(x) = \frac{ c_1 }{x} + \frac{ c_2 }{x+1}
\label{fisherwx}
\end{eqnarray}
that leads to the system

\begin{eqnarray}
p c_1 && =  \frac{({\mathring \alpha}_p-1)^2- (\alpha-1)^2}{4}
\nonumber \\
p c_2 && =\frac{(\alpha+\mu)^2 - ({\mathring \alpha}_p+{\mathring \mu}_p)^2}{4}
\nonumber \\
E_0(p) && = \frac{\mu^2 - {\mathring \mu}_p^2}{4}
\label{fishersusyrefactorsystem}
\end{eqnarray}
The two first equations allow to compute the two parameters ${\mathring \alpha}_p $ 
and ${\mathring \mu}_p $ of the conditioned force of Eq. \ref{fisherfpearsonring} as a function of $p$
\begin{eqnarray}
  {\mathring \alpha}_p  && = (\alpha-1)\sqrt{ 1 + \frac{4 p c_1}{(\alpha-1)^2} } +1
\nonumber \\
{\mathring \mu}_p   && = (\alpha+\mu)\sqrt{ 1 - \frac{4 p c_2}{(\alpha+\mu)^2} } 
-  {\mathring \alpha}_p
= (\alpha+\mu)\sqrt{ 1 - \frac{4 p c_2}{(\alpha+\mu)^2} } 
- (\alpha-1)\sqrt{ 1 + \frac{4 p c_1}{(\alpha-1)^2} } - 1
\label{fisherparametersring}
\end{eqnarray}
that can be plugged into the last equation of the system \ref{fishersusyrefactorsystem} 
to obtain the energy $E_0(p)$ as a function of $p$
\begin{eqnarray}
E_0(p) && = \frac{\mu^2 - {\mathring \mu}_p^2}{4}
= \frac{1}{4} \left[ \mu^2 - \left( (\alpha+\mu)\sqrt{ 1 - \frac{4 p c_2}{(\alpha+\mu)^2} } 
- (\alpha-1)\sqrt{ 1 + \frac{4 p c_1}{(\alpha-1)^2} } - 1\right)^2 \right]
\label{fisherE0p}
\end{eqnarray}

Let us now focus on the two interesting special cases $[c_1=1,c_2=0]$ and $[c_1=0,c_2=1]$.

%%%%%%%%%%%%%%%%%%%%%%%%%%%%%%%%%%%%%%%%%%%%

\subsubsection{ Case $c_1=1$ and $c_2=0$ : explicit large deviations for the time-average $W [x(0 \leq t \leq T) ]  \equiv  \frac{1}{T}   \int_0^T  dt \frac{1}{x(t) }$ when $\alpha>1$}

For the special case $c_1=1$ and $c_2=0$, the scaled cumulant generating function $E_0(p)$ of Eq. \ref{fisherE0p}
reduces to
\begin{eqnarray}
E_0(p) && 
=   \frac{1}{4} \left[ \mu^2 
- \left( \mu +(\alpha-1) - (\alpha-1)\sqrt{ 1 + \frac{4 p }{(\alpha-1)^2} } \right)^2 \right]
\label{fisherE0pw1}
\end{eqnarray}

For the Legendre transform of Eq. \ref{legendrereci}, one needs to invert
\begin{eqnarray}
W = E_0'(p) =    
\frac{ \mu +\alpha-1}{ (\alpha-1)\sqrt{ 1 + \frac{4 p }{(\alpha-1)^2} }}
- 1
\label{fisherw1legendrereci}
\end{eqnarray}
into
\begin{eqnarray}
p= \frac{1}{4} \left[ \left(\frac{\mu +\alpha-1}{W+1}\right)^2 - (\alpha-1)^2 \right]
\label{fisherw1legendrereciinvert}
\end{eqnarray}
that can be plugged into Eq. \ref{legendrereci} to obtain the rate function $I(W)$
\begin{eqnarray}
I(W)   && =  E_0(p)-W p
=   \frac{1}{4} \left[ \mu^2 
- (\mu +\alpha-1)^2\left(  \frac{W}{W+1} \right)^2 \right]
 - \frac{W}{4} \left[ \left(\frac{\mu +\alpha-1}{W+1}\right)^2 - (\alpha-1)^2 \right]
\nonumber \\
&&  = \frac{1}{ 4 (W+1)} \left[ \mu - (\alpha-1) W\right]^2 \ \ \ {\rm for } \ \ W \in ]0,+\infty[
\label{fisherw1Rate}
\end{eqnarray}
that vanishes only for the steady value
\begin{eqnarray}
W_*=  \frac{\mu}{\alpha-1} = m_{k=-1}^*
\label{fisherw1Ratevanish}
\end{eqnarray}
and that diverges at $(+\infty)$
\begin{eqnarray}
I(W)   &&  \opsimeq_{W \to +\infty}  \frac{(\alpha-1)^2}{ 4 } W
\label{fisherw1RateBoundaries}
\end{eqnarray}
while it remains finite at the boundary $W =0$
\begin{eqnarray}
I(W=0)   =  \frac{\mu}{ 4 } 
\label{fisherw1RateZero}
\end{eqnarray}

%%%%%%%%%%%%%%%%%%%%%%%%%%%%%%%%%%%

\subsubsection{ Case $c_1=0$ and $c_2=1$ : Rate function $I(W)$ for the time-average $W [x(0 \leq t \leq T) ]  \equiv  \frac{1}{T}   \int_0^T  dt \frac{1}{1+x(t)  }$ }

For the special case $c_1=0$ and $c_2=1$, 
the scaled cumulant generating function $E_0(p)$ of Eq. \ref{fisherE0p}
reduces to
\begin{eqnarray}
E_0(p) 
= \frac{1}{4} \left[ \mu^2 - \left( (\alpha+\mu)\sqrt{ 1 - \frac{4 p }{(\alpha+\mu)^2} } - \alpha \right)^2 \right]
\label{fisherE0pw2}
\end{eqnarray}

For the Legendre transform of Eq. \ref{legendrereci}, one needs to invert
\begin{eqnarray}
W = E_0'(p) = 
1- \frac{ \alpha }{ (\alpha+\mu) \sqrt{ 1 - \frac{4 p }{(\alpha+\mu)^2 }}}
\label{fisherw2legendrereci}
\end{eqnarray}
into
\begin{eqnarray}
p= \frac{1}{4} \left[ (\alpha+\mu)^2 - \frac{\alpha^2}{(1-W)^2}  \right]
\label{fisherw2legendrereciinvert}
\end{eqnarray}
that can be plugged into Eq. \ref{legendrereci} to obtain the rate function $I(W)$
\begin{eqnarray}
I(W)   && =  E_0(p)-W p 
=   \frac{1}{4} \left[ \mu^2 - \alpha^2 \frac{W^2}{(1-W)^2} \right]
- \frac{W}{4} \left[ (\alpha+\mu)^2 - \frac{\alpha^2}{(1-W)^2}  \right]
\nonumber \\
&& =  \frac{1}{ 4 (1-W)} \left[ \mu - (\alpha+\mu) W\right]^2 \ \ \ {\rm for } \ \ W \in ]0,1[ 
\label{fisherw2Rate}
\end{eqnarray}
that vanishes only for the steady value
\begin{eqnarray}
W_*= \frac{\mu}{\alpha +\mu} =
 \int_0^{+\infty} dx \frac{ P_*(x) }{1+x} 
% =   \int_0^{+\infty} dx   \frac{\Gamma(\alpha+\mu)}{ \Gamma(\alpha) \Gamma(\mu) } \frac{ x^{\alpha-1} }{(1+x)^{\alpha+\mu+1} }
\label{fisherw2Ratevanish}
\end{eqnarray}
and that diverges at $W\to 1^-$
\begin{eqnarray}
I(W)   &&  \opsimeq_{W \to 1^-}  \frac{\alpha^2}{ 4 (1-W)} 
\label{fisherw2RateBoundaries}
\end{eqnarray}
while it remains finite at the boundary $W =0$
\begin{eqnarray}
I(W=0)   
= \frac{\mu}{ 4 } 
\label{fisherw2RateZero}
\end{eqnarray}

%%%%%%%%%%%%%%%%%%%%%%%%%%%%%%%%%%%%%%%%%%%

\subsection{ Change of variables $x \to z$ towards the diffusion process $z(t)$ with constant diffusion coefficient $d(z)=1$ }

The change of variables of Eq. \ref{xtoz} 
towards a diffusion process $z(t)$ with constant diffusion coefficient $d(z)=1$ reads
\begin{eqnarray}
z(x) && = \int_0^x \frac{dy}{\sqrt{ D(y)} }= \int_0^x \frac{ dy}{\sqrt{ y(1+y)}} = 2 {\rm arcsinh } ( {\sqrt x } )  \in ]0,+\infty[
\nonumber \\
x(z) && = \sinh^2 \left(\frac{z}{2} \right) 
\label{xtozfisher}
\end{eqnarray}
The force of Eq. \ref{fokkerplanckzforce} involves hyperbolic functions of $z$
\begin{eqnarray}
 f(z) = \frac{F_S(x) }{\sqrt{ D(x)}} \bigg\vert_{x=x(z)}
% = \frac{(2 \alpha+\mu-1) - \mu \cosh z}{ \sinh z} 
 = \left( \alpha- \frac{1}{2} \right) \frac{\cosh \left(\frac{z}{2} \right)}{\sinh \left(\frac{z}{2} \right)}
 - \left( \alpha+\mu- \frac{1}{2} \right) \frac{\sinh \left(\frac{z}{2} \right)}{\cosh \left(\frac{z}{2} \right)}
 = - u'(z)
\label{fokkerplanckzforcefisher}
\end{eqnarray}
while the steady state of Eq. \ref{steadyeqz} reads
\begin{eqnarray}
  p_*(z)  = \frac{ e^{ - u(z)} }{\int_0^{+\infty} dz' e^{ - u(z')}} 
=  \frac{\Gamma(\alpha+\mu)}{ \Gamma(\alpha) \Gamma(\mu) } 
  \frac{ \left[ \sinh \left(\frac{z}{2} \right)  \right]^{2\alpha-1} }{ \left[ \cosh \left(\frac{z}{2} \right)  \right]^{2\alpha+2\mu-1}}
 \ \ \ {\rm for } \ z \in ]0,+\infty[
 \label{steadyeqzfisher}
\end{eqnarray}

The moments $m_k(t)$ of Eq. \ref{mktk} of the initial Pearson process translate into Eq. \ref{mktkz}
\begin{eqnarray}
m_k(t) = \int_0^{+\infty} dz  p_t(z) \left[  \sinh^2 \left(\frac{z}{2} \right)  \right]^k
 \label{mktkzfisher}
\end{eqnarray}

The quantum supersymmetric Hamiltonian of Eq. \ref{hamiltonienz}
\begin{eqnarray}
 h = h^{\dagger} && 
 % =\left(   - \frac{ d }{ d z}  + \frac{  \mu \cosh z- (2 \alpha+\mu-1)}{ 2 \sinh z} \right)
 %\left( \frac{ d }{ d z}  + \frac{  \mu \cosh z- (2 \alpha+\mu-1)}{ 2 \sinh z} \right)  
 = \left(   - \frac{ d }{ d z}
  +  \left( \alpha+\mu- \frac{1}{2} \right) \frac{\sinh \left(\frac{z}{2} \right)}{\cosh \left(\frac{z}{2} \right)}
 -\left( \alpha- \frac{1}{2} \right) \frac{\cosh \left(\frac{z}{2} \right)}{\sinh \left(\frac{z}{2} \right)}
  \right)
 \left( \frac{ d }{ d z}  +  \left( \alpha+\mu- \frac{1}{2} \right) \frac{\sinh \left(\frac{z}{2} \right)}{\cosh \left(\frac{z}{2} \right)}
 -\left( \alpha- \frac{1}{2} \right) \frac{\cosh \left(\frac{z}{2} \right)}{\sinh \left(\frac{z}{2} \right)}\right)  
\nonumber \\
&&  =  - \frac{d^2}{dz^2}  +v(z)
\label{hamiltonienzfisher}
\end{eqnarray}
involves the potential of Eq. \ref{quantumvz}
\begin{eqnarray}
v(z) 
&& = \frac{ f^2(z) }{4  } + \frac{f'(z)}{2}
= \frac{\mu^2}{4} 
+  \frac{\left( \alpha- \frac{1}{2} \right) \left( \alpha- \frac{3}{2} \right)}{ 4 \sinh^2 \left(\frac{z}{2} \right)}
+  \frac{ \frac{1}{4} - \left( \alpha+\mu \right)^2}{ 4 \cosh^2 \left(\frac{z}{2} \right)}
%old with \frac{ (2 \alpha+\mu-1)^2+\mu(\mu+2) }{ 4 \sinh^2 z}
%-  \frac{   (\mu+1)( 2 \alpha+\mu-1)  \cosh z}{ 2 \sinh^2 z} 
 \label{quantumvzfisher}
\end{eqnarray}
known as the generalized P\"oschl–Teller potential in the field of exactly solvable quantum supersymmetric potentials (see the review \cite{review_susyquantum} and references therein).

Its asymptotic behaviors for $z \to + \infty$
\begin{eqnarray}
  v(z)   \opsimeq_{z \to + \infty} \frac{   \mu^2   }{4  } 
  \label{largezquantumvzfisher}
\end{eqnarray}
yields that there is a continuous spectrum of the form of Eq \ref{continuum} with 
the lower boundary $v_{\infty}=\frac{   \mu^2   }{4  } $
\begin{eqnarray}
 \text{ Continuous spectrum :} \ \ \ \ E \in ] v_{\infty} = \frac{   \mu^2   }{4  },+\infty[
\label{continuumfisher}
\end{eqnarray}
besides the finite number of discrete levels of Eq. \ref{eigendiscretepos}
\begin{eqnarray}
 E_n = n \bigg( \mu -  n \bigg)  \ \ \ {\rm for } \ \  0 \leq n < \frac{\mu}{2}
  \label{enfisher}
\end{eqnarray}

Finally, the observables that have explicit large deviations for their time-averages
can be translated from Eq. \ref{fisherwx} via the change of variables of Eq. \ref{xtozfisher}
\begin{eqnarray}
 w(x) = \frac{ c_1 }{x} + \frac{ c_2 }{x+1} = \frac{ c_1 }{ \sinh^2 \left(\frac{z}{2} \right) } + \frac{ c_2 }{ \cosh^2 \left(\frac{z}{2} \right) }
\label{wxzfisher}
\end{eqnarray}
or equivalently, they correspond to the two functions of $z$ appearing
in the quantum potential $v(z)$ of Eq. \ref{quantumvzfisher}.

%%%%%%%%%%%%%%%%%%%%%%%%%%%%%%%%%%%%%%%%%%

\subsection{ Change of variables $x \to  y= x^{- \frac{1}{q} }$ involving the parameter $q>0$} 

Via the change of variables $ y= x^{- \frac{1}{q} }$ of Eq. \ref{xtoy},
the diffusion coefficient reads
\begin{eqnarray}
{\cal D} ( y  ) =\frac{y^{2}}{q^2} ( 1 +y^{q} )
\label{dyfisher}
\end{eqnarray}
while the three forces of Eqs \ref{fsy} \ref{ffiy} 
involve a linear contribution in $y$
and a non-linear contribution in $ y^{1+q}$ 
\begin{eqnarray}
{\cal F}_S ( y  ) && =\frac{\mu  }{q} y - \frac{ \left(\alpha- \frac{1}{2}\right)  }{q} y^{1+q}
\nonumber \\
  {\cal F} ( y  ) && = {\cal F}_S ( y  ) - \frac{{\cal D}' ( y  )}{2} 
  = \left( \frac{\mu  }{q} -  \frac{1}{q^2} \right) y 
  - \left[ \frac{ \left(\alpha- \frac{1}{2}\right)  }{q} +  \left( \frac{1}{q^2}+\frac{1}{2 q} \right)\right] y^{1+q}
\nonumber \\
  {\cal F}_I ( y  ) && = {\cal F}_S ( y  ) + \frac{{\cal D}' ( y  )}{2} 
  =  \left( \frac{\mu  }{q} +  \frac{1}{q^2} \right) y 
   - \left[ \frac{ \left(\alpha- \frac{1}{2}\right)  }{q} -  \left( \frac{1}{q^2}+\frac{1}{2 q} \right)\right] y^{1+q}
\label{fsyfisher}
\end{eqnarray}
The steady state
\begin{eqnarray}
{\cal P}_*(y) = P_*(x) \bigg\vert \frac{dx}{dy} \bigg\vert 
=  q \frac{\Gamma(\alpha+\mu)}{ \Gamma(\alpha) \Gamma(\mu) } \frac{ y^{q\mu-1} }
{(1+y^q)^{\alpha+\mu} } 
\ \ \ {\rm for } \ \ y \in ]0,+ \infty[
\label{ysteadyfisher}
\end{eqnarray}
reduces to the Fisher-Snedecor-distribution for $q=1$, but the process is different from the case studied in
the present section in the variable $x$, since the diffusion coefficient ${\cal D} ( y  ) $ and the forces are not linear in $y$.

The observables that have explicit large deviations for their time-averages
can be translated from Eq. \ref{kestenwx} via the change of variable $x=y^{-q}$
\begin{eqnarray}
 w(x) = \frac{ c_1 }{x} + \frac{ c_2 }{1+x} = c_1 y^q +  c_2 \frac{ y^q }{1+y^q}
\label{wxyfisher}
\end{eqnarray}

It is thus interesting to discuss the two special cases 

(1) for $q=1$, the diffusion coefficient is cubic and the force is quadratic
\begin{eqnarray}
{\cal D} ( y  ) && = y^{2} ( 1 +y )
\nonumber \\
{\cal F}_S ( y  ) && =\mu   y -  \left(\alpha- \frac{1}{2}\right)  y^2
\label{q1yfisher}
\end{eqnarray}
while the observables of Eq. \ref{wxyfisher} read
\begin{eqnarray}
 w(x) =  \frac{ c_1 }{x} + \frac{ c_2 }{x+1} = c_1 y +  c_2 \frac{ y }{1+y}
\label{q1wxyfisher}
\end{eqnarray}

(2) for $q=2$, the diffusion coefficient is quartic and the force is cubic 
\begin{eqnarray}
{\cal D} ( y  ) && =  \frac{y^2}{4} ( 1 +y^2 )
\nonumber \\
{\cal F}_S ( y  ) && = \frac{\mu  }{2} y - \frac{ \left(\alpha- \frac{1}{2}\right)  }{2} y^3
\label{q2yfisher}
\end{eqnarray}
and the observables of Eq. \ref{wxyfisher} read
\begin{eqnarray}
 w(x) =  \frac{ c_1 }{x} + \frac{ c_2 }{x+1} = c_1 y^2 +  c_2 \frac{ y^2 }{1+y^2}
\label{q2wxyfisher}
\end{eqnarray}

%%%%%%%%%%%%%%%%%%%%%%%%%%%%%%%%%%%%%%%%%%%

\subsection{ Explicit rate function for the inference of the two parameters $(\alpha,\mu) $ of the Pearson linear force}

The application of Eq. \ref{level2Infer} \ref{rateInferpara}
to the model of Eq. \ref{fisher}
yields that the rate function $I^{Infer} (\hat \alpha,\hat \mu) $ 
associated to the probability $ P_T^{[Infer]} (\hat \alpha,\hat \mu)  $ 
of the two inferred parameters $ (\hat \alpha,\hat \mu) $ reads
\begin{eqnarray}
I^{Infer} (\hat \alpha,\hat \mu)
&& =  \int_0^{+\infty} dx 
P_*^{[\hat \alpha , \hat \mu]} (x)  \frac{ \left[ (\hat \alpha- \alpha)- (\hat \mu - \mu) x \right]^2 }{ 4  x (x+1) }
\nonumber \\
&&
=  \int_0^{+\infty} dx 
\frac{\Gamma(\hat \alpha+\hat \mu)}{ \Gamma(\hat \alpha) \Gamma(\hat \mu) } 
\frac{ x^{\hat \alpha-1} }{(1+x)^{\hat \alpha+\hat \mu} }
\ \frac{(\hat \alpha- \alpha)^2 + (\hat \mu - \mu)^2 x^2
- 2  (\hat \alpha- \alpha) (\hat \mu - \mu) x  }{ 4  x (x+1) }
\nonumber \\
&& 
= \frac{1}{4} \int_0^{+\infty} dx 
\frac{\Gamma(\hat \alpha+\hat \mu)}{ \Gamma(\hat \alpha) \Gamma(\hat \mu) } 
\frac{ x^{\hat \alpha-1} }{(1+x)^{\hat \alpha+\hat \mu} }
\left[ (\hat \mu - \mu)^2 +\frac{(\hat \alpha- \alpha)^2}{x} - \frac{(\hat \alpha+\hat \mu - \alpha-\mu)^2}{1+x} \right]
\nonumber \\
&& =  \frac{1}{4} \left[ (\hat \mu - \mu)^2 
+(\hat \alpha- \alpha)^2\frac{ \hat \mu}{\hat \alpha -1} 
- (\hat \alpha+\hat \mu - \alpha-\mu)^2\frac{\hat \mu}{\hat \alpha+\hat \mu} \right]
\ \ \ {\rm for } \ \ \ \hat \alpha>1 
\label{rateInferparafisher}
\end{eqnarray}
while for $0 < \hat \alpha \leq 1$, the rate function $I^{Infer} (\hat \alpha,\hat \mu) $ is infinite unless $\hat \alpha = \alpha $.

%%%%%%%%%%%%%%%%%%%%%%%%%%%%%%%%%%%%%%%%%%

\section{ Case $D(x)=x (1-x)$ and Beta-distribution
for the steady state $P_*(x)$ on $ ]0,1[ $      }

\label{sec_jacobi}

In this section, we focus on the Pearson diffusion with the quadratic diffusion coefficient $D(x)=x(1-x)$ 
while the steady state $P_*(x)$ is the Beta-distribution 
\begin{eqnarray}
% D(x)  = x(1-x) \ \ \ {\rm and} \ \ 
P_*(x) =
\frac{\Gamma(\alpha+\beta)}{ \Gamma(\alpha) \Gamma(\beta) } x^{\alpha-1} (1-x)^{\beta-1}
 \ \ \ {\rm for } \ x \in ]0,1[
 \ \ \ {\rm with } \ \ \alpha>0 \ \ {\rm and } \ \ \beta>0
\label{jacobi}
\end{eqnarray}
where the two parameters $\alpha>0$ and $\beta>0$
parametrize the two coefficients of the corresponding linear forces of Eq. \ref{fpearson} and \ref{fokkerplancklangevinpearson}
\begin{eqnarray}
 F(x) && =D (x)  \frac{d \ln P_*( x) }{dx} =  (\alpha-1) - (\alpha+\beta-2) x
 \nonumber \\
    F_I(x) && = F(x)+ D' (x) =  \alpha - (\alpha+\beta) x
\nonumber \\
   F_S(x) && = F(x)+  \frac{D' (x)}{2} = \left(\alpha- \frac{1}{2}\right)- (\alpha+\beta-1) x
\label{forcesjacobi}
\end{eqnarray}

The moments $m_k^*$ 
of the steady state $P_*(x)$ of Eq. \ref{jacobi}
can be computed even for non-integer $k \in ]-\alpha,+\infty[ $
\begin{eqnarray}
 m_k^*  =  \int_0^{1} dx x^{k} P_*(x) = 
  \int_0^{+\infty} dx \frac{\Gamma(\alpha+\beta)}{ \Gamma(\alpha) \Gamma(\beta) } x^{k+\alpha-1} (1-x)^{\beta-1}
  = \frac{\Gamma(\alpha+k) \Gamma(\alpha+\beta)}{ \Gamma(\alpha) \Gamma(\alpha+\beta+k) }
  \ \ \ {\rm for } \ \ k \in ]-\alpha,+\infty[
   \label{mksteadyjacobi}
\end{eqnarray}
so that here all the integer moments are finite for $k=1,2,..,+\infty$.

The corresponding process is called the Jacobi diffusion
because the spectral decomposition of the propagator involves the Jacobi polynomials
(see  \cite{pearson_wong,pearson_class,pearson2012,PearsonHeavyTailed,pearson2018}
and references therein).

%%%%%%%%%%%%%%%%%%%%%%%%

\subsection{ Dynamical equations for the moments $m_k(t)$ 
 and for the Laplace transform ${\hat P}_t(s) $ 
}

The dynamical equation of Eq. \ref{dynPearsonmk} for the moment of order $k$
only involves the previous moment $m_{k-1}(t)$ of order $(k-1)$
\begin{eqnarray}
\partial_t m_k(t)  = k \bigg( 1 - k - \alpha-\beta \bigg) m_k(t)
+ k \bigg(   k  +  \alpha -1 \bigg) m_{k-1}(t)
 \label{OavdynPearsonkjacobi}
\end{eqnarray}
The convergence towards its finite steady value of Eq. \ref{mksteadyjacobi}
involves the $k$ relaxation rates $(\epsilon_1,..,\epsilon_k)$ of Eq. \ref{ekmk} 
\begin{eqnarray}
 \epsilon_k =  k \bigg(  k +\alpha+\beta -1  \bigg)
  \label{ekmkjacobi}
\end{eqnarray}

The Laplace transform evolves according to
\begin{eqnarray}
\partial_t {\hat P}_t(s) =  - s^2  \partial_s^2 {\hat P}_t(s)   + s  (\alpha+\beta  +s)  \partial_s {\hat P}_t(s) +  \alpha s   {\hat P}_t(s)
 \label{laplacedynjacobi}
 \end{eqnarray}
% limite incomplete gamma

%%%%%%%%%%%%%%%%%%%%%%%%%%%%%%%%%%%%%%%%%%%

%%%%%%%%%%%%%%%%%%%%%%%%%%%%%%%%%%%%%%%%%%%%

\subsection{ Observables $w(x)$ with explicit large deviations for the time-average $W [x(0 \leq t \leq T) ]  \equiv  \frac{1}{T}   \int_0^T  dt w(x(t)) $  }

 The quantum supersymmetric Hamiltonian of Eq. \ref{hamiltonien}
\begin{eqnarray}
 H = H^{\dagger} =  - \frac{ \partial  }{\partial x} x (1-x) \frac{ \partial  }{\partial x} +V(x)
\label{jacobihamiltonien}
\end{eqnarray}
involves the potential of Eq \ref{vfromu}
\begin{eqnarray}
V(x)  && = \frac{ F^2(x) }{4 D(x) } + \frac{F'(x)}{2} =
\nonumber \\
&& =  \frac{(\alpha-1)^2}{4x}+ \frac{(\beta-1)^2}{4(1-x)} - \frac{(\alpha+\beta)(\alpha+\beta-2)}{4}
\label{jacobivfromu}
\end{eqnarray}
so that the two functions introduced in Eq. \ref{vfromupearsonfracrtion}
are 
\begin{eqnarray}
V_1(x)  && = \frac{1}{x} 
\nonumber \\
V_2(x) && =\frac{1}{1-x}
\label{jacobiv1v2}
\end{eqnarray}

As discussed in detail in section \ref{subsec_pearsonw},
it is interesting to consider the linear conditioned force
\begin{eqnarray}
{\mathring F}^{[p]}(x)   = 
 ({\mathring \alpha}_p-1) - ({\mathring \alpha}_p+{\mathring \beta}_p-2) x
 \ \ \ \ \ {\rm with} \ \ {\mathring \alpha}_p>0 \ \ \ \ {\rm and} \ \  {\mathring \beta}_p >0
  \label{jacobifpearsonring}
\end{eqnarray}
with the restriction of Eq. \ref{restriction} that we repeat here for clarity : 
\begin{eqnarray}
\text {the possibility ${\mathring \alpha}_p \ne \alpha  $ can be considered only in the region $\alpha>1$
but not for $ 0< \alpha \leq 1$ }
\label{jacobirestriction}
\end{eqnarray}
and with the analog restriction concerning the other finite boundary $x_R=1$ :
\begin{eqnarray}
\text {the possibility ${\mathring \beta}_p \ne \beta  $ can be considered only in the region $\beta>1$
but not for $ 0< \beta \leq 1$ }
\label{jacobirestrictionbeta}
\end{eqnarray}

The potential ${\mathring V}^{[p]} (x) $ has the same form as Eq. \ref{jacobivfromu} with different coefficients
\begin{eqnarray}
{\mathring V}^{[p]} (x) = \frac{ \left( {\mathring F}^{[p]}(x)\right)^2 }{4 D(x) }
 + \frac{1}{2} \frac{ d {\mathring F}^{[p]}(x)}{dx}
=  \frac{({\mathring \alpha}_p-1)^2}{4x}
   + \frac{({\mathring \beta}_p-1)^2}{4(1-x)} 
   - \frac{({\mathring \alpha}_p+{\mathring \beta}_p)({\mathring \alpha}_p+{\mathring \beta}_p-2)}{4}
\label{jacobiRingv}
\end{eqnarray}
so that Eq. \ref{susyrefactor} reads
\begin{eqnarray}
p w(x) - E_0(p) && = {\mathring V}^{[p]} (x) - V(x)  
\nonumber \\
&& =  \frac{({\mathring \alpha}_p-1)^2- (\alpha-1)^2}{4x}
+ \frac{ ({\mathring \beta}_p-1)^2- (\beta-1)^2 }{4(1-x)} 
   + \frac{ (\alpha+\beta)(\alpha+\beta-2)- ({\mathring \alpha}_p+{\mathring \beta}_p)({\mathring \alpha}_p+{\mathring \beta}_p-2)}{4}
\label{jacobisusyrefactor}
\end{eqnarray}

This equation can be satisfied 
for the observables $w(x)$ corresponding to linear combinations of Eq. \ref{wexplicit}
with the two functions of Eq. \ref{fisherv1v2}
\begin{eqnarray}
 w(x) = c_1V_1(x)+c_2 V_2(x) = \frac{ c_1 }{x} + \frac{ c_2 }{1-x}
\label{jacobiwx}
\end{eqnarray}
that leads to the system
\begin{eqnarray}
p c_1 && = \frac{({\mathring \alpha}_p-1)^2- (\alpha-1)^2}{4}
\nonumber \\
p c_2 && = \frac{ ({\mathring \beta}_p-1)^2- (\beta-1)^2 }{4} 
\nonumber \\
E_0(p) && =  \frac{  ({\mathring \alpha}_p+{\mathring \beta}_p)({\mathring \alpha}_p+{\mathring \beta}_p-2)
- (\alpha+\beta)(\alpha+\beta-2)}{4}
\label{jacobisusyrefactorsystem}
\end{eqnarray}
The two first equations allow to compute the two parameters ${\mathring \alpha}_p $ 
and ${\mathring \beta}_p $ of the conditioned force of Eq. \ref{jacobifpearsonring} as a function of $p$
\begin{eqnarray}
 {\mathring \alpha}_p  && = (\alpha-1)\sqrt{ 1 + \frac{4 p c_1}{(\alpha-1)^2} } +1
\nonumber \\
 {\mathring \beta}_p && = (\beta-1)\sqrt{ 1 + \frac{4 p c_2}{(\beta-1)^2} } +1
\label{jacobiparametersring}
\end{eqnarray}
that can be plugged into the last equation of the system \ref{jacobisusyrefactorsystem} 
to obtain the energy $E_0(p)$ as a function of $p$
\begin{eqnarray}
&& E_0(p)  =  \frac{ 1}{4}
\bigg[- (\alpha+\beta)(\alpha+\beta-2)
\nonumber \\
&& +  \left((\alpha-1)\sqrt{ 1 + \frac{4 p c_1}{(\alpha-1)^2} } 
+(\beta-1)\sqrt{ 1 + \frac{4 p c_2}{(\beta-1)^2} } +2\right)
\left((\alpha-1)\sqrt{ 1 + \frac{4 p c_1}{(\alpha-1)^2} } 
+(\beta-1)\sqrt{ 1 + \frac{4 p c_2}{(\beta-1)^2} } \right)
 \bigg] \ \ \ \ 
\label{jacobiE0p}
\end{eqnarray}

Let us now focus on the two interesting special cases $[c_1=1,c_2=0]$ and $[c_1=0,c_2=1]$.

%%%%%%%%%%%%%%%%%%%%%%%%%%%%%%%%%%%%%%%%%%%%

\subsubsection{ Case $c_1=1$ and $c_2=0$ : explicit large deviations for the time-average $W [x(0 \leq t \leq T) ]  \equiv  \frac{1}{T}   \int_0^T  dt \frac{1}{x(t) } $ for $\alpha>1$}

For the special case $c_1=1$ and $c_2=0$, the scaled cumulant generating function $E_0(p)$ of Eq. \ref{jacobiE0p}
reduces to
\begin{eqnarray}
E_0(p) && 
= \frac{ 1}{4}
\bigg[- (\alpha+\beta)(\alpha+\beta-2)
+  \left((\alpha-1)\sqrt{ 1 + \frac{4 p }{(\alpha-1)^2} } 
+\beta+1\right)
\left((\alpha-1)\sqrt{ 1 + \frac{4 p }{(\alpha-1)^2} } 
+\beta-1 \right)
 \bigg]
\nonumber \\
&& = p + \frac{(\alpha-1) \beta}{2} \left[\sqrt{ 1 + \frac{4 p }{(\alpha-1)^2} } -1 \right]
\label{jacobiE0pw1}
\end{eqnarray}

For the Legendre transform of Eq. \ref{legendrereci}, one needs to invert
\begin{eqnarray}
W = E_0'(p) = 1 +  \frac{ \beta}{ (\alpha-1) \sqrt{ 1 + \frac{4 p }{(\alpha-1)^2} } }
\label{jacobiw1legendrereci}
\end{eqnarray}
into
\begin{eqnarray}
p= \frac{1}{4} \left[ \frac{\beta^2}{(W-1)^2} - (\alpha-1)^2 \right]
\label{jacobiw1legendrereciinvert}
\end{eqnarray}
that can be plugged into Eq. \ref{legendrereci} to obtain the rate function $I(W)$
\begin{eqnarray}
I(W)   && =  E_0(p)-W p
= \frac{1}{4 (W-1)} \left[ (\alpha+\beta-1 ) - (\alpha-1) W \right]^2 \ \ \ {\rm for } \ \ W \in ]1,+\infty[
\label{jacobiw1Rate}
\end{eqnarray}
that vanishes only for the steady value
\begin{eqnarray}
W_*=  \frac{\alpha+\beta-1}{ \alpha-1 } = m_{-1}^*
\label{jacobiw1Ratevanish}
\end{eqnarray}
and that diverges near the two boundaries as
\begin{eqnarray}
I(W)   &&  \opsimeq_{W \to 1^+}  \frac{\beta^2}{4 (W-1)}
\nonumber \\
I(W)   &&  \opsimeq_{W \to +\infty} \frac{(\alpha-1)^2}{4 } W
\label{jacobiw1RateBoundaries}
\end{eqnarray}

%%%%%%%%%%%%%%%%%%%%%%%%%%%%%%%%%%%

%%%%%%%%%%%%%%%%%%%%%%%%%%%%%%%%%%%%%%%%%%%%

\subsubsection{ Case $c_1=0$ and $c_2=1$ : explicit large deviations for the time-average $W [x(0 \leq t \leq T) ]  \equiv  \frac{1}{T}   \int_0^T  dt \frac{1}{1-x(t) } $ for $\beta>1$}

For the special case $c_1=0$ and $c_2=1$, the scaled cumulant generating function $E_0(p)$ of Eq. \ref{jacobiE0p}
reduces to
\begin{eqnarray}
E_0(p) && 
= 
 \frac{ 1}{4}
\bigg[- (\alpha+\beta)(\alpha+\beta-2)
\nonumber \\
&& +  \left(\alpha+1 
+(\beta-1)\sqrt{ 1 + \frac{4 p }{(\beta-1)^2} } \right)
\left(\alpha-1 
+(\beta-1)\sqrt{ 1 + \frac{4 p }{(\beta-1)^2} } \right)
 \bigg]
\nonumber \\
&&  = p + \frac{(\beta-1) \alpha}{2} \left[\sqrt{ 1 + \frac{4 p }{(\beta-1)^2} } -1 \right]
\label{jacobiE0pw2}
\end{eqnarray}

For the Legendre transform of Eq. \ref{legendrereci}, one needs to invert
\begin{eqnarray}
W = E_0'(p) = 1 +  \frac{ \alpha}{ (\beta-1) \sqrt{ 1 + \frac{4 p }{(\beta-1)^2} } }
\label{jacobiw2legendrereci}
\end{eqnarray}
into
\begin{eqnarray}
p= \frac{1}{4} \left[ \frac{\alpha^2}{(W-1)^2} - (\beta-1)^2 \right]
\label{jacobiw2legendrereciinvert}
\end{eqnarray}
that can be plugged into Eq. \ref{legendrereci} to obtain the rate function $I(W)$
\begin{eqnarray}
I(W)   && =  E_0(p)-W p
= \frac{1}{4 (W-1)} \left[ (\alpha+\beta-1 ) - (\beta-1) W \right]^2 \ \ \ {\rm for } \ \ W \in ]1,+\infty[
\label{jacobiw2Rate}
\end{eqnarray}
that vanishes only for the steady value
\begin{eqnarray}
W_*=  \frac{\alpha+\beta-1}{ \beta-1 } = \int_0^1 dx \frac{P_*(x) }{1-x}
\label{jacobiw2Ratevanish}
\end{eqnarray}
and that diverges near the two boundaries as
\begin{eqnarray}
I(W)   &&  \opsimeq_{W \to 1^+}  \frac{\alpha^2}{4 (W-1)}
\nonumber \\
I(W)   &&  \opsimeq_{W \to +\infty} \frac{(\beta-1)^2}{4 } W
\label{jacobiw2RateBoundaries}
\end{eqnarray}

%%%%%%%%%%%%%%%%%%%%%%%%%%%%%%%%%%%%%%%%%%%%%%

\subsection{ Change of variables $x \to z$ towards the diffusion process $z(t)$ with constant diffusion coefficient $d(z)=1$ }

The change of variables of Eq. \ref{xtoz} 
towards a diffusion process $z(t)$ with constant diffusion coefficient $d(z)=1$ reads
\begin{eqnarray}
z(x) && =\int_0^x \frac{dy}{\sqrt{ D(y)} }= \int_0^x \frac{ dy}{\sqrt{ y(1-y)}} = 2 \arcsin( {\sqrt x } )  \in ]0,\pi[
\nonumber \\
x(z) && = \sin^2 \left(\frac{z}{2} \right) 
\label{xtozjacobi}
\end{eqnarray}
The force of Eq. \ref{fokkerplanckzforce} involves trigonometric functions of $z$
\begin{eqnarray}
 f(z) = \frac{F_S(x) }{\sqrt{ D(x)}} \bigg\vert_{x=x(z)}
 = \left( \alpha- \frac{1}{2} \right) \frac{\cos \left(\frac{z}{2} \right)}{\sin \left(\frac{z}{2} \right)}
 - \left( \beta- \frac{1}{2} \right) \frac{\sin \left(\frac{z}{2} \right)}{\cos \left(\frac{z}{2} \right)} = -u'(z)
\label{fokkerplanckzforcejacobi}
\end{eqnarray}
while the steady state of Eq. \ref{steadyeqz} reads
\begin{eqnarray}
  p_*(z)    \frac{ e^{ - u(z)} }{\int_{0}^{\pi} dz' e^{- u(z') }}  
   = \frac{\Gamma(\alpha+\beta)}{ \Gamma(\alpha) \Gamma(\beta) } 
   \left[\sin \left(\frac{z}{2} \right) \right]^{2 \alpha-1}
   \left[ \cos \left(\frac{z}{2} \right) \right]^{2 \beta-1} \ \ \ {\rm for } \ \ z \in ]0,\pi[
 \label{steadyeqzjacobi}
\end{eqnarray}

The moments $m_k(t)$ of Eq. \ref{mktk} of the initial Pearson process translate into Eq. \ref{mktkz}
\begin{eqnarray}
m_k(t)   = \int_0^{\pi} dz  p_t(z) \left[ \sin^2 \left(\frac{z}{2} \right)  \right]^k
 \label{mktkzjacobi}
\end{eqnarray}

The quantum supersymmetric Hamiltonian of Eq. \ref{hamiltonienz}
\begin{eqnarray}
 h = h^{\dagger} && =\left(   - \frac{ d }{ d z} 
 + \left( \beta- \frac{1}{2} \right) \frac{\sin \left(\frac{z}{2} \right)}{\cos \left(\frac{z}{2} \right)}
 - \left( \alpha- \frac{1}{2} \right) \frac{\cos \left(\frac{z}{2} \right)}{\sin \left(\frac{z}{2} \right)}  \right)
 \left( \frac{ d }{ d z}  + \left( \beta- \frac{1}{2} \right) \frac{\sin \left(\frac{z}{2} \right)}{\cos \left(\frac{z}{2} \right)}
 - \left( \alpha- \frac{1}{2} \right) \frac{\cos \left(\frac{z}{2} \right)}{\sin \left(\frac{z}{2} \right)} \right) 
 \nonumber \\
 && =  - \frac{d^2}{dz^2}  +v(z)
\label{hamiltonienzjacobi}
\end{eqnarray}
involves the potential of Eq. \ref{quantumvz}
\begin{eqnarray}
v(z)  = \frac{ f^2(z) }{4  } + \frac{f'(z)}{2}
 = - \frac{ (\alpha+\beta-1)^2 }{4 }
 + \frac{\left( \alpha- \frac{1}{2} \right)\left( \alpha- \frac{3}{2} \right) }{4 \sin^2 \left(\frac{z}{2} \right)}
 +  \frac{ \left( \beta- \frac{1}{2} \right)\left( \beta- \frac{3}{2} \right) }{4 \cos^2 \left(\frac{z}{2} \right)}
  \label{quantumvzjacobi}
\end{eqnarray}
As discussed in \cite{us_kemeny}, the special cases ($\alpha=\frac{1}{2}$ or $\alpha=\frac{3}{2}$) with ($\beta=\frac{1}{2}$ or $\beta=\frac{3}{2}$)
where the quantum potential $v(z)$ reduces to a constant 
can be interpreted respectively as the pure diffusion in the presence of reflecting boundaries or 
in the presence of absorbing boundaries 
with conditioning to survive forever, known as Taboo processes 
\cite{refKnight,refPinskyTaboo,ref_GarbaczewskiTaboo,refAlainTaboo,ref_Adorisio}.

Finally, the observables that have explicit large deviations for their time-averages
can be translated from Eq. \ref{jacobiwx} via the change of variables of Eq. \ref{xtozjacobi}
\begin{eqnarray}
 w(x) = \frac{ c_1 }{x} + \frac{ c_2 }{1-x} =  \frac{ c_1 }{\sin^2 \left(\frac{z}{2} \right)} + \frac{ c_2 }{\cos^2 \left(\frac{z}{2} \right)}
\label{wxzjacobi}
\end{eqnarray}
or equivalently, they correspond to the two functions of $z$ appearing
in the quantum potential $v(z)$ of Eq. \ref{quantumvzjacobi}.

%%%%%%%%%%%%%%%%%%%%%%%%%%%%%%%%%%%%%%%%%%%

\subsection{ Explicit rate function for the inference of the two parameters $(\alpha,\beta) $ of the Pearson linear force}

The application of Eq. \ref{level2Infer} \ref{rateInferpara}
to the model of Eq. \ref{jacobi}
yields that the rate function $I^{Infer} (\hat \alpha,\hat \beta) $ 
associated to the probability $ P_T^{[Infer]} (\hat \alpha,\hat \beta)  $ 
of the two inferred parameters $ (\hat \alpha,\hat \beta) $ reads
\begin{eqnarray}
I^{Infer} (\hat \alpha,\hat \beta)
&& =  \int_0^{+\infty} dx 
P_*^{[\hat \alpha , \hat \beta]} (x)
\frac{ \left[ (\hat \alpha- \alpha)^2 - (\hat \alpha+\hat \beta - \alpha- \beta) x \right]^2 }{ 4  x (1-x) }
\nonumber \\
&& 
=  \int_0^{+\infty} dx 
\frac{\Gamma(\hat \alpha+\hat \beta)}{ \Gamma(\hat \alpha) \Gamma(\hat \beta) } 
 x^{\hat \alpha-1} (1-x)^{\hat \beta-1}
\frac{ (\hat \alpha- \alpha)^2 + (\hat \alpha+\hat \beta - \alpha- \beta)^2 x^2
- 2  (\hat \alpha- \alpha) (\hat \alpha+\hat \beta - \alpha- \beta) x }{ 4  x (1-x) }
\nonumber \\
&& 
=  \frac{1}{4} \int_0^{+\infty} dx 
\frac{\Gamma(\hat \alpha+\hat \beta)}{ \Gamma(\hat \alpha) \Gamma(\hat \beta) } 
 x^{\hat \alpha-1} (1-x)^{\hat \beta-1}
\left[ \frac{ (\hat \alpha- \alpha)^2}{x}+ \frac{ (\hat \beta- \beta)^2}{1-x}  - (\hat \alpha+\hat \beta - \alpha- \beta)^2 
\right]
\nonumber \\
&& =  \frac{1}{4} \left[  (\hat \alpha- \alpha)^2 \frac{\hat \alpha+\hat \beta-1}{\hat \alpha-1}
+  (\hat \beta- \beta)^2 \frac{\hat \alpha+\hat \beta-1}{\hat \beta-1}
 - (\hat \alpha+\hat \beta - \alpha- \beta)^2  \right]
 \ \ \ {\rm for } \ \ \ \hat \alpha>1  \ \ \ {\rm and } \ \ \ \hat \beta>1 
\label{rateInferparajacobi}
\end{eqnarray}
while for $0 < \hat \alpha \leq 1$, the rate function $I^{Infer} (\hat \alpha,\hat \beta) $ is infinite unless $\hat \alpha = \alpha $, and while for $0 < \hat \beta \leq 1$, the rate function $I^{Infer} (\hat \alpha,\hat \beta) $ is infinite unless $\hat \beta = \beta $.

%%%%%%%%%%%%%%%%%%%%%%%%%%%%%%%%%%%%%%%%%

\section{ Case $D(x)=1+x^2$ and generalized-Student-distribution
for $P_*(x)$ on $ ]-\infty,+\infty[ $      }

\label{sec_student}

In this section, we focus on the Pearson diffusion with the quadratic diffusion coefficient $D(x)=1+x^2$ 
while the steady state $P_*(x)$ is the generalized-Student-distribution 
\begin{eqnarray}
% D(x)  = 1+x^2  \ \ \ {\rm and} \ \ 
P_*(x) = \frac{\Gamma(\frac{\mu+1}{2}) }
{\Gamma(\frac{1}{2})\Gamma(\frac{\mu}{2}) \left( 1+x^2 \right)^{\frac{1+\mu}{2}}}
  \ \ \ {\rm } \ \ {\rm for } \ \ x \in ]-\infty,+\infty[
\ \ \ {\rm with } \ \ \mu>0
\label{student}
\end{eqnarray}
where the parameter $\mu>0$ 
parametrizes the corresponding linear forces of Eq. \ref{fpearson} and \ref{fokkerplancklangevinpearson}
\begin{eqnarray}
 F(x) && =D (x)  \frac{d \ln P_*( x) }{dx} = - (1+\mu) x
 \nonumber \\
    F_I(x) && = F(x)+ D' (x) = - (\mu -1) x
\nonumber \\
   F_S(x) && = F(x)+  \frac{D' (x)}{2} =  - \mu x 
\label{forcestudent}
\end{eqnarray}
Here it is important to stress that we have chosen to 
discuss only the case with no constant term $\lambda=0$ in these forces.
Indeed, the presence of a non-vanishing constant $\lambda \ne 0$ introduces
 the supplementary factor $e^{\lambda \arctan(x) } $
in the steady state $P_*(x)$ that complicates various analytical computations and 
somewhat obscures the physical meaning of various observables.
In addition, the case $\lambda=0$ seems more interesting for physical applications
 with its symmetry $x \to -x$
since the steady state $P_*(x)$ of Eq. \ref{student}
generalizes the Cauchy distribution corresponding to 
the special case $\mu=1$
\begin{eqnarray}
\text{ Case $\mu=1$} : \ \  
P_*(x) = \frac{1 }{ \pi ( 1+x^2 )}  \ \ \ {\rm } \ \ {\rm for } \ \ x \in ]-\infty,+\infty[
\label{cauchy}
\end{eqnarray}

The corresponding process is called the Student process for $\lambda=0$ 
and the skew-Student process
for $\lambda \ne 0$ (see \cite{pearson_wong,pearson_class,pearson2012,PearsonHeavyTailed,pearson2018,pearson_student_processes,pearson_student}
and references therein).

%%%%%%%%%%%%%%%%%%%%%%%%%%%%%%%%%%%%%

\subsection{ Dynamical equations for the moments $m_k(t)$ and for the Fourier transform ${\tilde P}_t(q) $ }

The dynamical equation of Eq. \ref{dynPearsonmk} for the moment $m_k(t)$ of order $k$
only involves the moment $m_{k-2}(t)$ of order $(k-2)$
\begin{eqnarray}
\partial_t m_k(t)  = k ( k- \mu ) m_k(t) +  k (k-1) m_{k-2}(t)
 \label{studentdynPearsonmk}
\end{eqnarray}
and the characteristic rate $\epsilon_k$ of Eq. \ref{ekmk} is positive only for $ k < \mu$
\begin{eqnarray}
 \epsilon_k = k \bigg( \mu -  k \bigg)
\ \ \ i.e. \ \  \begin{cases}
\text{ exponential relaxation as $e^{ - t k (\mu-k) } $  for $k<\mu$ } \\
 \text{ exponential growth as $e^{  t k (k-\mu) }  $  for $k>\mu$ }
\end{cases}
 \label{ekstudent}
\end{eqnarray}

For a non-vanishing initial condition $m_1(0) \ne 0$,
the dynamics for the first moment $m_1(t)$ of Eq. \ref{dynPearsonm1integ}
reduces to
\begin{eqnarray}
  m_1(t)    =    m_1(0)   e^{-  t (\mu-1)}  
\ i.e.   \begin{cases}
\text{ exponential relaxation as $e^{-  t (\mu-1)}  $ towards $m_1^*=0  $ for $\mu >1$ } \\
 \text{ exponential growth as $e^{  t (1-\mu)}  $  if $0<\mu <1$}
\end{cases}
 \label{studentdynPearsonm1integ}
\end{eqnarray}
More generally, the moment $m_k(t)$ for odd $k=2k'+1$ will converge towards zero for $k<\mu$
and towards infinity for $k>\mu$.

The dynamics for the second moment $m_2(t)$ of Eq. \ref{dynPearsonm2integ}
reduces to
\begin{eqnarray}
 m_2(t)  && =  e^{- 2 (\mu-2) t}  m_2(0)
 + \frac{1- e^{ -2 ( \mu-2 ) t} }{ ( \mu-2) } 
\ i.e. 
  \begin{cases}
\text{ exponential relaxation towards  $ m_2^* =\frac{1 }{ ( \mu-2) } $ if $\mu>2$}
 \\
\text{ exponential growth as $e^{  t 2 (2-\mu)}  $  if $0<\mu <2$ }
\end{cases}
 \label{studentdynPearsonm2integ}
\end{eqnarray}
More generally, the moment $m_k(t)$ for even $k=2 k'$ will converge towards 
its finite steady value only for $k<\mu$
\begin{eqnarray}
 m_{k=2 k'}^*  =  \int_{-\infty}^{+\infty} dx x^{2 k'} P_*(x) 
  \int_{-\infty}^{+\infty} dx x^{2k'}
 \frac{\Gamma(\frac{\mu+1}{2})}{\Gamma(\frac{1}{2})\Gamma(\frac{\mu}{2}) \left[ 1+x^2 \right]^{\frac{1+\mu}{2}}}
 =\frac{\Gamma(\frac{1}{2}+k')\Gamma(\frac{\mu}{2}-k')}{\Gamma(\frac{1}{2})\Gamma(\frac{\mu}{2}) }
 \ \ {\rm for } \ \ k=2k' < \mu
 \label{mksteadystudenteven}
\end{eqnarray}

The Fourier transform ${\tilde P}_t(q) $ satisfies the dynamical equation of Eq. \ref{fourierdyn}
\begin{eqnarray}
\partial_t {\tilde P}_t(q)
 =    q^2 \partial_q^2   {\tilde P}_t(q) + (1-\mu)   q \partial_q  {\tilde P}_t(q) - q^2    {\tilde P}_t(q)
 \label{fourierdynstudent}
\end{eqnarray}
and converges towards its steady state value 
\begin{eqnarray}
 {\tilde P}_{\infty}(q)
 =   \int_{-\infty}^{+\infty} dx e^{i q x} 
 \frac{\Gamma(\frac{\mu+1}{2})}{\Gamma(\frac{1}{2})\Gamma(\frac{\mu}{2}) \left[ 1+x^2 \right]^{\frac{1+\mu}{2}}}
 = \frac{2^{1-\frac{\mu}{2}}}{\Gamma(\frac{\mu}{2})} \vert q \vert^{\frac{\mu}{2}} K_{\frac{\mu}{2}} (\vert q \vert) 
 \label{fourierdynstudentsteady}
\end{eqnarray}
that simplifies for the Cauchy case $\mu=1$ of Eq. \ref{cauchy}
into
\begin{eqnarray}
\mu=1 \ \ \ : \ \ \  {\tilde P}_{\infty}(q)
 =   \int_{-\infty}^{+\infty} dx e^{i q x}  \frac{1}{\pi ( 1+x^2) }
 = e^{- \vert q \vert}
 \label{fourierdyncauchysteady}
\end{eqnarray}

%%%%%%%%%%%%%%%%%%%%%%%%%%%%%%%%%%%

\subsection{ Observable $w(x)$ with explicit large deviations for the time-average $W [x(0 \leq t \leq T) ]  \equiv  \frac{1}{T}   \int_0^T  dt w(x(t)) $  }

 The quantum supersymmetric Hamiltonian of Eq. \ref{hamiltonien}
\begin{eqnarray}
 H = H^{\dagger} =  - \frac{ \partial  }{\partial x} (1+x^2) \frac{ \partial  }{\partial x} +V(x)
\label{studenthamiltonien}
\end{eqnarray}
involves the potential of Eq \ref{vfromu}
\begin{eqnarray}
V(x)   = \frac{ F^2(x) }{4 D(x) } + \frac{F'(x)}{2}
=\frac{\mu^2-1}{4} - \frac{  (1+\mu)^2}{ 4 (1+x^2)}   
\label{studentvfromu}
\end{eqnarray}
The two functions introduced in Eq. \ref{vfromupearsonfracrtion} are 
\begin{eqnarray}
V_1(x)   =\frac{1}{1+x^2}
\label{studentv1}
\end{eqnarray}
that appears in Eq. \ref{studentvfromu} 
and $V_2(x)   =\frac{x}{1+x^2} $
that does not appear in Eq. \ref{studentvfromu} as a consequence of our choice $\lambda=0$ 
discussed after Eq. \ref{forcestudent}.

The potential ${\mathring V}^{[p]} (x) $ associated to the linear conditioned force
\begin{eqnarray}
{\mathring F}^{[p]}(x)   =  - (1+ {\mathring \mu}_p) x 
\label{studentfpearsonring}
\end{eqnarray}
has the same form as Eq. \ref{studentvfromu} 
\begin{eqnarray}
{\mathring V}^{[p]} (x) = \frac{ \left( {\mathring F}^{[p]}(x)\right)^2 }{4 D(x) }
 + \frac{1}{2} \frac{ d {\mathring F}^{[p]}(x)}{dx}
=  \frac{{\mathring \mu}_p^2-1}{4} - \frac{  (1+{\mathring \mu}_p)^2}{ 4 (1+x^2)}    
\label{studentRingv}
\end{eqnarray}
so that Eq. \ref{susyrefactor} reads
\begin{eqnarray}
p w(x) - E_0(p) = {\mathring V}^{[p]} (x) - V(x)  
=   \frac{{\mathring \mu}_p^2-\mu^2}{4} + \frac{ (1+\mu)^2 -(1+{\mathring \mu}_p)^2}{ 4 (1+x^2)}    
\label{studentsusyrefactor}
\end{eqnarray}
This equation can be satisfied for the observable corresponding to $V_1(x)$ of Eq. \ref{studentv1}
\begin{eqnarray}
 w(x) = V_1(x)= \frac{ 1 }{1+x^2}
\label{studentwx}
\end{eqnarray}
that leads to the system
\begin{eqnarray}
p  && =  \frac{ (1+\mu)^2 -(1+{\mathring \mu}_p)^2}{ 4 }
\nonumber \\
E_0(p) && =  \frac{\mu^2- {\mathring \mu}_p^2}{4} 
\label{studentsusyrefactorsystem}
\end{eqnarray}
The first equations allow to compute the parameter ${\mathring \mu}_p $ 
of the conditioned force of Eq. \ref{studentfpearsonring} as a function of $p$
\begin{eqnarray}
  {\mathring \mu}_p  && = (\mu+1)\sqrt{ 1 - \frac{4 p }{(\mu+1)^2} } -1
\label{studentparametersring}
\end{eqnarray}
that can be plugged into the last equation of the system \ref{studentsusyrefactorsystem} 
to obtain the energy $E_0(p)$ as a function of $p$
\begin{eqnarray}
E_0(p) && =  \frac{\mu^2- {\mathring \mu}_p^2}{4} 
= \frac{1}{4} \left[ \mu^2- \left(  (\mu+1)\sqrt{ 1 - \frac{4 p }{(\mu+1)^2} } -1\right)^2  \right]
\label{studentE0p}
\end{eqnarray}

For the Legendre transform of Eq. \ref{legendrereci}, one needs to invert
\begin{eqnarray}
W = E_0'(p) =    1- \frac{    1  }{ (1+\mu) \sqrt{ 1 - \frac{4 p }{(\mu+1)^2} } }
\label{studentw1legendrereci}
\end{eqnarray}
into
\begin{eqnarray}
p= \frac{1}{4} \left( (\mu+1)^2- \frac{1}{(1-W)^2} \right)
\label{studentw1legendrereciinvert}
\end{eqnarray}
that can be plugged into Eq. \ref{legendrereci} to obtain the rate function $I(W)$
\begin{eqnarray}
I(W)    =  E_0(p)-W p 
= \frac{1}{4 (1-W) } \left[ \mu-(1+\mu) W \right]^2 \ \ \ {\rm for } \ \ W \in ]0,1[
\label{studentw1Rate}
\end{eqnarray}
that vanishes only for the steady value
\begin{eqnarray}
W_*=  \frac{\mu}{ 1+\mu } = \int_{-\infty}^{+\infty} dx \frac{ P_*(x) }{1+x^2}
\label{studentw1Ratevanish}
\end{eqnarray}
and that diverges at $W\to 1^-$
\begin{eqnarray}
I(W)     \opsimeq_{W \to 1^-}  \frac{1}{4 (1-W) } 
\label{studentw2RateBoundaries}
\end{eqnarray}
while it remains finite at the boundary $W =0$
\begin{eqnarray}
I(W=0)  =  \frac{\mu^2}{4  } 
\label{studentw2RateZero}
\end{eqnarray}

%%%%%%%%%%%%%%%%%%%%%%%%%%%%%%%%%%%%%%%%%%%

\subsection{ Change of variables $x \to z$ towards the diffusion process $z(t)$ with constant diffusion coefficient $d(z)=1$ }

The change of variables of Eq. \ref{xtoz} 
towards a diffusion process $z(t)$ with constant diffusion coefficient $d(z)=1$ reads
\begin{eqnarray}
z && = \int_0^x \frac{dy}{\sqrt{ D(y)} }=\int_0^x \frac{dy}{\sqrt{1+y^2} } = {\rm arcsinh } (x) \in ]-\infty,+\infty[
\nonumber \\
x && = \sinh z
\label{xtozstudent}
\end{eqnarray}
The force of Eq. \ref{fokkerplanckzforce} involves the hyperbolic tangent of $z$
\begin{eqnarray}
 f(z)  =\frac{F_S(x) }{\sqrt{ D(x)}} \bigg\vert_{x=x(z)} =  -\mu \tanh z  = - u'(z)
\label{fokkerplanckzforcestudent}
\end{eqnarray}
while the steady state of Eq. \ref{steadyeqz}
\begin{eqnarray}
 p^*(z) = \frac{e^{-u(z)} }{ \int_{-\infty}^{+\infty} dz' e^{- u(z') }} = 
 \frac{\Gamma(\frac{\mu+1}{2}) }
{\Gamma(\frac{1}{2})\Gamma(\frac{\mu}{2}) [\cosh z]^{\mu} } \ \ \ {\rm for } \ \ z \in ]-\infty,+\infty[
\label{zsteadystudent}
\end{eqnarray}

The quantum supersymmetric Hamiltonian of Eq. \ref{hamiltonienz}
\begin{eqnarray}
 h = h^{\dagger} && =\left(   - \frac{ d }{ d z} + \frac{\mu}{2}  \tanh z  \right)
 \left( \frac{ d }{ d z} + \frac{\mu}{2}  \tanh z  \right) =  - \frac{d^2}{dz^2}  +v(z)
\label{hamiltonienzstudent}
\end{eqnarray}
involves the potential of Eq. \ref{quantumvz}
\begin{eqnarray}
v(z) 
&& = \frac{ f^2(z) }{4  } + \frac{f'(z)}{2}
 = \frac{ \mu^2}{4} - \frac{  \mu (\mu+2)    }{4\cosh ^2 z}
 \label{quantumvzstudent}
\end{eqnarray}
known as the symmetric Pöschl–Teller potential (see the review \cite{review_susyquantum} and references therein).

Its asymptotic behaviors for $z \to \pm \infty$
\begin{eqnarray}
  v(z)  && \opsimeq_{z \to \pm \infty} \frac{   \mu^2   }{4  } 
  \label{largezquantumvzstudent}
\end{eqnarray}
yields that there is a continuous spectrum of the form of Eq \ref{continuum} with 
the lower boundary $v_{\infty}=\frac{   \mu^2   }{4  } $
\begin{eqnarray}
 \text{ Continuous spectrum :} \ \ \ \ E \in ] v_{\infty} = \frac{   \mu^2   }{4  },+\infty[
\label{continuumstudent}
\end{eqnarray}
besides the finite number of discrete levels of Eq. \ref{eigendiscretepos}
\begin{eqnarray}
 E_n = n \bigg( \mu -  n \bigg)  \ \ \ {\rm for } \ \  0 \leq n < \frac{\mu}{2}
  \label{enstudent}
\end{eqnarray}

Finally, the observable that has explicit large deviations for its time-average
can be translated from Eq. \ref{studentwx} via the change of variables of Eq. \ref{xtozstudent}
\begin{eqnarray}
 w(x) = \frac{ 1 }{1+x^2} = \frac{ 1 }{\cosh ^2 z} 
\label{wxzstudent}
\end{eqnarray}
or equivalently, the function of $z$ appearing
in the quantum potential $v(z)$ of Eq. \ref{quantumvzstudent}.

%%%%%%%%%%%%%%%%%%%%%%%%%%%%%%%%%%%%%%%%%%%

\subsection{ Explicit rate function for the inference of the parameter $\mu $ }

The application of Eq. \ref{level2Infer} \ref{rateInferpara}
to the model of Eq. \ref{student}
yields that the rate function $I^{Infer} ( \hat \mu)  $ 
associated to the probability $ P_T^{[Infer]}( \hat \mu)  $ 
of the inferred parameter $ \hat \mu $ reads
\begin{eqnarray}
I^{Infer} ( \hat \mu) 
\int_{-\infty}^{+\infty} dx  P_*^{[\hat \mu]} (x ) \frac{(\hat \mu - \mu)^2 x^2  }{ 4  ( x^2+1 ) }
=   \int_{-\infty}^{+\infty} dx 
\frac{\Gamma(\frac{\hat \mu+1}{2}) }
{\Gamma(\frac{1}{2})\Gamma(\frac{\hat \mu}{2}) \left( 1+x^2 \right)^{\frac{1+\hat \mu}{2}}}
\ \ \frac{(\hat \mu - \mu)^2 x^2  }{ 4  ( x^2+1 ) }
 =   \frac{(\hat \mu - \mu)^2   }{ 4  (\hat \mu+1) }
\label{rateInferparastudent}
\end{eqnarray}

%%%%%%%%%%%%%%%%%%%%%%%%%%%%%%%%%%%%%%%%%%%%

\section{ Conclusions }

\label{sec_conclusion}

In this paper, after recalling the very specific properties of Pearson diffusions
for the dynamics of integer moments,
for the spectral decomposition of the propagator,
and for the associated quantum supersymmetric Hamilonians,
we have analyzed in detail their large deviations properties.
For time-averaged observables over the time-window $[0,T]$,
we have written the first rescaled cumulants for generic observables, with various simple examples,
and we have determined that the specific observables whose large deviations are explicit
involve the elementary functions that appear in the quantum potential of the associated quantum supersymmetric Hamiltonian.
Then the explicit large deviations at level 2 for the empirical density 
seen during a large time-window $[0,T]$ have been used to analyze
the statistics of the inferred parameters from the data of a long stochastic trajectory.
Finally, this general framework has been applied to the five representative examples of Pearson diffusions
with linear or quadratic diffusion coefficient $D(x)$, where the steady state
corresponds to the Gamma-distribution, the Beta-distribution, the heavy-tailed Inverse-Gamma-distribution, the heavy-tailed Fisher-Snedecor-distribution, and the heavy-tailed Student-distribution.

As a final remark, let us mention that the analysis via supersymmetric quantum mechanics
is also useful in one-dimension non-equilibrium diffusions,
either on the periodic ring \cite{c_lyapunov}
or for boundary-driven non-equilibrium models \cite{c_boundarydriven}.

%%%%%%%%%%%%%%%%%%%%%%%%%%%%%%%%%%%%%%%%%%%%%

\appendix

\section{ Pearson family : closed dynamics for the 
Laplace transform ${\hat P}_t(s) $ or the Fourier transform ${\tilde P}_t(q) $ }

\label{app_laplace}

In this Appendix, we mention another important specific property of Pearson diffusions
concerning the dynamics of the Laplace transform or the Fourier transform.

\subsection{Closed dynamics for the 
Laplace transform ${\hat P}_t(s) $ when $x \in ]0,+\infty[$ } 

When the variable $x$ remains positive as in many examples of Pearson diffusions,
it is convenient to consider the Laplace transform of parameter $s$ that 
corresponds to the average of observable $w(x)=e^{-s x}$ in Eq. \ref{Oav},
so that its series expansion in $s$ involves  all the integer moments $m_k(t)$ of Eq. \ref{mktk}
\begin{eqnarray}
{\hat P}_t(s) \equiv  \int_{x_L}^{x_R} dx e^{-s x} P_t(x) = \sum_{k=0}^{+\infty} \frac{(-s)^k}{k!} m_k(t)
 \label{laplace}
\end{eqnarray}
The dynamical equation of Eq. \ref{OavdynPearson}
\begin{eqnarray}
\partial_t {\hat P}_t(s)
&&  = \int_{x_L}^{x_R} dx    P_t(x)  \bigg[ (  \lambda_I - \gamma_I x \bigg) (-s e^{-s x}) 
+ \bigg(a x^2+b x +c  \bigg) (s^2 e^{-s x})  \bigg] 
 \nonumber \\
 && = \int_{x_L}^{x_R} dx    P_t(x)  \bigg[
-s (  \lambda_I + \gamma_I \partial_s \bigg)  e^{-s x} 
+ s^2 \bigg(a \partial_s^2-b \partial_s +c  \bigg)  e^{-s x}   \bigg]
  \nonumber \\
 && =    - s \bigg(     \lambda_I +  \gamma_I    \partial_s \bigg)  {\hat P}_t(s)
 +s^2 \bigg( a  \partial_s^2 - b  \partial_s + c  \bigg)  {\hat P}_t(s)
   \nonumber \\
 && =   a  s^2 \partial_s^2  {\hat P}_t(s)- (b s +\gamma_I) s \partial_s {\hat P}_t(s) + (c s-\lambda_I) s   {\hat P}_t(s)
 \label{laplacedyn}
\end{eqnarray}
involves the Laplace transform ${\hat P}_t(s) $ itself and its two first derivatives with respect to $s$.

In particular, the steady version of Eq. \ref{laplacedyn} yield that the 
Laplace transform of the steady state $P_*(x)$
\begin{eqnarray}
 {\hat P}_*(s) \equiv  \int_{x_L}^{x_R} dx e^{-s x} P_*(x)
 \label{laplacesteady}
\end{eqnarray}
satisfy the second-order differential equation 
\begin{eqnarray}
0 =  s \left[ a  s {\hat P}_*''(s)  - (b s +\gamma_I)  {\hat P}_*'(s) + (c s-\lambda_I)    {\hat P}_*(s) \right]
 \label{laplacesteadyeq}
\end{eqnarray}
% where the three coefficients contains terms in $s$ and $s^2$.

%%%%%%%%%%%%%%%%%%%%%%%%%%%%%%%%%

\subsection{Closed dynamics for the 
Fourier transform ${\hat P}_t(s) $ when $x\in ]-\infty,+\infty[$ }

When the variable $x \in ]-\infty,+\infty[$, 
it is convenient to replace the Laplace transform of Eq. \ref{laplace}
by the Fourier transform via $s=-i q$
\begin{eqnarray}
{\tilde P}_t(q) \equiv   \int_{x_L}^{x_R} dx e^{iqx} P_t(x)
 = \sum_{k=0}^{+\infty} \frac{(iq)^k}{k!} m_k(t)
 \label{fourier}
\end{eqnarray}
to obtain the dynamical equation
\begin{eqnarray}
\partial_t {\tilde P}_t(q)
=    a q^2 \partial_q^2  {\tilde P}_t(q) +(i b q - \gamma_I) q \partial_q {\tilde P}_t(q)+( i  \lambda_I - c q) q {\tilde P}_t(q)
 \label{fourierdyn}
\end{eqnarray}
with its steady version for the Fourier transform ${\tilde P}_t(q) $ of the steady state $P_*(x)$
\begin{eqnarray}
0
=   q \bigg[ a q {\tilde P}_*''(q) +(i b q - \gamma_I) {\tilde P}_*'(q) +( i  \lambda_I - c q) {\tilde P}_*(q)\bigg]
 \label{fouriersteadyeq}
\end{eqnarray}

%%%%%%%%%%%%%%%%%%%%%%%%%%%%%%%%%%%%%%%%%%%%%

\end{document}